\documentclass[letterpaper,11pt]{article}
\pdfoutput=1 

\usepackage{jheppub} 

\usepackage{graphicx}
\usepackage{epstopdf}
\usepackage{amsmath, amssymb}

\usepackage{tikz}
\usepackage{verbatim}
\usepackage{bm}
\usepackage{bbold}
\usepackage{caption}
\usepackage{subcaption}

\usepackage{youngtab}
\usepackage{ytableau}

\newcommand{\be}{\begin{eqnarray}}
\newcommand{\ee}{\end{eqnarray}}
\newcommand{\bea}{\begin{eqnarray}}
\newcommand{\eea}{\end{eqnarray}}

\newcommand{\nn}{\nonumber}
\newcommand{\bn}{\begin{enumerate}}
\newcommand{\en}{\end{enumerate}}

\def\IZ{\mathbb{Z}}

\def\CC{{\cal C}}

\def\CF{{\cal F}}

\def\CL{{\cal L}}
\def\CM{{\cal M}}
\def\CN{{\cal N}}

\def\CS{{\cal S}}
\def\CT{{\cal T}}

\def\a{\alpha}
\def\b{\beta}

\def\e{\epsilon}

\def\s{\sigma}

\def\half{\frac{1}{2}}

\def\vev#1{\langle #1 \rangle}

\def\Tr{{\rm Tr}}
\def\tr{{\rm Tr}}

\newcommand\SU{{\rm SU}}

\def\vec#1{\bm{#1}}
\def\fp{\mathfrak{p}}
\def\fq{\mathfrak{q}}

\title{Quiver Tails and $\CN=1$ SCFTs from M5-branes}

\author[a]{Prarit Agarwal,}
\author[b, c]{Ibrahima Bah,}
\author[d]{Kazunobu Maruyoshi}
\author[a]{and Jaewon Song}

\affiliation[a]{Department of Physics, University of California, San Diego \\La Jolla, CA 92093, USA}
\affiliation[b]{Department of Physics and Astronomy, University of Southern California\\ Los Angeles, CA 90089, USA}
\affiliation[c]{Institut de Physique Th\'eorique, CEA/Saclay, 91191 Gif-sur-Yvette, France}
\affiliation[d]{California Institute of Technology, Pasadena, CA 91125, USA}

\emailAdd{pagarwal@physics.ucsd.edu}
\emailAdd{bah@usc.edu}
\emailAdd{maruyosh@caltech.edu}
\emailAdd{jsong@physics.ucsd.edu}

\preprint{UCSD-PTH-14-04, CALT-TH-2014-155}

\abstract{
We study a class of four-dimensional $\CN=1$ superconformal field theories obtained by wrapping M5-branes on a Riemann surface with punctures. We identify four-dimensional UV descriptions of the SCFTs corresponding to curves with a class of punctures.  The quiver tails appearing in these UV descriptions differ significantly from their $\CN=2$ counterpart.  We find a new type of object that we call the `Fan'.  We show how to construct new $\CN=1$ superconformal theories using the Fan. Various dual descriptions for these SCFTs can be identified with different colored pair-of-pants decompositions. For example, we find an $\CN=1$ analog of Argyres-Seiberg duality for the $SU(N)$ SQCD with $2N$ flavors. We also compute anomaly coefficients and superconformal indices for these theories and show that they are invariant under dualities. 
}

\begin{document} 
\maketitle
\flushbottom


\section{Introduction}
 Six-dimensional $(2,0)$ theory, as the low energy effective theory on the M5-brane worldvolume, 
 plays a crucial role in studying lower dimensional supersymmetric gauge theories.
  In particular, a large class of four-dimensional $\CN=2$ superconformal theories, 
  which are called class $\CS$ theories, have been discovered in \cite{Gaiotto:2009we, Gaiotto:2009hg} 
  as a compactification of the $(2,0)$ theory on a Riemann surface with a partial twist.  Class $\CS$ theories
  turn out to be related to various objects in different dimensional theories 
  \cite{Alday:2009aq, Gadde:2009kb}, bridged by the (2,0) theory picture.

  $\CN=2$ class $\CS$ theories are included in a larger class of theories with $\CN=1$ supersymmetry
  associated to compactifications of the $(2,0)$ theory \cite{Bah:2012dg}.
  This latter class, which we will call $\CN=1$ class $\CS$, 
  has been investigated
  in \cite{Bah:2011je, Bah:2011vv,Bah:2012dg, Beem:2012yn,Gadde:2013fma, Xie:2013gma, Bah:2013aha, Agarwal:2013uga} in field theory and in \cite{Bah:2011vv,Bah:2012dg,Bah:2013qya,Bah:2013wda} in AdS/CFT
  (see \cite{Maruyoshi:2009uk, Benini:2009mz} for the mass deformed $\CN=2$ class $\CS$ theories).
  The theories in this class flow to superconformal fixed points in the IR.
  See also \cite{Maruyoshi:2013hja, Bonelli:2013pva, Xie:2013rsa, Yonekura:2013mya}
  for theories in Coulomb and confining phases.
  
  The $\CN=1$ theories of class $\CS$ are specified through the following data 
    \begin{itemize}
    \item The choice of `gauge group' $\Gamma = A, D, E$. 
    \item A Riemann surface $\CC_{g, n}$ of genus $g$ with $n$ punctures called UV curve. 
    \item Two integers $p, q$ with a constraint $p+q = 2g-2+n$. 
    \end{itemize}
  From the M-theory point of view, this class of theories is obtained by wrapping M5-branes on
  $\CC_{g, n}$ inside the total space of two line bundles over $\CC_{g, n}$.
  Then, $p$ and $q$ are the degrees of the two line bundles.\footnote{In general, to preserve supersymmetry, the normal bundle over the Riemann surface needs to be a rank-2 bundle whose determinant line bundle is the canonical bundle. Here we restrict ourselves to a particular case where the normal bundle simply decomposes as a sum of two line bundles.}
  
  In addition, we assign data to each puncture. 
  A class of punctures, called the regular colored $\CN=2$ punctures,
  are specified by the following data:
    \begin{itemize}
    \item For each puncture, the choice of $\rho_i$ which is an embedding of $SU(2)$ into $\Gamma$. 
    \item The choice of $\IZ_2$-valued `color' $\s_i = \pm$. 
    \end{itemize}
  When $\Gamma = A_{N-1}$ which we will focus on, the choice of $\rho_i$ is in one-to-one correspondence 
  with the choice of partition of $N$ or a Young diagram of $N$ 
  boxes\footnote{Punctures can also be twisted by an outer-automorphism group of $\Gamma$. 
              This will affect the choice of $\rho_i$.
              We will not consider the twist in this paper.}
  with $N = \sum_{k} n_{k} k$.
  The monicker `colored $\CN=2$ puncture' stems from the fact that locally these punctures are the same 
  as those of $\CN=2$ theories except that we have the freedom to choose one of the two normal directions 
  to the M5-branes.\footnote{While we will not study in this paper, the $\CN=1$ punctures should be given by the $\frac{1}{4}$-BPS codimension-2 defects of the 6d $\CN=(2, 0)$ theory. Upon dimensional reduction these yield the $\frac{1}{4}$-BPS boundary conditions of $\CN=4$ super Yang-Mills theory. 
  This problem has been studied recently by \cite{Hashimoto:2014vpa, Hashimoto:2014nwa} generalizing the work of \cite{Gaiotto:2008sa, Gaiotto:2008ak} who studied the $\half$-BPS boundary conditions.}
 
  A four-dimensional UV theory can be associated to every pair-of-pants decomposition of $\CC_{g,n}$.\footnote{Here by UV theory or UV description we do not mean the underlying six-dimensional theory. By partial topological twist and dimensional reduction, we are looking at the four-dimensional theory below the Kaluza-Klein scale given by the size of the UV curve. Here we are interested in various different four-dimensional gauge theories (which may also have non-Lagrangian building blocks) that flow to the SCFT in the same conformal manifold. We refer to these gauge theories as UV descriptions or duality frames.}
 These UV theories are in the same class, 
  in the sense that the theories corresponding to the different pants decompositions of the same $\CC_{g, n}$, 
  flow to fixed points that are connected by exactly marginal deformations. 
  This provides a nice geometric picture of the duality of $\CN=1$ class $\CS$ theories 
   \cite{Bah:2012dg,Beem:2012yn,Agarwal:2013uga,Gadde:2013fma} generalizing the well-known Seiberg duality \cite{Seiberg:1994pq}.

  Among these theories, linear quiver gauge theories form an important subset
  describing characteristic features of class $\CS$. 
  A linear quiver theory has two tails 
  each of which is composed of a product of gauge groups whose ranks are non-decreasing.
  In $\CN=2$ theories, the quiver tail has been fully understood to be related to
  a sphere with a maximal puncture ($N=1+1+\dots+1$), a number of minimal punctures ($N=1+(N-1)$),
  and a generic puncture \cite{Gaiotto:2009we}. 
  The purpose of this paper is to identify the $\CN=1$ version of quiver tails 
  associated with a similar sphere but with colors.
  
  It turns out that the $\CN=1$ quiver tails have an important ingredient, 
  which we will call the Fan.
  The Fan is composed of a collection of various chiral multiplets
  coupled by a specific superpotential that preserve the global symmetry 
  $SU(N) \times SU(N') \times \prod_{k} U(n_{k}) \times U(1)$.
  The quiver tail is constructed by gauging some of the global symmetries.
  When $N'$ is absent, the Fan is shown to be associated to a pair-of-pants
  whose three punctures are: one maximal, one minimal, and a third generic puncture specified by a partition of $N=\sum_{k} k n_{k}$.
  (The color of the former two punctures are the same as that of the pair-of-pants,
  and are different from that of the generic puncture.)
  
  We obtain the $\CN=1$ quiver tail, and in particular the Fan, by the nilpotent Higgsing
  which was first studied in \cite{Heckman:2010qv} from the different point of view
  and in \cite{Tachikawa:2013kta,Gadde:2013fma} from the class $\CS$ point of view.
  We start from the linear quiver theory where all gauge groups are $SU(N)$, 
  and give a nilpotent vev to the quark bilinear at the end of quiver.
  This produces a quiver tail.
  In $\CN=2$ linear quiver theories, the nilpotent Higgsing propagates to neighboring gauge nodes
  of the quiver because of the F-term equations \cite{Tachikawa:2013kta}, which we also discuss in detail in appendix \ref{sec:N2higgs}.
  On the other hand, if there is an $\CN=1$ gauge group in the quiver, 
  the Higgsing stops at that node 
  and does not propagate further.
  This indicates the main characteristic difference of the Higgsing between $\CN=1$ and $\CN=2$ theories.
  We will confirm this in different ways by using multiple Seiberg dualities.
  
  The Fan can be used as a new building block to construct not only the quiver tail, but more general $\CN=1$ gauge theories 
  in class $\CS$.
  Moreover, the Fan plays a crucial role in the study of the dualities in class $\CS$ theories.
  As a remarkable example, we find that the Fan coupled to an $\CN=1$ vector multiplet appears 
  as a dual description of the $\CN=1$ supersymmetric QCD with $N_{f}=2N$ flavors.
  The precise description is an $\CN=1$ $SU(N)$ gauge theory coupled to the Fan, a $T_{N}$ theory \cite{Gaiotto:2009we} and an adjoint chiral multiplet,
  with a particular superpotential.
  From the UV curve viewpoint, this duality can be seen as a pair-of-pants decomposition that exchanges maximal and minimal punctures, 
  and therefore is an $\CN=1$ analog of the Argyres-Seiberg duality \cite{Argyres:2007cn}, which was first discussed in \cite{Agarwal:2013uga} for the case of $SO/Sp/G_2$ gauge theories.

  The organization of this paper is as follows.
  In section \ref{sec:Higgs}, we first review the $\CN=1$ linear quiver gauge theories of class $\CS$ \cite{Bah:2013aha}, 
  and the nilpotent Higgsing.
  In section \ref{sec:fan}, the Fan is introduced.
  We will see that the $\CN=1$ quiver tail in which the Fan plays a central role can be obtained
  by the nilpotent Higgsing of the $\CN=1$ linear quiver gauge theory.
  In section \ref{sec:duality}, we consider the application of the Fan to dualities.
  We first show that the Fan appears in an $\CN=1$ quiver theory with an $\CN=2$ quiver tail
  by successive application of Seiberg duality.
  We then consider the duality of $\CN=1$ SQCD with $N_{f}=2N$ flavors. 
  In section \ref{sec:anomaly}, we study the 't Hooft anomaly coefficients of the $\CN=1$ class $\CS$ theories,
  in particular the Fan.
  We then present formulae of the anomalies in terms of the UV curve.
  In section \ref{sec:index}, we calculate the superconformal index of the class $\CS$ theories
  involving the Fan.
  This is the strongest check of the duality conjecture in section \ref{sec:duality}. 
  In appendix \ref{sec:appFanW}, we derive the superpotential of the Fan from nilpotent Higgsing. 
  We also discuss the nilpotent Higgsing in the $\CN=2$ linear quiver theories in appendix \ref{sec:N2higgs}.

\section{$\CN=1$ quiver theories of class $\CS$ and nilpotent Higgsing}
\label{sec:Higgs}
  Our main object is the class of theories, in particular quiver tails, obtained by giving nilpotent vevs 
  to $\mathcal{N}=1$ linear quiver gauge theories of class $\CS$ \cite{Bah:2013aha}.
  We first discuss our criteria for constructing $\CN=1$ class $\CS$ theories in section \ref{subsec:criteriaS} and then describe $\CN=1$ linear quiver gauge theories of class $\CS$
  in section \ref{subsec:linearquiver}.
  We then study the generic features of nilpotent Higgsing of the quiver theory
  in section \ref{subsec:Higgsing},
  focusing on the differences between $\CN=1$ and $\CN=2$ quiver theories.
  
  \subsection{Generic features of $\CN=1$ class $\CS$}\label{subsec:criteriaS}
  
  There is no complete classification of $\CN=1$ class $\CS$ field theories from compactifications of the six-dimensional $(2,0)$ theory.  
  But there are two prevalent features of the existing constructions of class $\CS$ theories.  
  In our explorations, we impose these conditions as criteria for class $\CS$.  They are:
 
  \paragraph*{Criterion I: R-symmetry} $\CN=1$ class $\CS$ theories admit a $U(1)_+ \times U(1)_- $ global symmetry, whose generators will be denoted by $(J_+,J_-)$. This corresponds to the generic subgroup of the $SO(5)$ $R$-symmetry of the $(2,0)$ theory that can be preserved after a partial topological twist on a UV curve. From the point of view of M5-branes, this symmetry corresponds to the rotations of the two line bundles fibered over the UV curve.  One combination of this symmetry will become the superconformal $R$-symmetry and the other will be a global symmetry of the four-dimensional $\CN=1$ SCFT. 
  
  Another notation for the global symmetry $U(1)\times U(1)$ is $(R_0, \mathcal{F})$ defined as
  \begin{equation} \label{eq:RandJ}
    R_0 = \frac{1}{2} \left(J_+ + J_- \right), \qquad \CF = \frac{1}{2} \left(J_+ - J_- \right) \ .
    \end{equation} 
 This latter notation is more convenient when computing central charges and anomalous dimensions.  
    The superconformal $R$-symmetry is
    \begin{equation}
    R_{\CN=1} = R_0 + \epsilon \CF \ , 
    \end{equation} 
 where $\epsilon$ is fixed by a-maximization \cite{Intriligator:2003jj}.
    
   In order to satisfy the $R$-symmetry criterion, we impose the condition: All additional $U(1)$ symmetries, $F_I$, are baryonic; i.e., they cannot mix with the $R$-symmetry.  In the class $\CS$ theories, there are flavor symmetries associated to the punctures on the UV curve. We assume these are all baryonic symmetries hence do not mix with the $R$-symmetry; this is the case for all known theories.\footnote{The flavor symmetry associated with a puncture for a Lagrangian theory comes from a pair of chiral multiplets. The axial symmetries are usually anomalous, and we only see the baryonic part of the symmetry. In fact, for a given puncture with global symmetry $G_F$, we generally expect the theory has $G_F \times G_F$ symmetry at some point in the conformal manifold, which is broken in a general point. }
   
\paragraph*{Criterion II: Marginal Coupling} 
For every gauge coupling, there is an associated exactly marginal direction.  
In the construction of class $\CS$, the number of gauge groups is given by the dimension of the complex structure moduli space of the UV curve. The addition of gauge groups maps to the addition of punctures or handles on the UV curve and therefore increases the dimension of the conformal manifold \cite{Maruyoshi:2009uk, Gaiotto:2009we,Benini:2009mz,Bah:2012dg}.  

This condition is not entirely correct if the UV curve has an irregular puncture. For example, one can realize $SU(N)$ gauge theory with $N_f < 2N$ flavors by a three-punctured sphere with irregular punctures. This theory flows to a conformal fixed point with no marginal direction. There is no complex structure deformation associated to this UV curve, nevertheless it has a gauge group. In this paper, we aim to find theories with regular punctures only, where the number of gauge groups is the same as the dimension of complex structure moduli space of the UV curve. 

These criteria are surprisingly constraining and generic quiver gauge theories do not satisfy them.  They are satisfied in $\CN=1$ class $\CS$ linear quivers and all theories constructed so far.  As we will find, they are always preserved by nilpotent Higgsing.

\subsection{Linear quiver theory}
\label{subsec:linearquiver} 
\begin{figure}[t]
	\centering 
	\begin{subfigure}[b]{6.0in}
	\quad~~
	\includegraphics[width=5.5in]{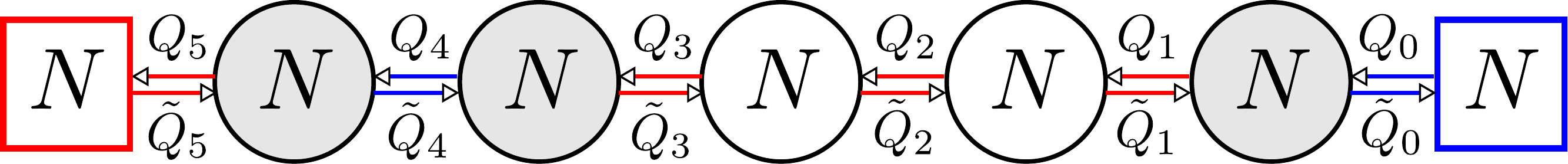}
	\caption{The quiver diagram for a generic class $\CS$ linear quiver gauge theory. The black and white node corresponds to $\CN=1$ and $\CN=2$ gauge nodes respectively. The blue/red arrows denote the bifundamental matter fields with $\s=1$/$\s=-1$ respectively.}
	\label{fig:GenLinearQuiver}
	\end{subfigure}
	\begin{subfigure}[b]{6.0in}
	\centering 
	\includegraphics[width=5.0in]{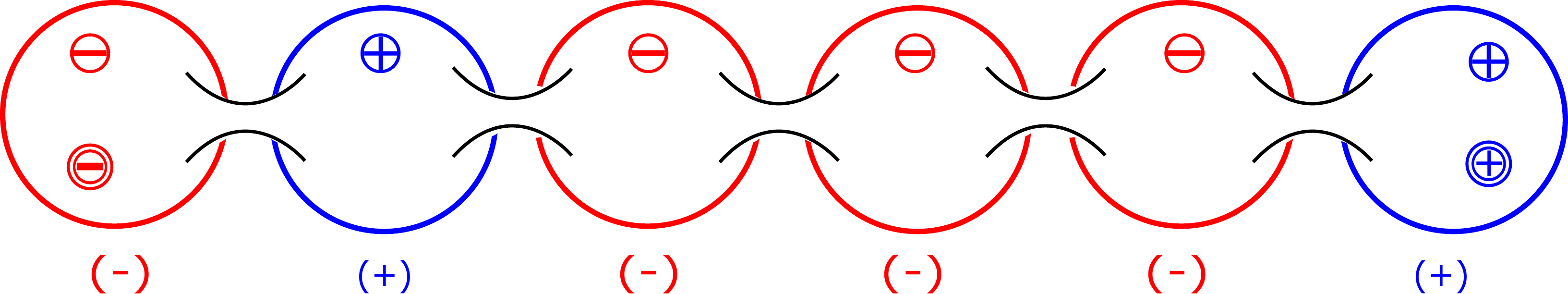}
	\caption{The UV curve and its colored pair-of-pants decomposition corresponding to the quiver \ref{fig:GenLinearQuiver}. The symbols $\oplus, \ominus$ denote the minimal punctures of each color, and the ones with extra circle denote the maximal punctures. The $(+), (-)$ below each pair-of-pants denote the coloring of the pair-of-pants itself. }
	\label{fig:GenLinearQuiverPoP}
	\end{subfigure}
	\caption{An example of a generic $SU(N)$ quiver theory corresponding to the UV curve given by a sphere with two maximal and a number of minimal punctures.  
	Note that the colored pair-of-pants mapped to the bifundamentals, and the tubes mapped to the gauge nodes. }
\end{figure}

  Let us consider a linear quiver theory given as follows. 
 It has $\ell$ gauge groups labelled as $SU(N)_i$, which can be $\mathcal{N}=2$ or $\mathcal{N}=1$.
  The former is an $\mathcal{N}=1$ vector multiplet with a chiral multiplet transforming in the adjoint representation of the gauge group. 
  The gauge nodes, $SU(N)_{i+1}$ and $SU(N)_i$, are linked by hypermultiplets, 
  $H_i = (Q_i, \widetilde{Q}_{i})$, 
  transforming in the bifundamental representation of $SU(N)_{i+1}$ and $SU(N)_i$.  
  Our conventions are such that $(Q_i,\widetilde{Q}_i)$ transforms in 
  $(\bf{N\otimes \bar{N}, \bar{N} \otimes N})$ of $SU(N)_{i+1} \times SU(N)_i$.  
  The right-most and left-most hypermultiplets are denoted by $H_0, H_\ell$ respectively and they transform 
  in the bifundamental representations of $SU(N)_1\times SU(N)_0$ and  $SU(N)_{\ell+1} \times SU(N)_\ell$ where $SU(N)_0, SU(N)_{\ell+1}$ are flavor symmetries.  
  See figure \ref{fig:GenLinearQuiver} for the $\ell = 5$ case.
  
  As mentioned above, 
  the theory preserves distinguished anomaly-free $U(1)$ symmetries, $U(1)_+ \times U(1)_-$.  
We denote the charge of fields under this symmetry as $(j_+,j_-)$;
  the charge of any gaugino is $(1,1)$.  
  We fix the charges of the matter fields and a theory 
  by giving the sequence $(\sigma_{-1},\sigma_{0}, \sigma_1, \cdots, \sigma_{\ell},\sigma_{\ell+1})$ 
  with $\sigma_i^2 =1$.  
  Each hypermultiplet $H_i$ also comes with a baryonic $U(1)_i$, whose generators we denote as $J_{i}$.  
  The charges of the $H_i$ are given as
    \bea
    J_{\pm}(Q_{i})
     =     \frac{1 \pm \sigma_i}{2}, ~~~~
    J_{j}(Q_{i})
     =     \delta_{ij}.
    \eea
  Note that the $J_{j}$ charge of the anti-fundamental $\widetilde{Q}_{i}$ has an opposite sign.
 
  Each gauge group can come with an $\mathcal{N}=2$ or with an $\mathcal{N}=1$ vector multiplet.
  When $\s_i = \s_{i-1} = \pm 1$, the $SU(N)_{i}$ 
  gauge group has a chiral field $\phi_i^{\mp}$ transforming in the adjoint representation and we add the superpotential terms
 
    \be
    W_i
     =  \sigma_i   \Tr \left[ \phi^\mp_i ( Q_{i-1} \widetilde{Q}_{i-1}  - \widetilde{Q}_{i} Q_{i} ) \right].
           \label{N=2superpotential}
    \ee
  For $\s_i = -\s_{i-1}$, there is no adjoint chiral field. However we can add the quartic superpotential terms
    \be
    W_i
     =     \Tr \left( Q_{i-1} \widetilde{Q}_{i-1}\widetilde{Q}_{i} Q_{i} \right)
         - \frac{1}{N} \Tr (Q_{i-1} \widetilde{Q}_{i-1} ) \Tr(\widetilde{Q}_{i} Q_{i}). 
           \label{N=1superpotential}
    \ee
  Let us note that these can be uniformly written as
    \begin{align}
    W_i 
    &=\Tr \left[  \widetilde{Q}_i Q_i \left(\frac{1-\sigma_i}{2} \phi_i^+ - \frac{1+\sigma_i}{2} \phi_i^-\right) 
        +  Q_{i-1} \widetilde{Q}_{i-1}  \left(\frac{1+\sigma_{i-1}}{2} \phi_i^- - \frac{1-\sigma_{i-1}}{2} \phi_i^+\right)  \nn \right. \\
    &{ }~~ \left. + m_i \left(\frac{1-\sigma_{i}}{2} \phi_i^- - \frac{1+\sigma_{i}}{2} \phi_i^+\right) 
      \left(\frac{1-\sigma_{i-1}}{2} \phi_i^- - \frac{1+\sigma_{i-1}}{2} \phi_i^+\right) \right] \ , 
    \end{align} 
  where the trace is over the gauge group $SU(N)_i$. 
  Below the energy scale $m_{i}$, some of adjoint fields are integrated out, 
  giving \eqref{N=2superpotential} or \eqref{N=1superpotential} depending on $\sigma_{i}$ and $\sigma_{i-1}$.
The total superpotential is given as $W = \sum_{i=1}^\ell W_i$.

  Since the fields $H_{\ell+1}$ and $H_{-1}$ do not exist, 
  and $SU(N)_{\ell+1}$ and $SU(N)_0$ are flavor groups, 
  the choices $\sigma_{-1}$ and $\sigma_{\ell+1}$ attaches or turns off adjoint chiral multiplets 
  to the end of hypermultiplets.
  Namely, if $\sigma_{-1} = \sigma_{0} = \pm$, we attach the adjoint $\phi_{0}^{\mp}$ 
  with $W_{0} = \Tr \widetilde{Q}_{0} Q_{0} \phi_{0}^{\mp}$;
  if $\sigma_{-1} = - \sigma_{0}$, we do not have any adjoints.
  The $U(1)_{\pm}$ charges of the fields are
    \bea
    J_{\pm}(\phi_{i}^{\pm})
     =     \frac{2+\sigma_i + \sigma_{i-1}}{2} \ , \qquad
    J_{\pm}(\phi_{i}^{\mp})
     =     \frac{2-\sigma_i -    \sigma_{i-1}}{2} \ .
    \eea

   \begin{figure}[t]
    \centering
    \begin{subfigure}[b]{6.0in}
	\quad ~~
	\includegraphics[width=5.5in]{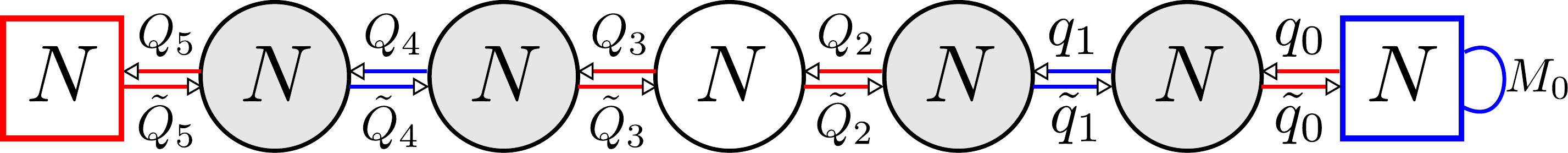}
	\caption{The quiver diagram corresponding to the Seiberg dual of figure \ref{fig:GenLinearQuiver}. }
	\label{fig:GenQuiverDual}
    \end{subfigure}

    \begin{subfigure}[b]{6.0in}
	\centering 
	\includegraphics[width=5.0in]{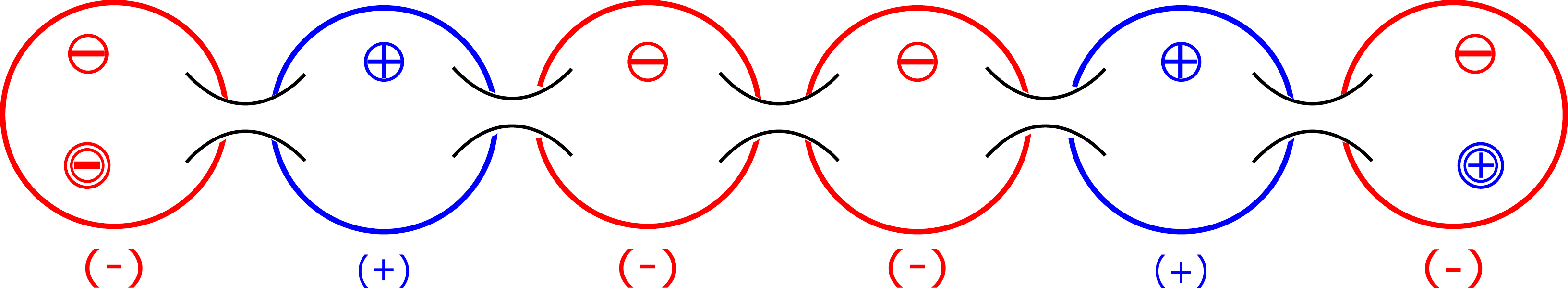}
	\caption{The UV curve and its colored pair-of-pants decomposition corresponding to the quiver \ref{fig:GenQuiverDual}. }
	\label{fig:GenQuiverDualCurve}
    \end{subfigure}
    \caption{The Seiberg dual of the quiver given by figure \ref{fig:GenLinearQuiver} 
             and its colored pair-of-pants decomposition. 
             Here we dualized the right-most gauge group $SU(N)_1$. 
             Note that the second gauge group $SU(N)_2$ became $\CN=1$ because of the meson dual 
             to $Q_1 \widetilde{Q}_1$ behaves as an extra adjoint chiral, 
             which generates a mass term for the adjoint chiral. 
             From the UV curve viewpoint, this is represented by that the colors of the second and third
             pairs-of-pants are different.}
    \label{fig:GenLinQuiverDual}
\end{figure}
    
  Let us now briefly describe the connection with the UV curve picture.
  The linear quiver gauge theory is in class $\CS$
  and is associated to the sphere with $\ell+1$ minimal punctures and two maximal punctures \cite{Bah:2013aha}.
  See figure \ref{fig:GenLinearQuiverPoP} for illustration.
  The sphere is decomposed into $\ell+1$ pairs-of-pants, each of which has a color.
 Note that the color of pair-of-pants is the same as that of the minimal puncture it contains. 
  Locally each unit preserves $\CN=2$ supersymmetry 
  and corresponds to bifundamental hypermultiplet $H_{i}$.
  The $\sigma_{i}$ ($i=0,1,\ldots,\ell$) is exactly the color of the $i$-th pair-of-pants.
  The $\CN=1$ vector multiplet appears when two pairs-of-pants with different colors are connected by a tube;
  the $\CN=2$ vector multiplet appears when two pairs-of-pants with the same colors are connected.
  The $\sigma_{-1}$ and $\sigma_{\ell+1}$ are associated with the colors of the maximal punctures.
  If the color of the maximal puncture is different form that of the pair-of-pants,
  an adjoint chiral multiplet is attached.
  See figures \ref{fig:GenQuiverDual} and \ref{fig:GenQuiverDualCurve}.

 It is important to consider Seiberg duality in this class of theories.
  Given a quiver where $SU(N)_{i}$ gauge group is $\CN=1$ with $\sigma_i = -\sigma_{i-1}$, 
  we can dualize at this node.  
  This will map a linear quiver to another linear quiver since each gauge node satisfies $N_f=2N_c$.
  Dualizing at $SU(N)_i$ will have the effect $\sigma_i \to -\sigma_{i}$ and $\sigma_{i-1} \to -\sigma_{i-1}$.  
  From the perspective of the UV curve, this is equivalent to exchanging neighboring two minimal punctures 
  of different colors and at the same time inverting the colors of pair-of-pants,
  as in figures \ref{fig:GenQuiverDual} and \ref{fig:GenQuiverDualCurve}.  The Seiberg duality preserves the parameters $p$ and $q$ which correspond to the number of pairs-of-pants or $\sigma_{i=0, 1, \cdots, \ell}$'s with $+$ and $-$, respectively.

\subsection{Nilpotent Higgsing}
\label{subsec:Higgsing}

\paragraph{$\CN=2$ Higgsing}
  Before discussing nilpotent Higgsing in $\CN=1$ theories, we summarize the effect in the case of $\CN=2$ theories. 
  We elaborate more in the appendix \ref{sec:N2higgs}. This was also discussed in \cite{Tachikawa:2013kta}. 
    \begin{figure}[t]
	\centering
	\includegraphics[width=4.0in]{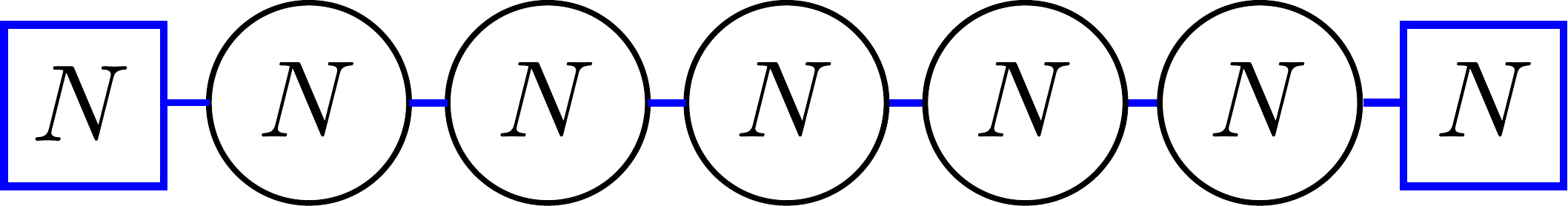}
	\caption{An $\CN=2$ linear quiver theory.}
	\label{fig:N2linearQuiver}
    \end{figure}

  Consider a linear quiver theory as in figure \ref{fig:N2linearQuiver} 
  with gauge group $G = \prod_{i=1}^{\ell} SU(N)_i$.
  This is the special case of the quiver introduced in the section \ref{subsec:linearquiver} by setting 
  all the colors of punctures and pairs-of-pants to be the same.
  From the superpotential \eqref{N=2superpotential}, we get the F-term equation for the $\phi_i$
    \be \label{eq:N2Fterm}
    F_{\phi_i} = Q_{i-1} \widetilde{Q}_{i-1} - \widetilde{Q}_i Q_i = 0 \ . 
    \ee
   
    \begin{figure}[t]
	\centering 
	\includegraphics[width=4.0in]{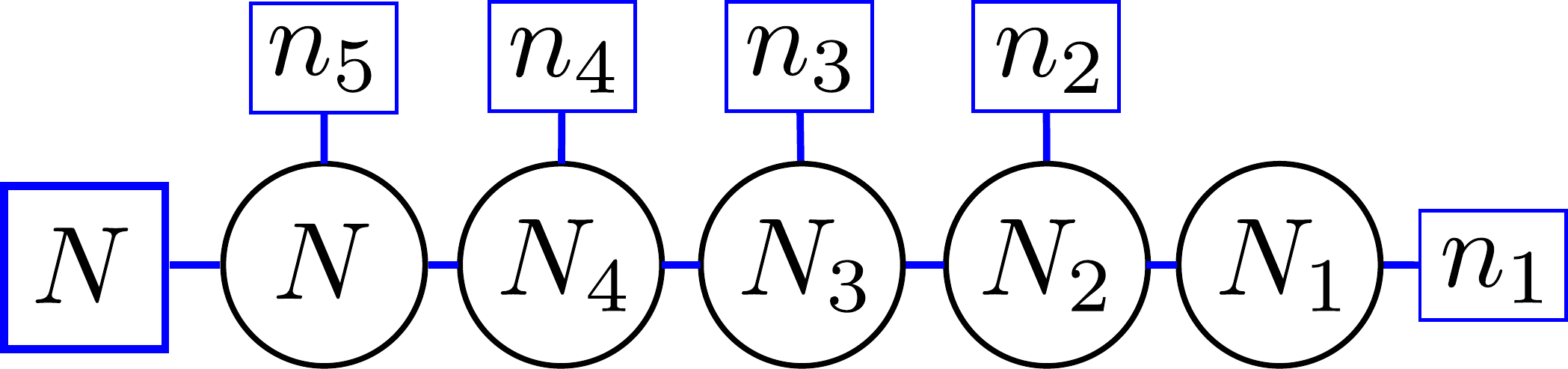}
	\caption{An $\CN=2$ quiver theory obtained after Higgsing specified by the partition $N = \sum_{k=1}^5 n_k k$. The ranks of the gauge groups are fixed by conformality condition $2N_i = N_{i-1} + N_{i+1} + n_i$. }
	\label{fig:N2linearQuiverHiggs}
    \end{figure}

  Now, let us consider a Higgsing of $H_0$ 
  by giving a nilpotent vev to $\mu_0 = \widetilde{Q}_0 Q_0 - \frac{1}{N} \Tr \widetilde{Q}_0 Q_0$, 
  which partially closes the maximal puncture. 
  For a given partition of $N = \sum_k n_k k $, we give the vev $\vev{\mu_0} = \bigoplus_k  J_k^{\oplus n_k}$, 
  where $J_k$ is the Jordan cell of size $k$
    \be
    J_k = \begin{pmatrix}
            0 & 1 & \phantom{0} & \phantom{0} &\phantom{0}\\
            \phantom{0}& 0 & 1 & \phantom{0} & \phantom{0} \\
            \phantom{0} & \phantom{0} & \ddots & \ddots & \phantom{0} \\
            \phantom{0} & \phantom{0} & \phantom{0} & 0 & 1 \\
            \phantom{0} & \phantom{0} & \phantom{0} & \phantom{0} &0
            \end{pmatrix}.
    \ee
  The matrix $J_k$ is the $k$-dimensional representation 
  of the raising operator $\s^+ = \s^1 + i \s^2$ of $SU(2)$. 
  A crucial observation here is that from the F-term for the adjoint chirals \eqref{eq:N2Fterm}, 
  the vev of $Q$'s are propagated to the neighboring node. 
  As it propagates, the operator $\widetilde{Q}_i Q_i$ will have smaller rank than that 
  of $\widetilde{Q}_{i-1} Q_{i-1}$ until it hits zero at some finite length. 
  From this way, we can explicitly derive the quiver tails corresponding 
  to a given partition of $N$ labeling the puncture, as in figure \ref{fig:N2linearQuiverHiggs}.

  Before going to $\CN=1$ theories, let us make a comment on the Higgsing 
  through a diagonal vev such as $Q_0 = \widetilde{Q}_0 = \textrm{diag}(v_1, v_2, 0, \cdots, 0)$. 
  It is certainly possible to solve the F-term equation \eqref{eq:N2Fterm} by such a diagonal vev
  for all the bifundamental hypermultiplets $Q_0 = Q_1 = \cdots Q_\ell$. 
  Therefore all the gauge symmetries are broken by the same amount. 
  We will not discuss these cases.   

\paragraph{$\CN=1$ Higgsing}
  Suppose every gauge node we described above is replaced by $\CN=1$ gauge nodes. 
  Let us Higgs the theory by giving the vev to $\mu_0$ as before. 
  This time, from the superpotential \eqref{N=1superpotential}, the F-term equation for $Q_i, \widetilde{Q}_i$ 
    \be
    F_{Q_i}
     = Q_{i-1} \widetilde{Q}_{i-1} \widetilde{Q}_i + \widetilde{Q}_{i} \widetilde{Q}_{i+1} Q_{i+1} = 0 \ , 
    \ee
  does not give us a propagating effect to the neighboring node. 
  The F-term can be simply solved by taking all the other $Q_i, \widetilde{Q}_i$ to be zero. 
  Therefore, the Higgsing happens completely locally on the first node. 
  There is no propagation of vev contrary to the case of $\CN=2$. 
  Generally if we have a number of $\CN=2$ nodes on the right, 
  the propagation continues until it hits the $\CN=1$ node and then stop. 
  In the next section, we will describe how Higgsing creates an $\CN=1$ version of the quiver tail. 

  In the case of a diagonal vev, the D-term equations for the quiver theories can be solved. 
  The effect of diagonal Higgsing has been thoroughly studied and has been used 
  to test the consistency of the Seiberg duality in $\CN=1$ $SU(N_{c})$ SQCD with $N_{f}$ flavors 
  \cite{Seiberg:1994pq}: the gauge symmetry and the flavor symmetry go down by a same amount, say $k$. 
  Then the gauge symmetry will be $SU(N_c - k)$ and the flavor symmetry will be $SU(N_f - k)$. 
  On the dual side, the gauge group remains the same, 
  but only the dual quarks become massive and reduces the number of flavors by the same amount $k$. 
  From the magnetic theory perspective, mass terms for the dual quarks are generated through the superpotential
  $W = \left( \vev{M} + \delta M \right) q \tilde{q}$, 
  where $\vev{M}$ is of rank $k$. 
  Once we integrate out the massive (dual) quarks, 
  we generate $M^2 q \tilde{q}$ term in the superpotential which is irrelevant in the IR. 
  The Higgsed theory will have $SU(N_f - k)$ flavor symmetry which is the same as the electric theory. 

  On the other hand, as we have seen in the $\CN=2$ case, 
  the nilpotent vevs can deform the theory in an interestingly different way. 
  The number of flavors will be reduced, but the superpotential terms generated are quite different 
  from the diagonal Higgsing. 
  Depending on the choice of nilpotent vevs, we can generate various types of flavor symmetry of the form
    \be
    G_F
     =     S\left( \prod_{i=1}^\ell U(n_k) \right) \ . 
    \ee 
  We will see how the nilpotent Higgsing works for $\CN=1$ theories in detail in the next section. 
  There will be various seemingly irrelevant terms in the superpotential generated through this procedure. 
  But, we will argue that  all of these terms become exactly marginal in the IR SCFT. 
  This kind of operators in the superpotential which looks irrelevant in the UV but not in the IR are called 
  dangerously irrelevant operators. See \cite{Kutasov:1995ss} for example. 

\section{Higgsing, Fan and quiver tails}
\label{sec:fan}
  In this section, we give an $\CN=1$ version of the quiver tails. 
  First, we define the Fan in section \ref{subsec:fan}. 
  Then in section \ref{subsec:fanSeiberg} we describe its Seiberg duality. 
  Then in section \ref{subsec:quivertail}, we will summarize the $\CN=1$ quiver tail obtained 
  by the nilpotent Higgsing of the linear quiver, where the Fan appears as an important ingredient.
  Finally in section \ref{subsec:derivation} we show that the Fan is indeed obtained by
  Higgsing the linear quiver with the adjoint fields attached to the end.
  
\subsection{Description of the Fan} 
\label{subsec:fan} 

\begin{figure}[t]
	\centering 
	\includegraphics[width=2.1in]{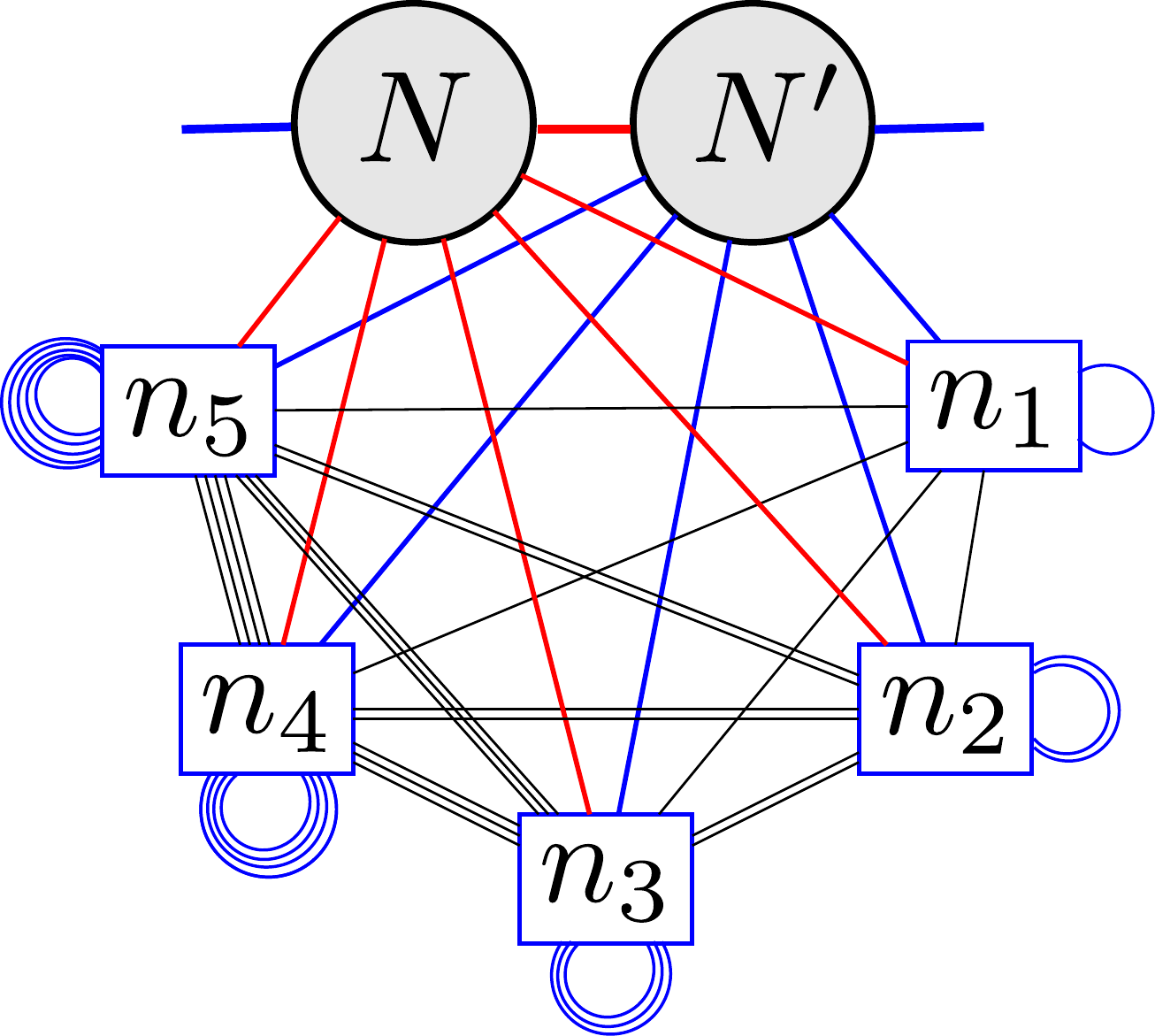}
	\caption{A generic form of the Fan given by $(N, N')$ and the partition $N-N' = \sum_{k=1}^5 k n_k$.}
	\label{fig:Fan}
    \end{figure}

  The Fan is a collection of free chiral multiplets with certain global symmetries and superpotential. 
 It is labelled by two integers $N, N'$ with $N > N'$ and an $\ell$-partition 
    \be
    N - N' = \sum_{k=1}^\ell k n_k \ . 
    \ee       
 We will refer to $\ell$ as its size.  The matter content is displayed in table \ref{Table:FanMatter}.  We also have a choice of a color, $\sigma$; that we pick to be $\s = -1$ for simplicity.  The other choice, $\sigma=1$, corresponds to swapping $J_+$ and $J_-$ in table \ref{Table:FanMatter}.  It has the global symmetry 
\begin{equation}
 SU(N) \times SU(N')  \times U(1)_B \times \prod_{i=1}^{\ell} U(n_i)  \times U(1)_{+} \times U(1)_{-} \ .
\label{flavorsymfan}
\end{equation} 
Figure \ref{fig:Fan} is a representation of the Fan with size $\ell =5$.  Each line corresponds to a bifundamental hypermultiplet and each loop corresponds to an adjoint chiral multiplet.  
    \begin{table}[t] 
\begin{equation*}
\begin{array}{|c|c|c|c|c|c|c||c|c|} 
\hline
 & SU(N) & SU(N') & U(n_i) & U(n_j) &U(1)_B & J_+ & J_- \\
\hline
\hline
(Q, \widetilde{Q}) & (\Box,\bar{\Box}) & (\bar{\Box},\Box) & \cdot &\cdot & (1,-1)& 0 &1  \\ 
(Z_i, \widetilde{Z}_i)&  (\Box,\bar{\Box}) & \cdot  &  (\bar{\Box},\Box)  &\cdot & (1,-1) &1-i & 1  \\
(Y_i, \widetilde{Y}_i) & \cdot & (\Box, \bar{\Box}) & (\bar{\Box}, \Box)& \cdot & \cdot  & i+1 & 0 \\ 
M_{ii}^{(p)} & \cdot & \cdot & \mbox{adj} & \cdot &\cdot  & 2(i-p) & 0  \\
(M_{ij}^{(p)},M_{ji}^{(p)}) & \cdot & \cdot & (\Box,\bar{\Box}) & (\bar{\Box},\Box) &\cdot & i+j - 2p & 0  \\
\hline
\end{array} 
\end{equation*}
\caption{The Fan contains many fields organized in representation of the flavor symmetry.  The indices $i,j$ range in the interval $[1,\ell]$ and are ordered as $i<j$.  The index $p$ labels a tower of fields in the same representation of the flavor symmetry, its range is $0\leq p \leq i-1$.  }  
\label{Table:FanMatter} \label{table:fan}
\end{table} 

The Fan appears in quiver gauge theories with the $SU(N)\times SU(N')$ symmetries gauged.  When the fan is glued, chiral anomalies at the  $SU(N)\times SU(N')$ gauge groups of $J_\pm$ must be cancelled.  This will restrict the matter content that can appear on either side.  The contributions of the Fan to the anomaly coefficient are:
 \begin{align}
 SU(N):& \qquad \mbox{Tr} J_+ T^a T^b = -N \delta^{ab}\ ,  \quad~ \mbox{Tr} J_- T^a T^b = 0\ , \\
  SU(N'):& \qquad \mbox{Tr} J_+ T'^a T'^b = -N' \delta^{ab} \ , \quad  \mbox{Tr} J_- T'^a T'^b = - \sum_{i=1}^{\ell} n_i  \delta^{ab}\ ,
 \end{align}
where $T^a$ and $T'^a$ are the generators of $SU(N)$ and $SU(N')$ respectively. 
The anomaly at $SU(N)$, when it is gauged with an $\CN=1$ vector multiplet, can be cancelled by coupling the Fan to $N$ $(1,0)$-fundamental hypermultiplets.\footnote{When we say $(m, n)$-operators/fields, $(m, n)$ are their $(J_+, J_-)$ charges.}  
When it is gauged with an $\CN=2$ vector, the anomaly is cancelled by coupling $N$ $ (0,1)$-fundamental hypermultiplets to the Fan.  This provides $\CN=1$ and $\CN=2$ gluing of the Fan at the $SU(N)$ gauge group.   

\begin{figure}[t]
\centering
\begin{subfigure}[b]{2.5in}
\centering
\includegraphics[width=2.5in]{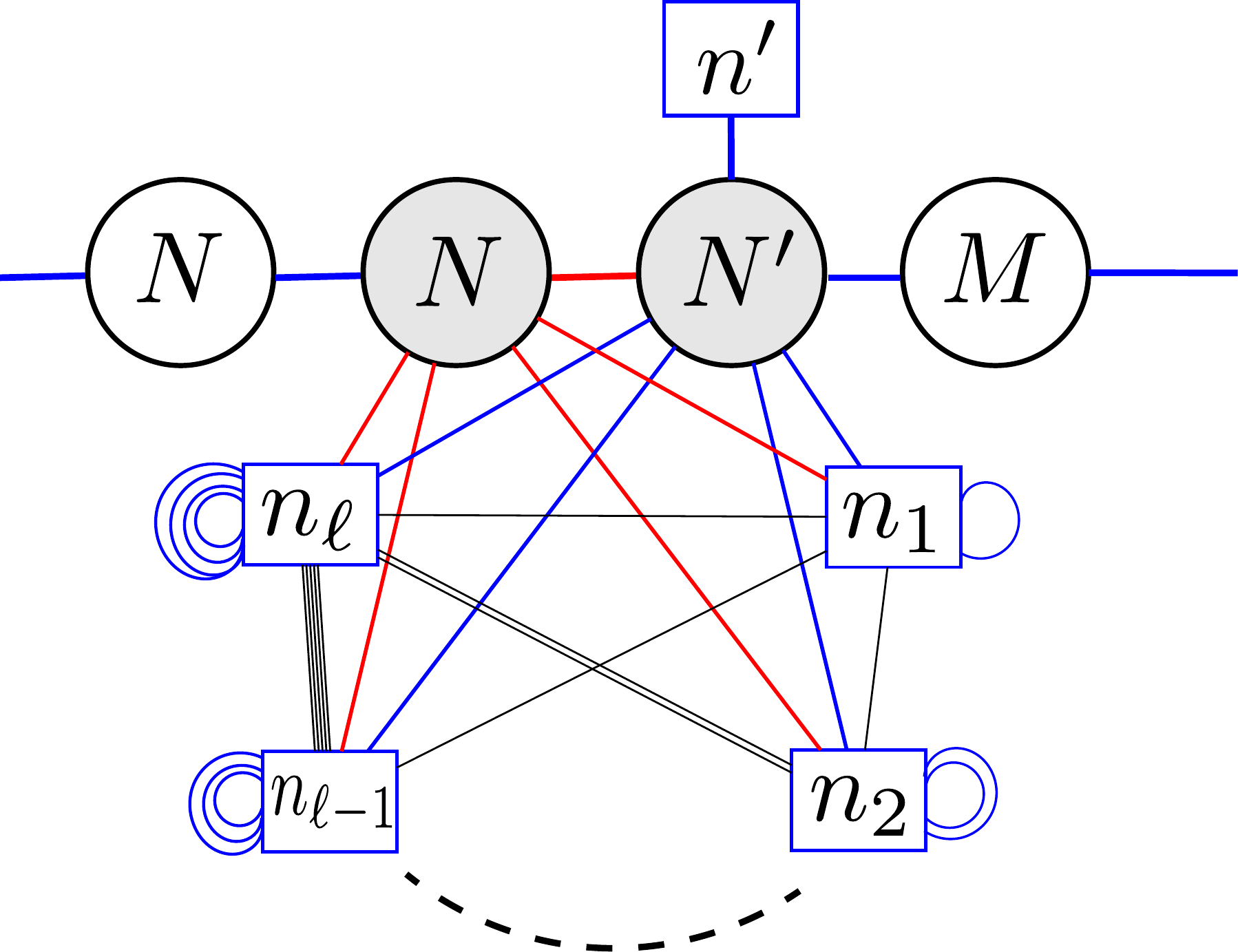}
\caption{A quiver tail with the Fan}\label{fig:FanG}
\end{subfigure}
\begin{subfigure}[b]{2.5in}
\centering
\includegraphics[width=2.0in]{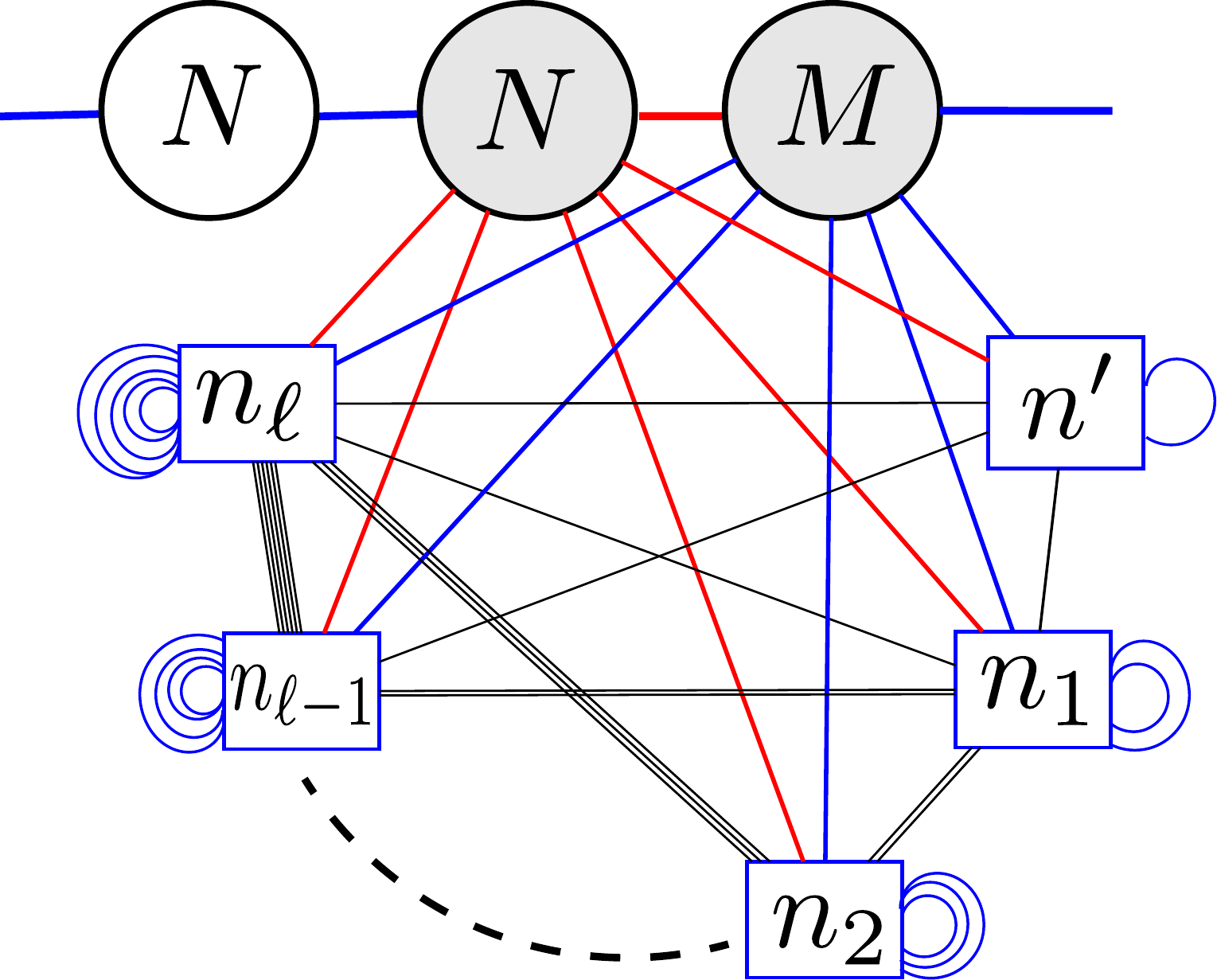}
\caption{Seiberg dual}\label{fig:FanGd}
\end{subfigure}
\caption{Seiberg dualizing at $SU(N')$ in \ref{fig:FanG} yields another quiver \ref{fig:FanGd} with the new Fan.  The $U(n')$ group is absorbed into the new Fan, labelled by $(N, M)$ and the partition $N-M = \sum_k k n_k'$ with $n_1' = n', n_{i+1}' = n_i$. }
\end{figure}

When the $SU(N')$ is gauged with an $\CN=1$ vector multiplet, the anomaly at the $SU(N')$ can be cancelled by adding $(N' - \sum_{i=1}^{\ell} n_i)$ $(1,0)$ fundamental hypermultiplets.  Unlike the $SU(N)$ side, we cannot gauge $SU(N')$ with an $\CN=2$ vector multiplet because the anomaly cannot be cancelled with either $(1,0)$ or $(0,1)$ hypermultiplets only.  We can glue the Fan to an $\CN=2$ quiver tail labelled by a partition of $N'$
by an $\CN=1$ $SU(N')$ vector multiplet.    
In figure \ref{fig:FanG} we illustrate the Fan glued to general quivers with $\mathcal{N}=1$ gluing at the $SU(N)$ gauge group.  

\paragraph*{Superpotential}

When the Fan appears in a larger quiver, we can write a superpotential by considering all possible gauge invariant $(2,2)$-operators that preserve the flavor symmetry.  We decompose it into three contributions
\begin{equation}
W_F = W_0 + W_R + W_L
\label{s}
\end{equation}   where $W_0$ is composed of fields in the Fan only, $W_R$ comes from gluing at $SU(N')$ and $W_L$ comes from gluing at $SU(N)$.  Now we describe them.  

If we consider the matter content of the Fan, the only superpotential terms we can write are
\begin{equation}
W_0 = \sum_{i=1}^\ell \left[\lambda_i^0  \mbox{Tr} \left(Z_i \widetilde{Q} \widetilde{Y}_i\right) + \tilde{\lambda}_i^0 \mbox{Tr} \left(\widetilde{Z}_i Q Y_i\right) \right] \label{suzero}
\end{equation} where the $\lambda$'s are complex coupling constants.  

The next class of operators comes from the coupling of the quiver tail to the Fan through the $SU(N')$.  To write these terms we consider the set of $(2,0)$-operators, $\mu'$ and $\mu_t$, constructed from the $U(n')$ and $SU(M)$ quarks in figure \ref{fig:FanG}.  The superpotential is
\begin{equation}
W_R = \lambda' \mbox{Tr}\left(Q \widetilde{Q} \mu' \right) + \lambda_t \mbox{Tr}\left(Q \widetilde{Q} \mu_t \right). \label{suright}
\end{equation}  

The last class of operators come from gluing the Fan at the $SU(N)$.  To write these terms, we consider the tower operators, $\mu_\alpha^{(p)}$, $\left(\mathcal{M}_{ij}^{(p),\alpha}, \mathcal{M}_{ji}^{(p),\alpha}\right)$, and $\mathcal{M}_{ii}^{(p), \alpha}$.  The $\mu$'s are constructed from fields to the left of the Fan.  The $\mathcal{M}$'s are constructed from the $M_{ij}$ fields of the Fan.  Their charges are written in the table \ref{table:MMcharges}. 
\begin{table}[t]
\begin{equation}
\begin{array}{|c|c|c|c|c|c|c||c|c|} 
\hline
 & SU(N)  & U(n_i) & U(n_j) & J_+ & J_- \\
\hline
\hline
\mu_\alpha^{(p)} & \mbox{adj} & \cdot &\cdot & 2p & 0  \\ 
\mathcal{M}_{ii}^{(p),\alpha} & \cdot & \mbox{adj} &\cdot  & 2(i-p) & 0  \\
\left(\mathcal{M}_{ij}^{(p),\alpha},\mathcal{M}_{ji}^{(p),\alpha} \right) & \cdot & (\Box,\bar{\Box}) & (\bar{\Box},\Box) & i+j - 2p & 0  \\
\hline
\end{array} \nn
\end{equation}
\caption{Charges of the $\CM$ and $\mu$ operators used in \eqref{suleft}. }
\label{table:MMcharges}
\end{table}
When we glue at the $SU(N)$, we obtain the superpotential
\begin{align}
W_L &= \lambda^\alpha \mbox{Tr}  \left(\mu_\alpha^{(1)} \widetilde{Q} Q \right) +\sum_{i=1}^\ell\sum_{p=0}^{i-1} \lambda_{i,p}^{\alpha, \beta} \mbox{Tr} \left(\mu_{\alpha}^{(p)} \widetilde{Z}_i Z_i \mathcal{M}_{ii}^{(p),\beta} \right) \nonumber \\ 
&+ \sum_{i=1}^\ell\sum_{p=0}^{i-1} \lambda_{i j,p}^{\alpha, \beta} \mbox{Tr} \left(\mu_{\alpha}^{(p)} \widetilde{Z}_i Z_j \mathcal{M}_{ji}^{(p),\beta} \right)+\sum_{i=1}^\ell\sum_{p=0}^{i-1} \lambda_{ji,p}^{\alpha, \beta} \mbox{Tr} \left(\mu_{\alpha}^{(p)} \widetilde{Z}_j Z_i \mathcal{M}_{ij}^{(p),\beta} \right). \label{suleft}
\end{align}
To illustrate the $\mathcal{M}$ operators, we consider the set $\mathcal{M}_{ij}^{(p),\alpha}$.  The simplest examples in this class are $M_{ik}^{(p_1)} M_{kj}^{(p_2)}$ with $p_1+p_2 -k = p$ where we trace over the $U(n_k)$ group.
  
In the case of $\mathcal{N}=2$ gluing at $SU(N)$, the $\mu_\alpha^{(p)}$ operators are entirely given by the chiral adjoint $\phi$ in the $\CN=2$ vector multiplet, as $\mu^{(p)} = \phi^p$.  The index $\alpha$ is trivial in this case.  On the other hand, if we consider $\CN=1$ gluing, then the $\mu$ operators are more complicated.  To illustrate this, we consider gluing the $\mathcal{N}=2$ linear quiver in figure \ref{fig:N2linearQuiver} with the box $N$ identified with the $SU(N)$ in the Fan gauged with an $\CN=1$ vector.  In this case, the set $\mu_\alpha^{(p)}$ corresponds to the chain operators that can be constructed from the products of the quarks.  To give an explicit example, we label the bifundamentals as $(Q_a, \widetilde{Q}_a)$ with $a=1$ corresponding to the one attached to the Fan.  The operators, $\mu_\alpha^{(2)}$ are $( \widetilde{Q}_1 Q_1 )_{\textrm{adj}}^2$ and $(\widetilde{Q}_1 \widetilde{Q}_2 Q_2  Q_1)_{\textrm{adj}} $. 

\subsection{Seiberg duality and Fans} 

Under the Seiberg duality, a quiver with the Fan maps to another quiver with the Fan.  To illustrate this, we consider the quiver in \ref{fig:FanG} and dualize at $SU(N')$ to obtain \ref{fig:FanGd}.  Under the duality, the $U(n')$ flavor group is absorbed into the new Fan and thereby increasing its size to $\ell+1$.  We denote the $U(n')$ and $SU(M)$ hypermultiplets as $(Q',\widetilde{Q}')$ and $(Q_t, \widetilde{Q}_t)$.  We also denote the fields of the new Fan as
$(q,\tilde{q})$, $(z,\tilde{z})$, $(y,\tilde{y})$ and $(m,\tilde{m})$.  
\begin{itemize}
\item Firstly we need to replace $SU(N')$ with its magnetic dual, $SU(N_f-N')$.  The total number of flavors coming into this gauge group is $N_f=N+N'$; the contributions are $N$ $Q$'s, $\sum_{i=1}^\ell n_i$ $Y$'s, and $n' + M$ $(1,0)$ fields
where $n'+M=N'-\sum n_{i}$. 
\item The superpotential terms in \eqref{suzero} and \eqref{suright} become mass terms under the duality.  In the magnetic theory, we replace the meson operators $QY_i$, $\widetilde{Q} \widetilde{Y}_i$, $Q \widetilde{Q}'$, $Q' \widetilde{Q}$, $Q \widetilde{Q}_t$, and $Q_t \widetilde{Q}$ with their dual chiral superfields.  The cubic terms in \eqref{suzero} become mass terms for the $Z$'s while the quartic terms in \eqref{suright} become mass terms for the new chiral fields.  Integrating out the $Z$'s decouples the $SU(N)$ gauge group from the Fan.  
\item The chiral superfield dual to $\widetilde{Q}Q$ is an adjoint of the first $SU(N)$ group.  If we have $\mathcal{N}=2$ gluing, the first term in equation \eqref{suleft} will become a mass term for the chiral adjoint in the vector multiplet.  Integrating out the massive chirals yields an $\mathcal{N}=1$ vector multiplet.  On the other hand if the gluing is $\mathcal{N}=1$, the vector multiplet will become $\mathcal{N}=2$ with the addition of the chiral fields dual to $\widetilde{Q}Q$.  
\item The cubic superpotential involving the chiral adjoint of $SU(M)$ becomes a mass term when we replace the meson $\widetilde{Q}_t Q_t$ with its dual chiral superfield.  Thus the $SU(M)$ gauge group becomes an $\CN=1$.  
\item The fields of the Fan in figure \ref{fig:FanGd} come from three different sectors, which are listed as in the table \ref{table:newFields}. 
\begin{table}[h]
\begin{equation*}
\begin{array}{|c||c|}
\hline
\mbox{New fields} & \mbox{Electric dual}\\
\hline
\hline
 m_{i+1, i+1}^{(p+1)} &   M_{ii}^{(p)} \\
\left( m_{i+1, j+1}^{(p+1)}, m_{j+1, i+1}^{(p+1)} \right) & (M_{ij}^{(p)},M_{ji}^{(p)})  \\ 
\hline \hline
 (q,\tilde{q}) & (Q_t,\widetilde{Q}_t)\\ 
 (z_1,\tilde{z}_1) &  (Q',\widetilde{Q}') \\ 
 (z_{i+1}, \tilde{z}_{i+1}) &   (Y_i,\widetilde{Y}_i) \\
  \hline \hline
(y_1,\tilde{y}_1) & \left( \mbox{Tr}_g(q\widetilde{Q}_t), \mbox{Tr}_g(\tilde{q}Q_t)\right) \\ 
 (y_{i+1}, \tilde{y}_{i+1})  &  \left( \mbox{Tr}_g(Y_i\widetilde{Q}_t), \mbox{Tr}_g(\widetilde{Y}_iQ_t)\right) \\
 m_{1,1}^{(0)} & \mbox{Tr}_g (q \tilde{q}) \\
 m_{i+1,i+1}^{(0)}  & \mbox{Tr}_g (Y_i \widetilde{Y}_i) \\
 (m_{1,j+1}^{(0)},m_{j+1,1}^{(0)}) & \left(\mbox{Tr}_g (q \widetilde{Y}_j),\mbox{Tr}_g (Y_j \tilde{q}) \right) \\
 (m_{i+1,j+1}^{(0)},m_{j+1,i+1}^{(0)})  & \left(\mbox{Tr}_g (Y_i \widetilde{Y}_j),\mbox{Tr}_g (Y_j \widetilde{Y}_i) \right) \\
\hline
\end{array}
\end{equation*} 
\caption{The set of new fields appears upon dualizing the Fan.}
\label{table:newFields}
\end{table}
The first set of fields is inherited from the old Fan. 
And the second set of fields consists of the dual quarks of the $SU(N')$ gauge group. 
The last set of fields consists of the ones dual to the mesons of the old quiver tail. 
\item The flavor group $U(n')$ is absorbed into the Fan as the first flavor group $U(n_1')$, and the labeling of the rest is shifted by 1 to $n'_{i+1} = n_i$. This yields the Fan labelled by $(N, M)$ and the partition $N-M = \sum_k k n'_k $. 
\end{itemize}
\label{subsec:fanSeiberg} 

The superpotential of the dual theory is constructed by considering all possible gauge invariant $(2,2)$-operators that preserve the global symmetry.  The same superpotential is reproduced under the Seiberg duality.

\subsection{Fan as a quiver tail}
\label{subsec:quivertail}
  In this section, we describe how the Fan and quiver tails appear 
  in class $\CS$ theories. 
  A quiver tail associated to the partition $Y$ of $N$ is given by a punctured sphere  
  with one maximal, a number of minimal punctures and a puncture labeled by $Y$. Here $Y$ corresponds to the partition $N=\sum_{k=1}^\ell k n_k$. 

  Starting from the linear quiver given in section \ref{subsec:linearquiver}, 
  we can get the quiver tail by Higgsing one of the maximal punctures to $Y$. 
  When the puncture has the same color as that of the pair-of-pants, 
  this is same as giving a nilpotent vev to the quark bilinear 
  $\mu_0 = \widetilde{Q}_0 Q_0 - \frac{1}{N} \Tr \widetilde{Q}_{0} Q_{0} $.
  When the color of the puncture is different from that of the pair-of-pants,
  we give a vev to the adjoint chiral multiplet.
  In both cases, the $U(1)_0 \times SU(N)_0$ flavor symmetry of the quiver is broken down 
  to $\left(\prod_{i=1}^\ell U(n_i) \right)$. 
  
\begin{figure}[t]
	\centering
	\includegraphics[width=5in]{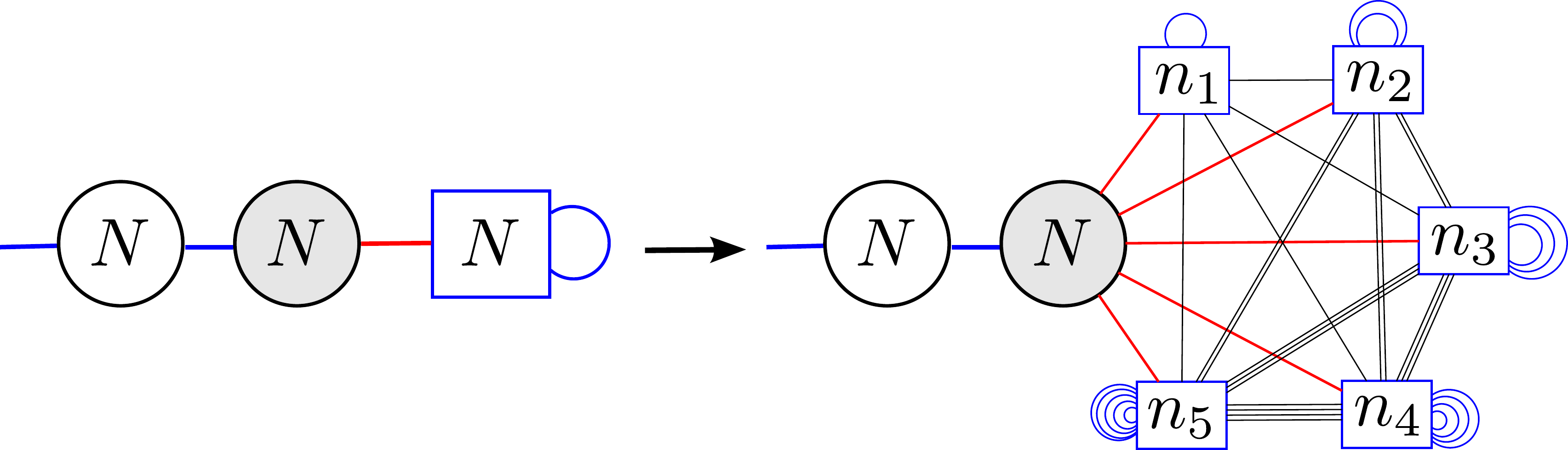}
	\caption{A Nilpotent vev to the adjoint chiral gives a Fan attached to the end of 
	the quiver with $N = 1 n_1 + 2n_2 + \cdots 5 n_5$ and $N'=0$.}
	\label{fig:suNfan0}
    \end{figure}

  Now, let us describe the quiver tail associated to the partition above. 
  If the color of the puncture we Higgs is different from that of the pair-of-pants, 
  the theory we obtain is given by attaching the Fan with $(N, N'=0)$ as in the figure \ref{fig:suNfan0}. 

\begin{figure}[t]
    \centering
    \begin{subfigure}[b]{5.0in}
	\centering
	\includegraphics[width=4.0in]{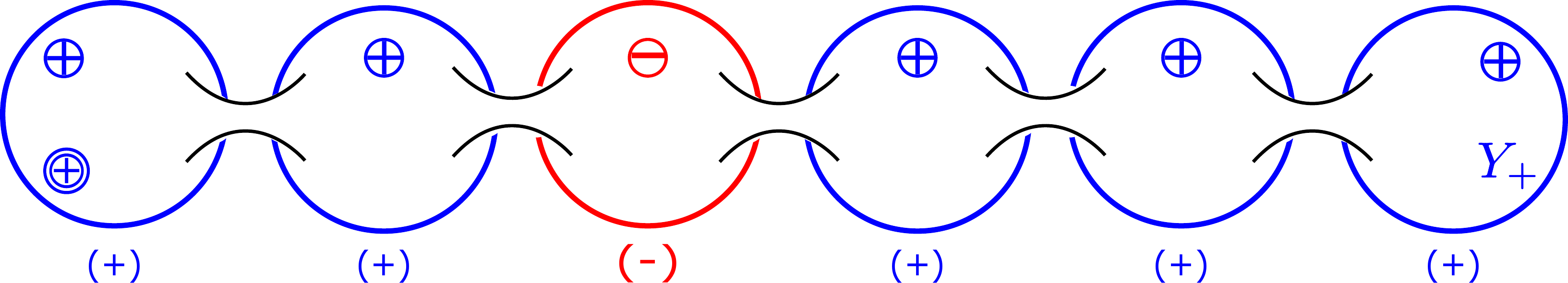}
	\caption{A colored pair-of-pants decomposition corresponding to the quiver tail.}
    \end{subfigure}

    \begin{subfigure}[h]{5.0in}
	\centering
	\includegraphics[width=3.2in]{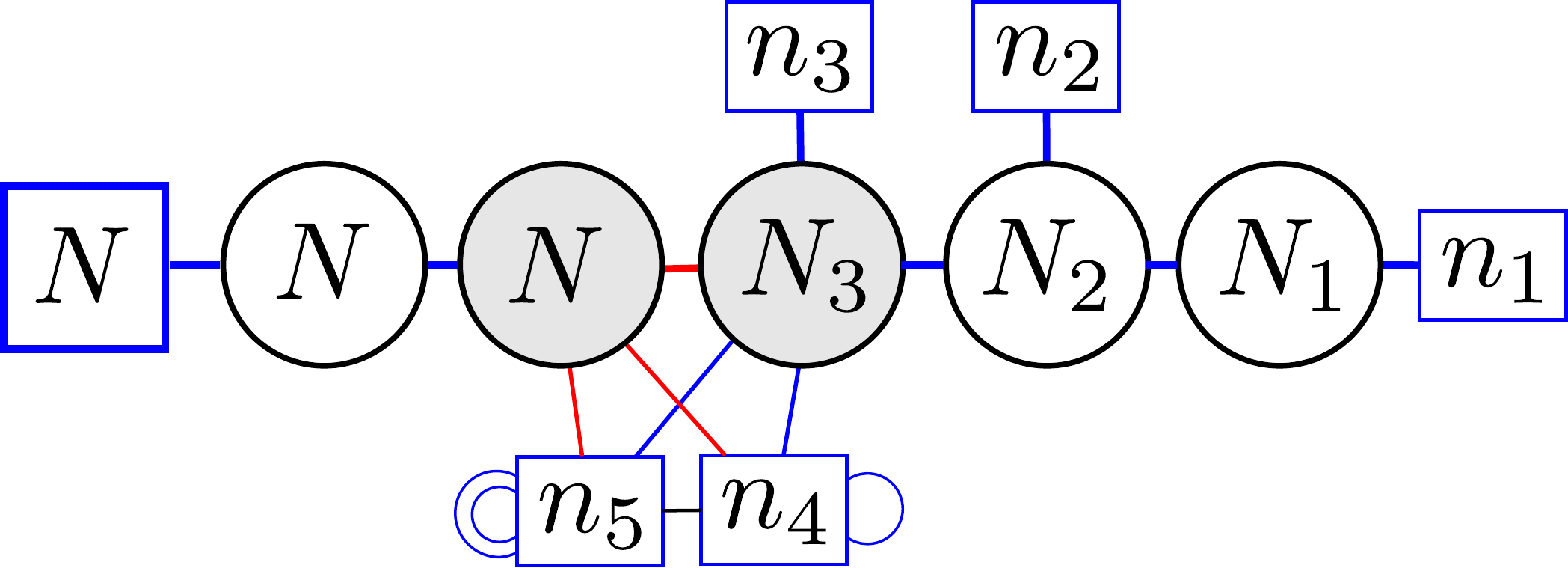}
	\caption{The quiver tail corresponding to the above colored pair-of-pants decomposition.}
    \end{subfigure}
	\caption{The quiver tail given by the partition $N = 1 n_1 + 2n_2 + \cdots 5 n_5$. $\CN=2$ Higgsing propagated until we hit $k=3$. Then the gauge group of the next node becomes $SU(N)$, and we have the Fan between $SU(N)$ and $SU(N_3)$. The Fan is given by $(N, N_3)$ and the partition $N-N_3 = n_4 + 2n_5$.  }
	\label{fig:suNquiverTail}
    \end{figure}

  If the color of the puncture is the same as the pair-of-pants, we proceed as follows. 
    \begin{enumerate}
    \item When the neighboring gauge node of $Q_0$ is $\CN=2$, 
          the flavor node becomes $n_1$ and the gauge node becomes $N_1 = \sum_{i=1}^\ell n_i$. If it is $\CN=1$, then go to step 3. 
    \item When the next neighboring gauge node is again $\CN=2$, 
          the gauge group becomes $N_2 = N_1 + \sum_{i=2}^\ell n_i$, and add $n_2$ fundamental flavors to it. 
          If it is $\CN=1$, then go to step 3. 
    \item Proceed until we hit an $\CN=1$ gauge node. 
          In this case, the neighboring gauge node remains to be $SU(N)$, 
          since the Higgsing stops propagating. 
          Suppose we hit the $\CN=1$ node at step $k$. 
          In this case, the remaining flavor boxes $n_{i}$ with $ k < i < \ell$ should be attached 
          to the gauge node of $N_k$. 
          Therefore we get the Fan labelled by $(N, N_k)$ with partition $N-N_k = \sum_{m=1}^{\ell-k} m n_{m+k}$. 
    \end{enumerate}
  See figure \ref{fig:suNquiverTail} for the case with $\ell=5$ and $k=3$. 
  We see that the Fan serves as a role of gluing $\CN=1$ nodes with different ranks in the quiver tail. 

Let us remark on the flavor symmetry of the quiver tail with the Fan. Even though the Fan itself has the flavor symmetry $U(1) \times \prod_{k} U(n_k)$, the flavor symmetry of the whole quiver tail does not include the overall $U(1)$ piece of $\prod_{k} U(n_k) $. The global symmetry of the quiver tail associated to the puncture $Y$ does not contain the extra $U(1)$. We can see this directly in the case of figure \ref{fig:suNfan0}. In this case, we see that the overall $U(1)$ can be identified with $U(1)_B$ symmetry of the Fan.

\subsection{Nilpotent Higgsing and Fan}
\label{subsec:derivation}

 In this section, we give a derivation of the Fan for the case when $N'=0$. 
  Let us now consider the linear quiver theory as in figure \ref{fig:GenLinQuiverDual}. 
  It has a chiral adjoint $M_0$ attached at the flavor $SU(N)$ node. 
  The superpotential is $W = \Tr M_0 \mu_0$,
  where $\mu_0$ is the quark bilinear $\mu_0 = \tilde{q}_{0} q_{0} - \frac{1}{N} \Tr \tilde{q}_{0} q_{0} $ 
  with $(J_+, J_-) = (0, 2)$. 
  Here we choose the color of the pair-of-pants corresponding to $q_0$ to be $\s=-1$. 
  We Higgs the flavor $SU(N)$ by a nilpotent vev corresponding to the partition $N=\sum_{k} k n_{k}$ to $M_0$.
  In the following, we omit the subscript of $\mu$ and $M$ for simplicity.
  
  Under the $SU(2)$ embedding $\rho$ labelled by the partition of $N$, the fundamental representation of $SU(N)$ decomposes as follows:
    \be
    {\bf N} \rightarrow \bigoplus_{i=1}^{\ell} V_{\frac{i-1}{2}} \otimes {\bf n}_{i}  \ , 
    \ee
  where $V_{j}$ is the spin $j$ representation of $SU(2)$ and ${\bf n}_i$ is the fundamental representation of $SU(n_i) \subset S[\prod_{i=1}^{\ell} U(n_i)]$.  The residual flavor symmetry $S[\prod_{i=1}^{\ell} U(n_i)]$ is given by the commutant of the embedding. 
  The adjoint representation of $SU(N)$ decomposes as 
    \bea \label{eq:adjDecomp}
    {\rm adj} 
    & \rightarrow &
           \bigoplus_{i,j=1}^\ell (V_{\frac{i-1}{2}} \otimes {\bf n}_{i}) 
           \otimes (V_{\frac{j-1}{2}} \otimes {\bf n}_{j}) - V_{0} 
           \nonumber \\
    &=&    \bigoplus_{i<j} \bigoplus_{k=1}^{i} V_{\frac{j-i+2k-2}{2}} 
           \otimes \left( {\bf n}_{i} \otimes {\bf \bar{n}}_{j} 
           \oplus {\bf \bar{n}}_{i} \otimes {\bf n}_{j} \right)
           \oplus \bigoplus_{i=1}^{\ell} \bigoplus_{k=1}^{i} V_{k-1} 
           \otimes {\bf n}_{i} \otimes {\bf \bar{n}}_{i} - V_{0} \ .
           \label{adjointdecomposition}
    \eea
  This decomposition gives us the quantum numbers of the various elements of the $SU(N)$-adjoint $M$. 
  
  We now use the decoupling argument of \cite{Gadde:2013fma}. 
  Due to the vev of $M$ the superpotential is written as
    \bea
    W
     =     \mu_{1,-1, 1} + \sum_{J,m,f} {M}_{J,-m,f} \mu_{J,m,f},
    \eea 
  where ${M}_{J, m, f}$ is the fluctuation from the vev, 
  and $J$, $m$ and $f$ labels the spins, $\sigma_{3}$-eigenvalues 
  and the representations of the flavor symmetry $\prod_i SU(n_{i})$ 
  appearing in the decomposition \eqref{adjointdecomposition}.
  By the presence of the first term the $SU(N)$ current is not conserved anymore,
  and becomes non-BPS by absorbing the components of $\mu$ except for the $m=J$.
  The components of $M$ which coupled to the absorbed $\mu$ will be decoupled 
  and the remaining components are $M_{J,-J,k}$.
  Namely the $m=-J$ component of each term of \eqref{adjointdecomposition}.
  Also we should note that due to the first term of the superpotential
  the $U(1)_{\pm}$ symmetries are shifted as
    \bea
    J_+ \rightarrow J_+ -2 \rho(\sigma^3)  , \qquad J_- \rightarrow J_- \ , 
    \label{chargeshift}
    \eea
  (or $R_{0} \rightarrow R_{0} - \rho(\sigma^{3})$ and $\CF \rightarrow \CF -  \rho(\sigma^{3})$)
  in order to keep the first term to be $J_{+} = J_{-} = 2$ ($R_{0}=2$, $\CF=0$).
  
  This gives us the gauge neutral components of the Fan in the low energy theory. 
  We saw that there are $i$ gauge neutral chiral multiplets $(M_{ij}^{(p)}, {M}_{ji}^{(p)})$, $ 0 \leq p < i$ 
  transforming as bifundamentals of $U(n_i) \times U(n_j)$, $i \leq j$. 
  We identify these chirals with the component of ${M}$
  \eqref{adjointdecomposition} with $m=-J$ (and $k=i-p$).
  As a  consequence of \eqref{chargeshift}, the $(J_{+}, J_-)$ charges of $(M_{ij}^{(p)}, {M}_{ji}^{(p)})$ become 
  $(i+j-2p, 0)$, which indeed match with table \ref{table:fan}.

  Some elements of the (anti-)quark multiplet transforming 
  in the (anti-)fundamental representation of the $SU(N)$ flavor symmetry become massive due to the Higgsing 
  and will be integrated out. 
  Since $\langle M \rangle = \rho(\sigma^+)$ which is $J=1$, $m=1$ component, it implies 
  that the (anti-)quarks $Z_{i}$ ($\widetilde{Z}_{i}$) that remain massless are 
  the components with $m=\frac{i-1}{2}$ in $V_{\frac{i-1}{2}} \otimes {\bf n}_{i}$
  ($V_{\frac{i-1}{2}} \otimes {\bf \bar{n}}_{i}$).
  $Z_i$ and $\widetilde{Z}_i $ together form a hypermultiplet whose $(J_+, J_-)$ charges 
  are $(1-i, 1)$ by using \eqref{chargeshift}.

  In addition we have the superpotential \eqref{s}.
  We give a derivation of it in Appendix \ref{sec:appFanW}.

\paragraph{The Goldstone multiplets}

In any field theory we expect the spontaneous breaking of global symmetries to be accompanied by the presence of massless Goldstone bosons whose number is equal to the number of broken generators of the global symmetry. In supersymmetric theories these Goldstone bosons will form the scalar components of massless chiral multiplets which we will call Goldstone multiplets. 

However, the number of Goldstone multiplets is not necessarily equal to the number of broken generators of the global symmetry. For example, consider the linear quiver of figure \ref{fig:GenLinQuiverDual} with gauge group being $SU(3)$. Upon nilpotent Higgsing (giving a nilpotent vev to $M_0$) of the $SU(3)$ linear quiver by the partition $3=2+1$, the $SU(3)$ symmetry gets broken down to $U(1)$. The chiral fields that decouple from the low energy theory are expected to be the Goldstone multiplets. But there are only 4 such chiral multiplets while the number of broken generators is 7. 

The reason behind the discrepancy in this counting is that the scalar in a Goldstone multiplet is complex. Thus it might be that a Goldstone multiplet is either made up of two Goldstone bosons or a single Goldstone boson that gets paired up with a non-Goldstone scalar. In view of this we see that the number of Goldstone multiplets will always be less than or equal to the number broken generators of the global symmetry. The correct number of Goldstone multiplets is obtained by observing that the superpotential is holomorphic. This implies we should count the number of broken generators of the complexified global symmetry \cite{Lerche:1983qa}. Using this we now show that the number of decoupled chirals indeed matches with the number of expected Goldstone multiplets.  

In the theories of interest here, we want to consider the breaking of $G=SU(N)$ down to $H= S[ \prod_{i=1}^{\ell}U(n_i) ]$. The complexification of $G$ is $\bar{G} = SL(N, \mathbb{C})$. Since the breaking of global symmetries is achieved through $\vev{M} = \rho^+$, we therefore look for generators $X$ of $SL(N,\mathbb{C})$ which satisfy 
\be
[ \rho^+ , X] \neq 0 \ .
\label{eq:Goldstone}
\ee
Note that any generator of $SL(N,\mathbb{C})$ can be thought of as a complex matrix transforming in the adjoint representation of $SU(N)$. We can therefore label each element of $X$ by its $SU(2) \hookrightarrow SU(N)$ quantum numbers. In fact we can also simultaneously label them by the $S\Big( \prod_{i=1}^\ell U(n_i)\Big) $ symmetries that commute with the $SU(2)$ embedding. The components of $X$ are therefore classified as in \eqref{adjointdecomposition}. 
In terms of $X_{J, m, k}$, we see that (\ref{eq:Goldstone}) is satisfied if $X$ has a non-zero component with $m \neq J$.   
The Goldstone multiplet corresponding to such an $X$ will be the quantum fluctuation proportional to $[\rho^+,X] $. These fluctuations therefore correspond to the components in \eqref{adjointdecomposition} that have $\sigma_3$-egenvalues, $m \neq -J$.
 This is same as the quantum numbers of the decoupled chiral multiplets. We thus establish a one-to-one correspondence between the expected Goldstone multiplets and the decoupled chirals.

\section{$\CN=1$ dualities} \label{sec:duality}

\begin{figure}[h]
	\centering
	\includegraphics[width=3.5in]{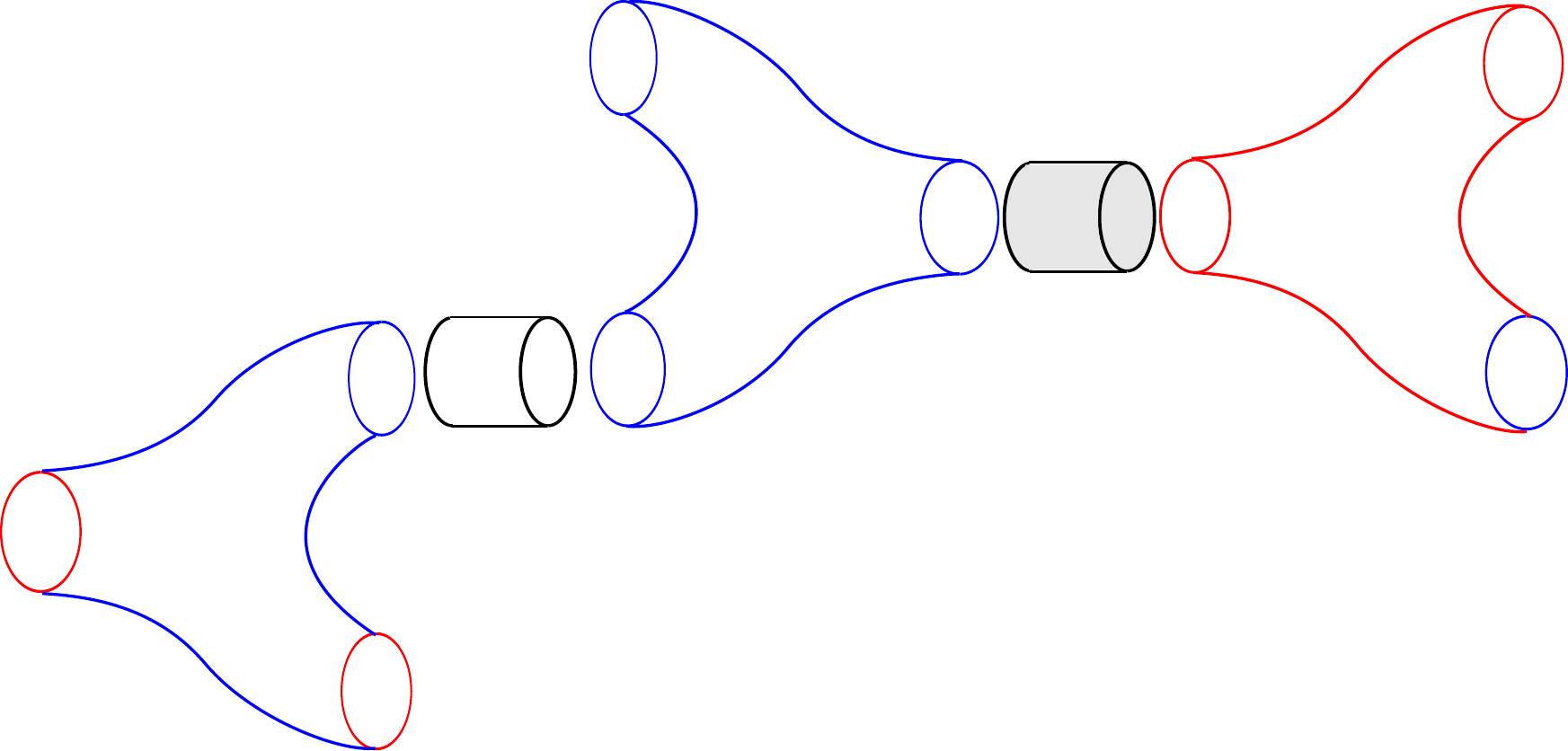}
	\caption{An example of colored pair-of-pants decomposition for $(p, q) = (2, 1)$. The shaded cylinder corresponds to an $\CN=1$ vector multiplet and unshaded one correspond to an $\CN=2$ vector multiplet. We have 3 punctures of opposite color. There is an adjoint chiral attached to each of them. }
	\label{fig:coloredPoP}
\end{figure}

In this section, we discuss various duality frames for an SCFT associated to a UV curve. 
In order to give a UV description of the theory, we need to specify a colored pair-of-pants decomposition. Any Riemann surface with negative Euler number can be decomposed in terms of pairs-of-pants. We assign $\IZ_2$-valued colors to each pairs-of-pants so that the number of $(+, -)$-colored pants are the degrees of the normal bundles $(p, q)$. Different colored pair-of-pants decompositions give rise to different UV descriptions of the same SCFT in the IR. See figure \ref{fig:coloredPoP} for an example. 

\begin{figure}[h]
	\centering
	\includegraphics[width=3.7in]{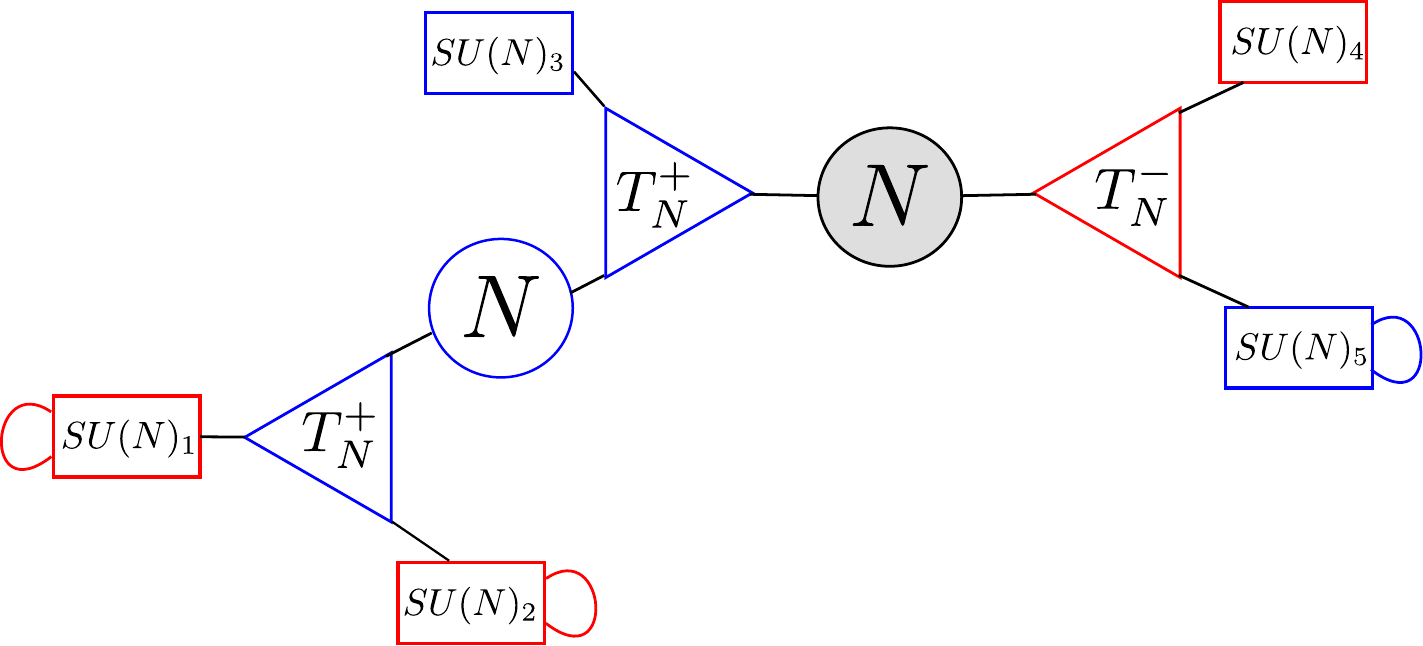}
	\caption{The UV description corresponding to the colored pair-of-pants description of figure \ref{fig:coloredPoP}. Here we assumed all punctures to be maximal. }
	\label{fig:PoPquiver}
\end{figure}

Let us assume all the punctures to be maximal for the moment. For a given colored pair-of-pants, we associate the $T_N$ theory found in \cite{Gaiotto:2009we} which we will review in \ref{sec:ASduality}. For each puncture, we have an operator $\mu_i$ transforms as the adjoint of $SU(N)_i$. When the puncture has a different color from the pair-of-pants itself, we add chiral field $M_i$ transforming as the adjoint of $SU(N)_i$ and also a superpotential $W = \Tr(M_i \mu_i)$. When we glue two pair-of-pants with the same color, we gauge the flavor symmetry with an $\CN=2$ vector multiplet. When gluing two different colored pair-of-pants, we gauge the flavor symmetry by an $\CN=1$ vector multiplet. See figure \ref{fig:PoPquiver}, which is the UV description corresponding to the pair-of-pants decomposition of figure \ref{fig:coloredPoP}. 

\begin{figure}[h]
	\centering
	\includegraphics[width=3.2in]{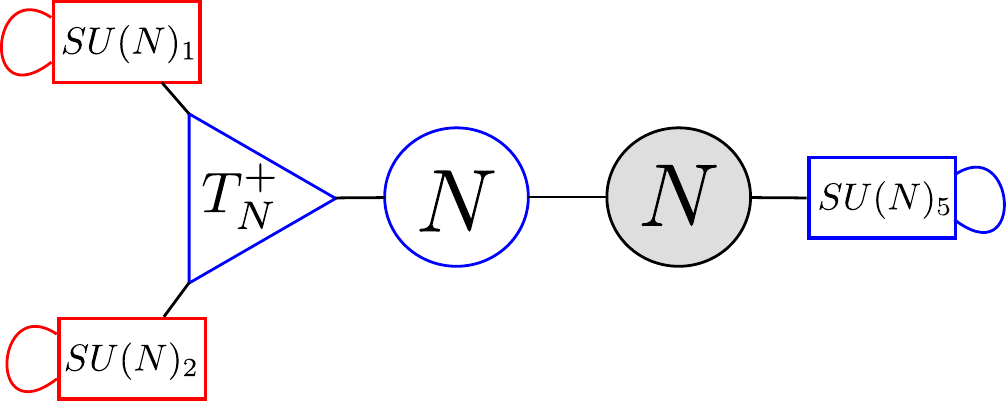}
	\caption{A UV description obtained from partially closing $SU(N)_{3, 4}$ punctures to the minimal punctures.}
	\label{fig:PoPquiverHiggs}
\end{figure}

Non-maximal punctures can be obtained by Higgsing or partially closing the puncture. Let us call $\rho_i$ to be the $SU(2)$ embedding into $\Gamma$ that is used to label the punctures. For a puncture having the same color as the pair-of-pants, Higgsing is implemented through giving a nilpotent vev $\rho_i (\s^+)$ to the operator $\mu_i$, and for an opposite colored puncture, we give a vev to $M_i$ instead. For example, consider the UV description of figure \ref{fig:PoPquiver}. When we Higgs $SU(N)_3$ and $SU(N)_4$ to minimal punctures, we get the theory as in figure \ref{fig:PoPquiverHiggs}. 
Since we closed the punctures that have the same color as the pair-of-pants, we can simply use $\CN=2$ results of \cite{Gaiotto:2009we, Tachikawa:2009rb, Chacaltana:2010ks, Chacaltana:2011ze, Chacaltana:2012ch, Chacaltana:2013oka, Chacaltana:2014jba, Chacaltana:2012zy} to identify the theory corresponding to the pair-of-pants. This is really the same as choosing $\CN=2$ building block and gluing through the $\CN=1$ or $\CN=2$ vector multiplets. 

Things are different when we close the punctures with opposite colors. When we close $SU(N)_1$ to minimal puncture, the theory (in this duality frame) is still non-Lagrangian, but we can identify decoupled operators and global symmetry \cite{Gadde:2013fma}. When we close $SU(N)_5$, we give a vev $\rho_5(\s^+)$ to the chiral superfield $M_5$, from which the quarks acquire nilpotent masses. This theory has a Lagrangian description. As we have seen, this kind of Higgsing yields the Fan labelled by $(N, N'=0)$ and the partition corresponding to $\rho_5$. 

We see that there are many different colored pair-of-pants decompositions for a given UV curve. 
From the six-dimensional perspective, four-dimensional physics in the IR has to be independent from the specific choice of colored pair-of-pants. Therefore we can give equivalent descriptions for the same IR theory from the UV curve and its colored pair-of-pants decompositions. This generalizes the usual Seiberg duality for the $\CN=1$ theories and also Argyres-Seiberg-Gaiotto duality of $\CN=2$ class $\CS$ theories. 

In the rest of this section, we discuss two particular examples. In section \ref{subsec:fanduality}, we study successive application of Seiberg duality on the $\CN=2$ quiver tail connected by an $\CN=1$ gauge node. This illustrates the appearance of the Fan in $\CN=1$ quiver tail. 
In section \ref{sec:ASduality}, we discuss duality of $SU(N)$ SQCD with $2N$ fundamental flavors. We find a dual frame involving the $T_N$ theory and the Fan, which is similar to the strong coupling dual of $\CN=2$ SQCD discovered by Argyres and Seiberg \cite{Argyres:2007cn}. 

\subsection{$\CN=1$ quiver tails} 
\label{sec:n2higgsdual} \label{subsec:fanduality}

Let us consider a UV curve with 5 minimal punctures of $+$ color, 1 minimal puncture of $-$ color, one $+$ colored maximal puncture and one $+$ colored generic puncture labelled by a partition $N=\sum_k k n_k$. We also pick the degrees of normal bundles to be $(p, q)=(5, 1)$. This theory has many different dual frames. We start with a dual frame which resembles the more familiar $\CN=2$ quiver tail and then dualize multiple times to see the various dual frames for the $\CN=1$ quiver tail. 
\begin{figure}[t]
	\centering
	\begin{subfigure}[t]{5in}
	  \centering
	  \includegraphics[width=4.3in]{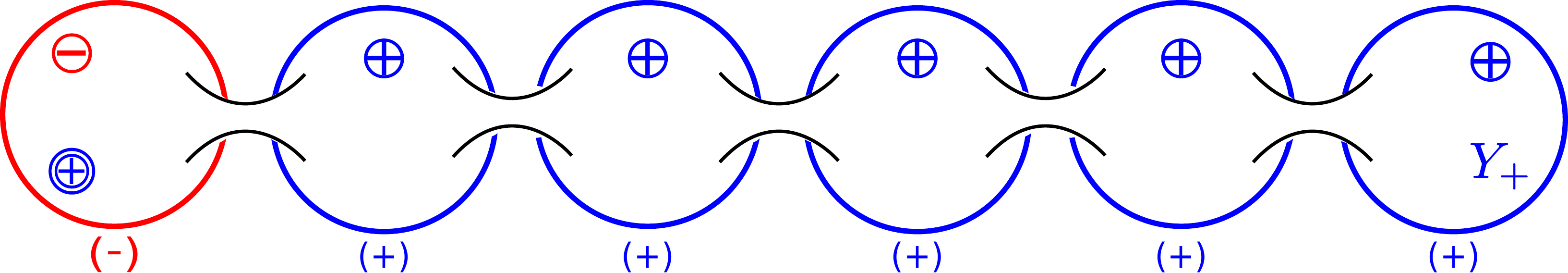}
	  \caption{A colored pair-of-pants decomposition corresponding to the quiver.}
  	  \label{fig:suNN2HiggsCurve0}
	\end{subfigure}
	
	\begin{subfigure}[t]{5in}
	  \centering
	  \includegraphics[width=3.0in]{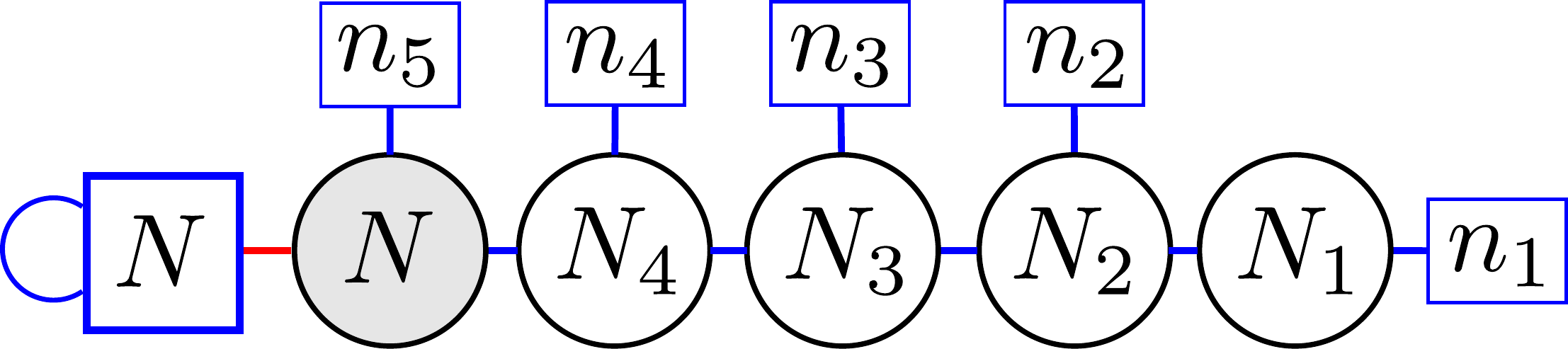}
	  \caption{The quiver tail corresponding to the above colored pair-of-pants decomposition.}
	  \label{fig:suNN2Higgs0}
	\end{subfigure}
	\caption{The quiver tail obtained from $\CN=2$ Higgsing for the partition $N=1  n_1 + 2  n_2 + \ldots +5  n_5$. The rank of gauge group is fixed by $2 N_i = N_{i+1} + N_{i-1} + n_i$. }
\end{figure}

Consider the dual frame given by the colored pair-of-pants decomposition of figure \ref{fig:suNN2HiggsCurve0}. This is essentially the same as the $\CN=2$ quiver tail, so that we get the \ref{fig:suNN2Higgs0}. Only the very last node is gauged via an $\CN=1$ vector multiplet. 

\begin{figure}[h]
	\centering
	\begin{subfigure}[t]{5in}
	  \centering
	  \includegraphics[width=4.3in]{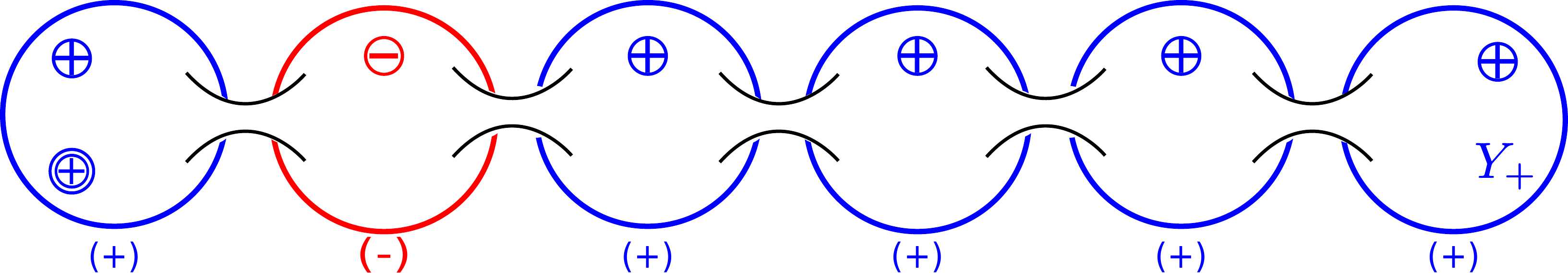}
	  \label{fig:suNN2HiggsCurve1}
	  \caption{A colored pair-of-pants decomposition corresponding to the quiver.}
	\end{subfigure}

	\begin{subfigure}[t]{5in}
	  \centering
	  \includegraphics[width=3.0in]{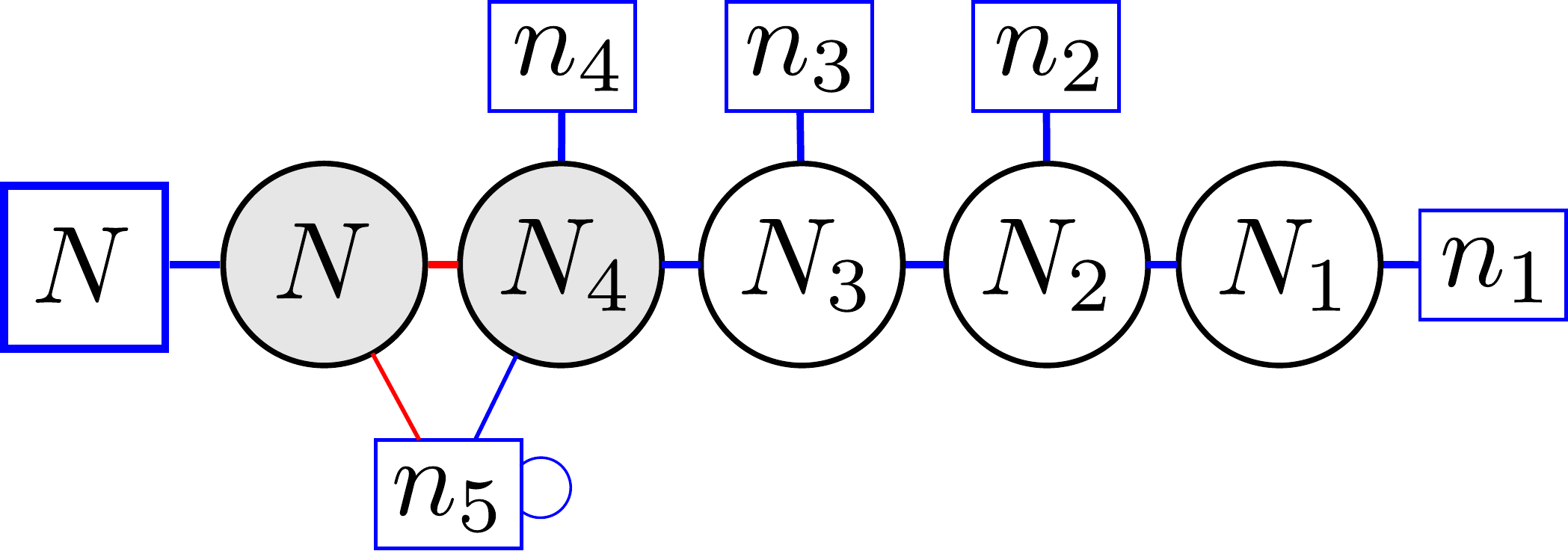}
	  \caption{The quiver tail corresponding to the above colored pair-of-pants decomposition.}
	  \label{fig:suNN2Higgs1}
	\end{subfigure}

	\caption{The quiver tail consists of the $\CN=2$ tail of length 4 and the Fan labelled by $(N, N_4)$ and the partition $N-N_4 = 1 n_5$. }
	\label{fig:suNHiggsDual3}
\end{figure}

Now, if we Seiberg dualize the $\CN=1$ node, we get the 	quiver as shown in figure \ref{fig:suNHiggsDual3}. We see that there is a chiral multiplet dual to the meson formed from the quarks attached at node $n_5$. The dual quarks will have the opposite $\CF$ charge which is depicted by red. Also, there is an additional blue edge connecting $N_4$ and $n_5$ which is the dual to the quark bilinear formed from the $SU(N) \times SU(N_4)$ bifundamental and the fundamental attached at $n_5$ node in figure \ref{fig:suNN2Higgs0}. The rest of the dual mesons become massive from the superpotential.  
\begin{figure}[t]
	\centering
	\begin{subfigure}[t]{5in}
	  \centering
	  \includegraphics[width=4.3in]{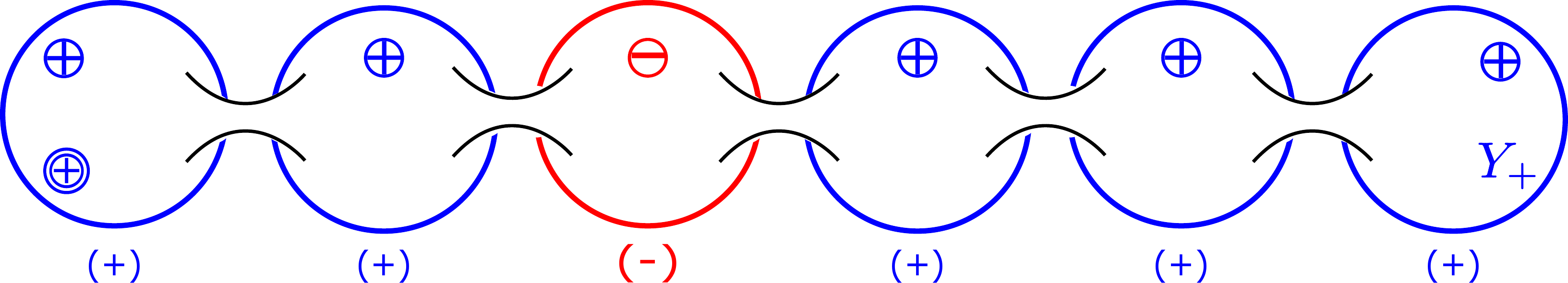}
	  \label{fig:suNN2HiggsCurve}
	  \caption{A colored pair-of-pants decomposition corresponding to the quiver.}
	\end{subfigure}
	\begin{subfigure}[t]{5in}
	  \centering
	  \includegraphics[width=3.0in]{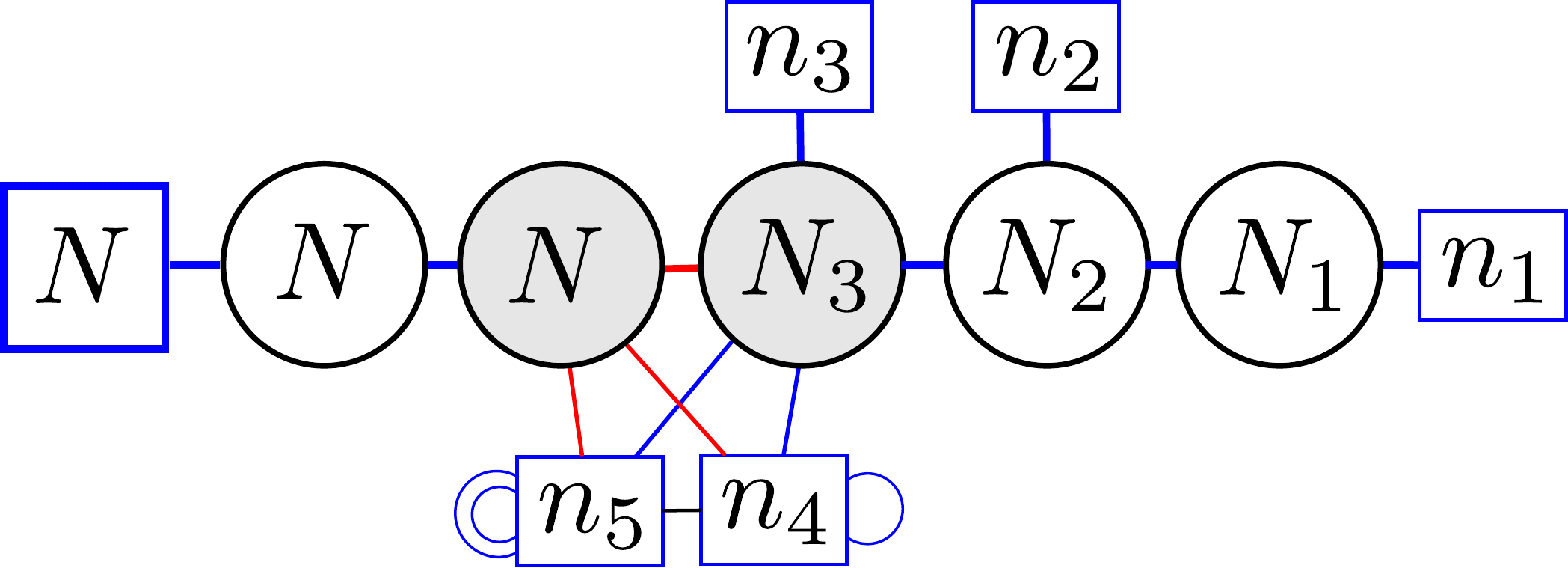}
	  \caption{The quiver tail corresponding to the above colored pair-of-pants decomposition.}
	  \label{fig:suNN2Higgs}
	\end{subfigure}
	\caption{The quiver tail consists of the $\CN=2$ tail of length 3 and the Fan labelled by $(N, N_3)$ and the partition $N-N_3 = 1 n_4 + 2 n_5$. }
	\label{fig:suNHiggsDual2}
\end{figure}
In this frame, we see that there is the Fan labelled by $(N, N_4)$ and the partition $N-N_4 = 1 \cdot n_5$, connecting a shorter $\CN=2$ quiver tail of length 4 and the left-hand segment of the quiver. In terms of nilpotent Higgsing of the linear quiver, the propagation of vev is terminated at the $\CN=1$ node $N_4$, giving us the Fan that glues to the $SU(N)$ gauge node. 

\begin{figure}[h]
	\centering
	\begin{subfigure}[t]{5in}
	  \centering
	  \includegraphics[width=4.3in]{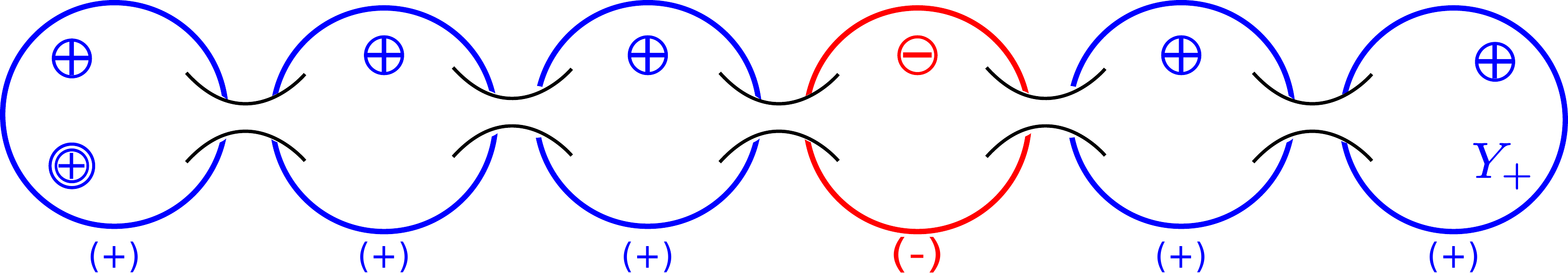}
	  \label{fig:suNN2HiggsCurve}
	  \caption{A colored pair-of-pants decomposition corresponding to the quiver.}
	\end{subfigure}

	\begin{subfigure}[t]{5in}
	  \centering
	  \includegraphics[width=2.7in]{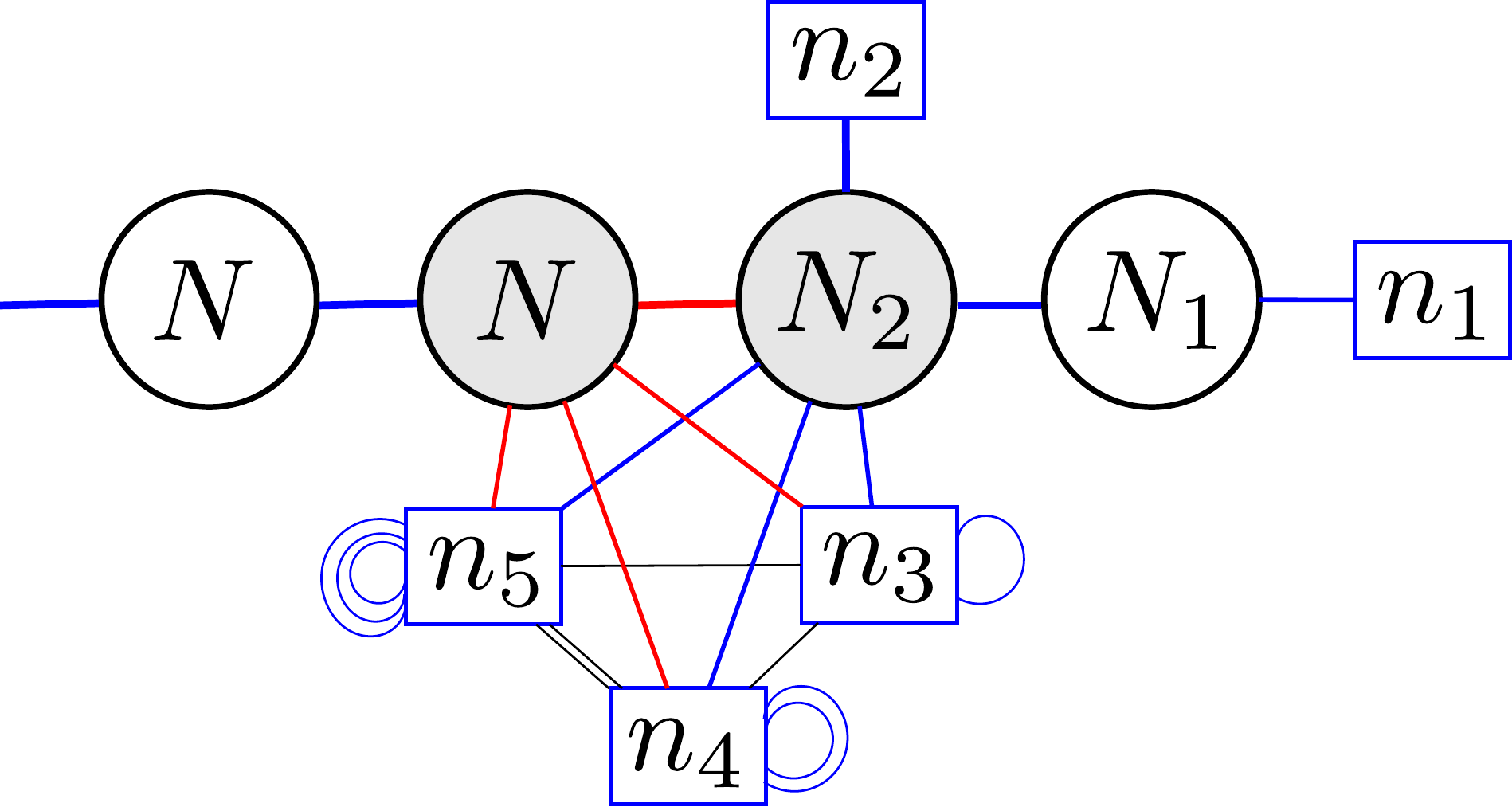}
	  \caption{The quiver tail corresponding to the above colored pair-of-pants decomposition.}
	  \label{fig:suNN2Higgs}
	\end{subfigure}

	\caption{The quiver tail consists of the $\CN=2$ tail of length 2 and the Fan labelled by $(N, N_2)$ and the partition $N-N_2 = 1 n_3 + 2 n_4 + 3 n_5$. }
	\label{fig:suNHiggsDual1}
\end{figure}
Now, we dualize the gauge group $SU(N_4)$ node to get the quiver depicted in figure \ref{fig:suNHiggsDual2}. The flavor node $n_4$ becomes part of the new Fan, which is labelled by $(N, N_3)$ and the partition $N-N_3 = 1 n_4 + 2 n_5$. We see that there is an extra dual meson attached to the $n_5$ node. 

Further dualizing the $SU(N_3)$ node, we get the quiver of figure \ref{fig:suNHiggsDual1}. The flavor node $n_3$ now becomes the part of the Fan, and we get extra dual mesons for each of the preexisting nodes in the Fan. Note that we also have additional chiral multiplets transforming as the bifundamental of $U(n_4) \times U(n_5)$. 
\begin{figure}[h]
	\centering
	\begin{subfigure}[t]{5in}
	  \centering
	  \includegraphics[width=4.3in]{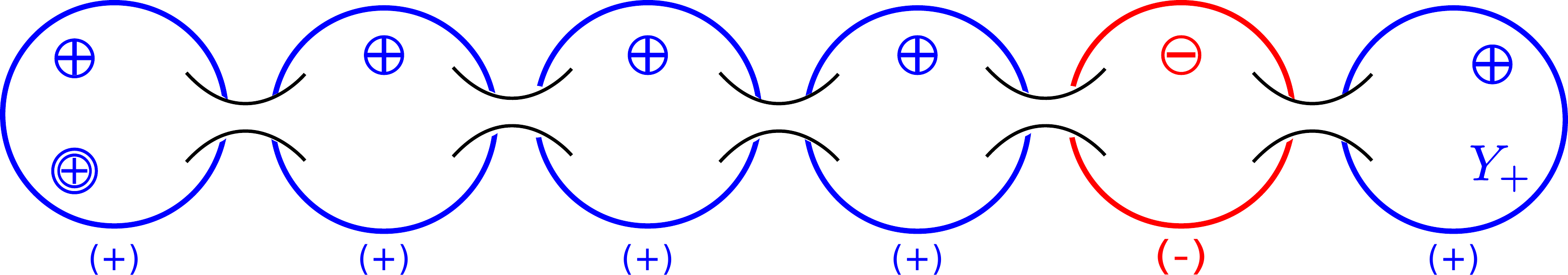}
	  \label{fig:suNN2HiggsCurve}
	  \caption{A colored pair-of-pants decomposition corresponding to the quiver.}
	\end{subfigure}

	\begin{subfigure}[t]{5in}
	  \centering
	  \includegraphics[width=2.2in]{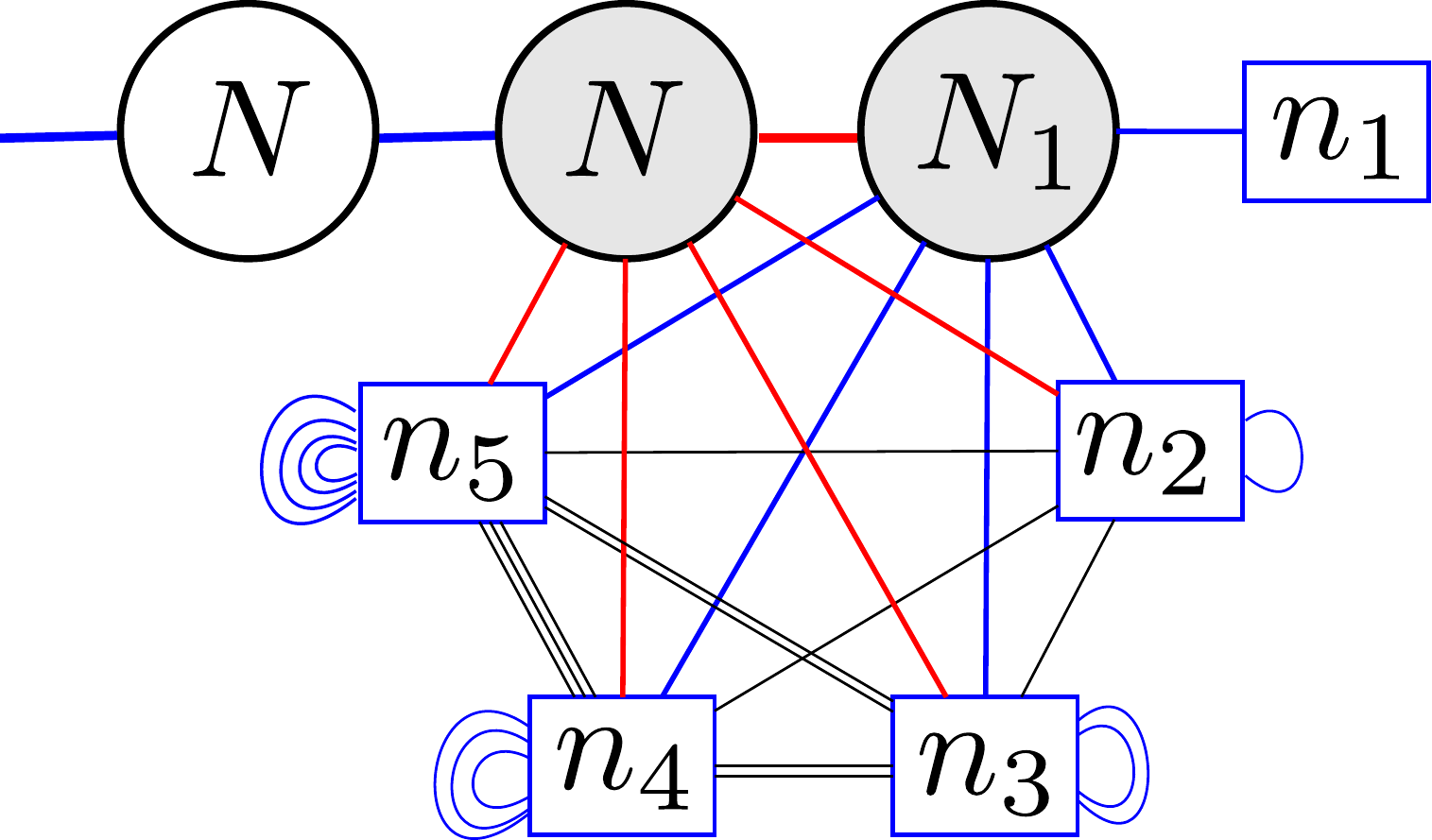}
	  \caption{The quiver tail corresponding to the above colored pair-of-pants decomposition.}
	  \label{fig:suNN2Higgs}
	\end{subfigure}

	\caption{The quiver tail consists of the $\CN=2$ tail of length 1 and the Fan labelled by $(N, N_1)$ and the partition $N-N_1 = 1 n_2 + 2 n_3 + 3 n_4 + 4n_5$. }
	\label{fig:suNdirectHiggs}
\end{figure}

Dualizing once again, we get the quiver tail of figure \ref{fig:suNdirectHiggs}. Once again, the flavor node $n_2$ becomes a part of the Fan, and chiral multiplets get added. This quiver tail can also be obtained from starting with the linear quiver and Higgsing $\mu_0 = ( \widetilde{Q}_0 Q_0 )_{\textrm{adj}}$ directly by a nilpotent vev associated to the partition $N=\sum_k k n_k$. We see that the Higgsing does not propagate beyond $N_1$. All the flavor nodes are attached to $N_1$ and its neighbor $N$. 

\begin{figure}[h]
	\centering
	\begin{subfigure}[t]{6in}
	  \centering
	  \includegraphics[width=4.3in]{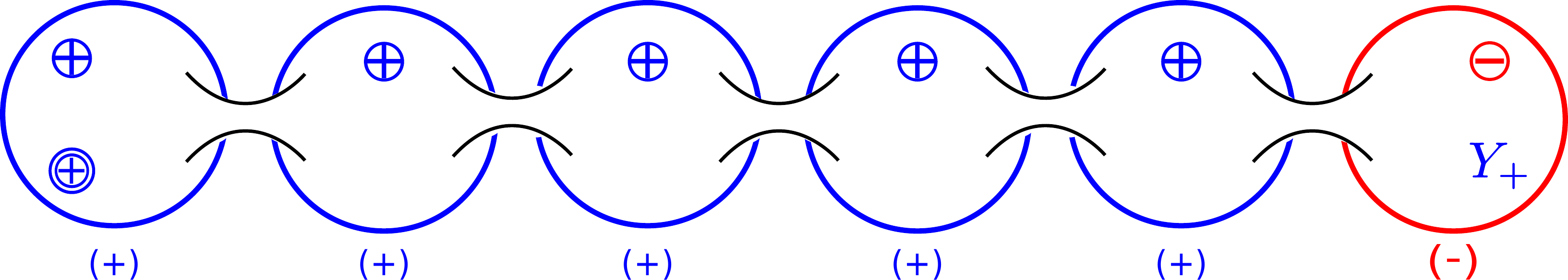}
	  \label{fig:suNN2HiggsCurve}
	  \caption{A colored pair-of-pants decomposition corresponding to the quiver.}
	\end{subfigure}

	\begin{subfigure}[t]{6in}
	  \centering
	  \includegraphics[width=2.8in]{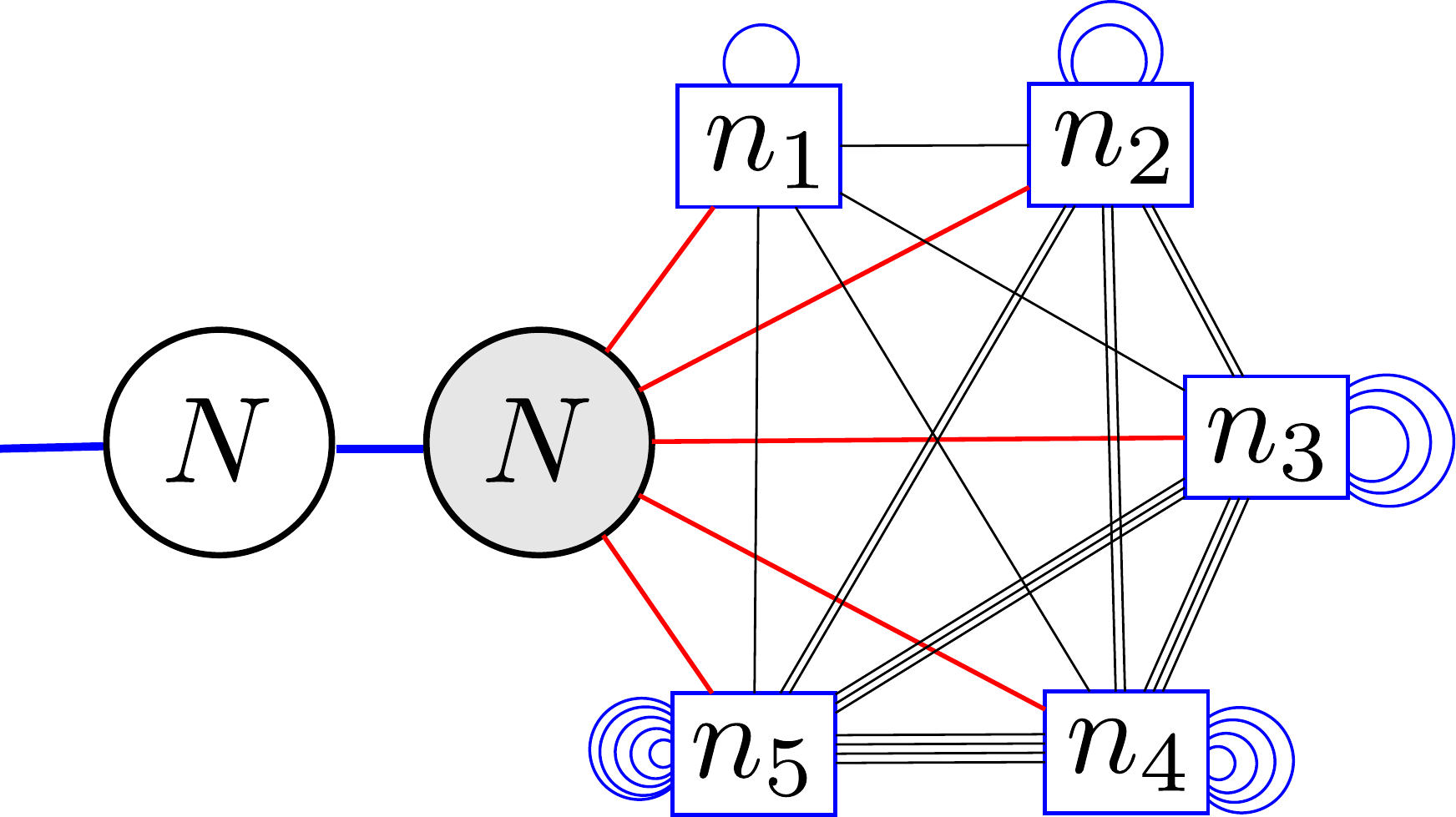}
	  \caption{The quiver tail corresponding to the above colored pair-of-pants decomposition.}
	  \label{fig:suNN2Higgs}
	\end{subfigure}

	\caption{The quiver tail consists of the maximal Fan of size $\ell=5$, labelled by $(N, 0)$ and the partition $N=\sum_k k n_k$. }
	\label{fig:suNdualHiggs}
\end{figure}

Now finally, upon dualizing the $SU(N_1)$ gauge node, we get the theory as in the figure \ref{fig:suNdualHiggs}. This gives us the Fan of size $\ell=5$ labelled by $(N, 0)$ and the partition $N=\sum_{k=1}^5 k n_k$ attached to the right end of the quiver. 

We see that there are many different quiver tail descriptions for a given choice of punctures in $\CN=1$ class $\CS$ theories. In the above example, we have only described UV frames that have Lagrangian descriptions. For these cases, all the pairs-of-pants have the same color as the minimal puncture inside. In general, one can also consider a dual frame which has a different colored puncture inside its pair-of-pants. Then the dual frame has a sector with no Lagrangian description. We will discuss such a case in the next section. 

\subsection{$\CN=1$ analog of Argyres-Seiberg duality}
\label{sec:ASduality}

\begin{figure}[h]
	\centering
	\includegraphics[width=2.5in]{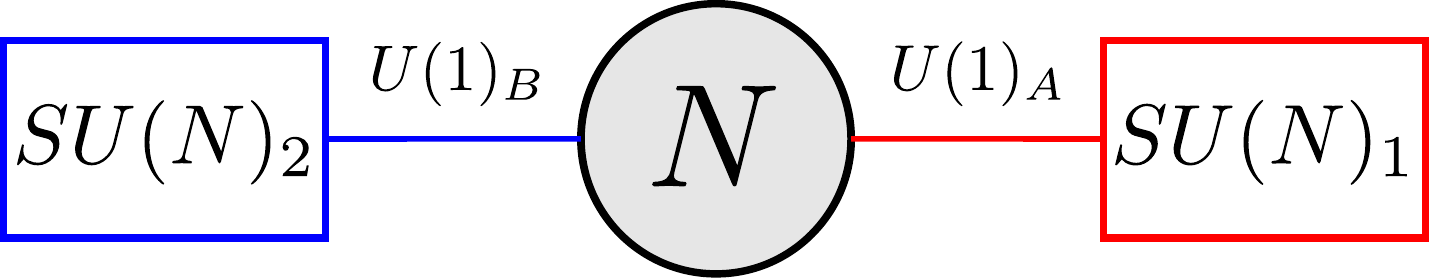}
	\caption{The quiver diagram of $SU(N)$ SQCD with $2N$ flavors.} \label{fig:ElectricSQCDQuiver}
\end{figure}

 In this section we use the Fan to provide a new dual description of $\CN=1$ $SU(N)$ SQCD with $2N$ flavors
  with the quartic coupling \eqref{N=1superpotential} with $i=1$.  This is the $(\s_{-1}, \s_0, \s_1, \s_2) = (-1,1,-1,1)$ linear quiver as described in section \ref{subsec:linearquiver}.   
 The flavor symmetry of the theory is $SU(N)_{1} \times SU(N)_{2} \times U(1)_{A} \times U(1)_{B}$.  We summarize the matter content in table \ref{table:ele} and quiver in figure \ref{fig:ElectricSQCDQuiver}. 
 In this section it is more convenient to use the symmetries $R_0$ and $\mathcal{F}$ defined in \eqref{eq:RandJ}.  
  
    \begin{table}[h]
    \begin{center}
    \begin{tabular}{|c||c|c|c|c|c|c|c|}
    \hline
        & $SU(N)_{g}$ & $SU(N)_{1}$ & $SU(N)_{2}$ & $U(1)_{R_{0}}$ & $U(1)_{\CF}$ & $U(1)_{A}$ & $U(1)_{B}$ \\ \hline \hline
    $(Q_{0},\widetilde{Q}_0)$ & $(\Box,\bar{\Box})$ & $(\bar{\Box},\Box)$& $\cdot$  & 1/2 & $-1/2$ & $(1,-1)$ & $\cdot$ \\ \hline
    $(Q_{1},\widetilde{Q}_1)$ & $(\bar{\Box},\Box)$& $\cdot$ & $(\Box,\bar{\Box})$ & 1/2 & $1/2$ & $\cdot$ & $(1,-1)$ \\ \hline
    \end{tabular}
    \caption{Charges of matter multiplets in SQCD.}
    \label{table:ele}
    \end{center}
    \end{table}    
    
\begin{figure}[b]
\centering
	\begin{subfigure}[b]{2.8in}
	\centering
	\includegraphics[width=2.0in]{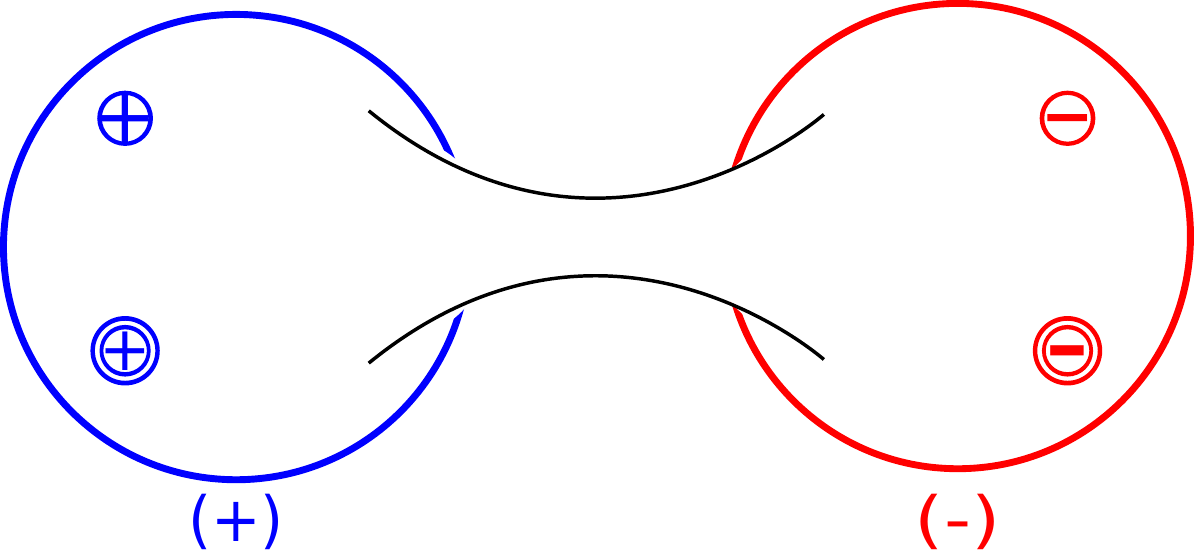}
	\caption{Electric SQCD} \label{fig:ElectricSQCD}
	\end{subfigure}
	\begin{subfigure}[b]{2.8in}
	\centering
	\includegraphics[width=2.0in]{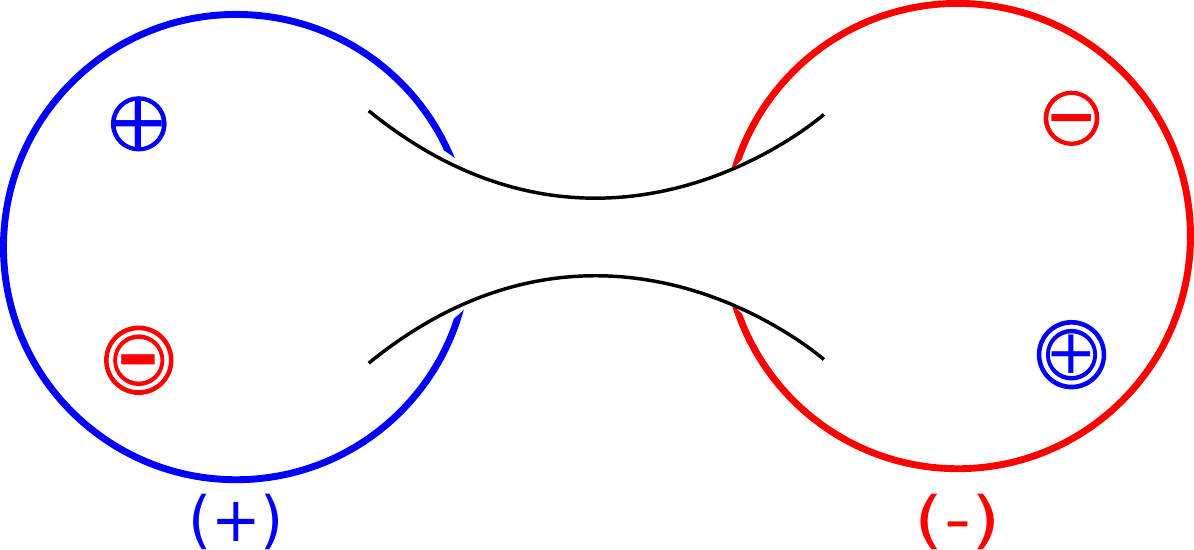}
	\caption{Magnetic SQCD} \label{fig:MagneticSQCD}
	\end{subfigure}

	\begin{subfigure}[b]{2.8in}
	\centering
	\includegraphics[width=2.0in]{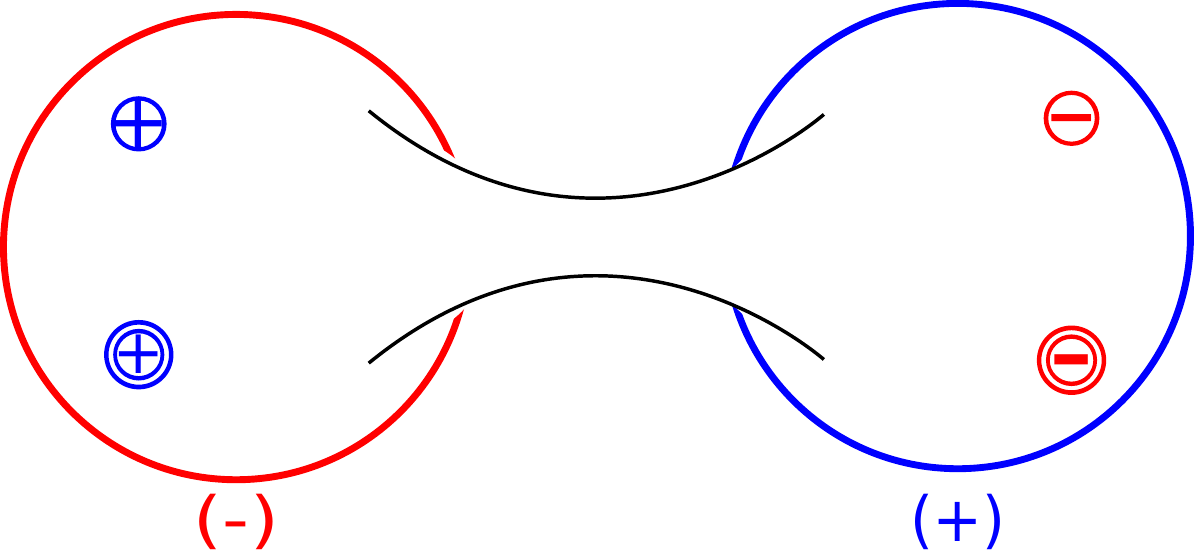}
	\caption{Swapped SQCD} \label{fig:SwappedSQCD}
	\end{subfigure}
	\begin{subfigure}[b]{2.8in}
	\centering
	\includegraphics[width=2.0in]{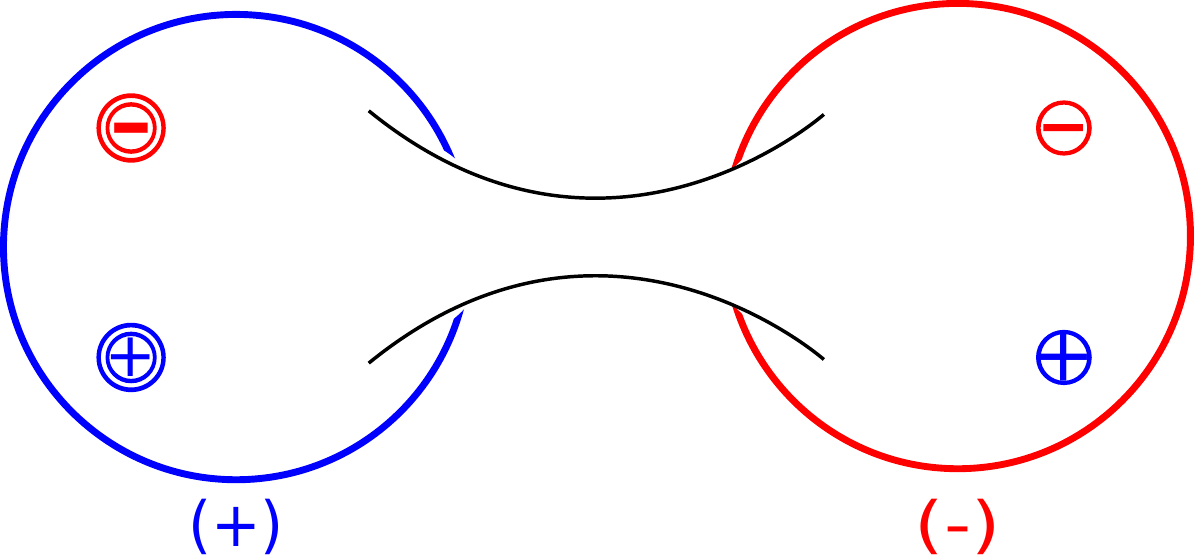}
	\caption{Argyres-Seiberg dual} \label{fig:ASDualSQCD}
	\end{subfigure}
	\caption{Colored pair-of-pants decompositions of the UV curve corresponding to the SQCD with $SU(N)$ gauge group and $2N$ flavors and its dual descriptions.}
	\label{fig:SQCDCurve}
\end{figure}

  It has been pointed out in \cite{Gadde:2013fma} that there are two dual descriptions of the SQCD.
  Let us shortly explain these here.
  One of them is $\CN=1$ $SU(N)$ SQCD with $2N$ flavors with a chiral multiplet in the adjoint representation 
  of $SU(N)_{1}$ and a chiral multiplet in the adjoint of $SU(N)_{2}$
  coupled by the cubic interaction with quarks.
  This is indeed the Seiberg dual theory of the original SQCD with the quartic coupling. 
  In terms of the Riemann surface this is understood as the exchange of the maximal punctures 
  as in figures \ref{fig:MagneticSQCD}.
  Other dual description whose Lagrangian is not known corresponds 
  to the exchange of the minimal punctures as in figure \ref{fig:SwappedSQCD}.
  To obtain this theory, we first consider an $\CN=1$ $SU(N)$ gauge theory coupled to two $T_{N}$ theories 
  \cite{Gaiotto:2009we} (which will be reviewed below) and to two chiral multiplets, 
  which are the adjoints of $SU(N)_{A}$ and $SU(N)_{B}$ flavor symmetries of the two $T_{N}$ theories respectively.
  This is associated to the Riemann surface where all the punctures are maximal, 
  but the color assignment is same as in \ref{fig:SwappedSQCD}.
  Then the dual description is obtained by Higgsing of $SU(N)_{A}$ and $SU(N)_{B}$ symmetries 
  down to $U(1)_{A}$ and $U(1)_{B}$.
  
  In this section we will find a third dual description of the SQCD 
  corresponding to the figure \ref{fig:ASDualSQCD}.
  Since the UV description involves the $T_{N}$ theory, we will review relevant details first.
  
  The $T_{N}$ theory is obtained by compactifying $N$ coincident M5-branes, with $\CN=2$ twist, on a sphere with three maximal punctures. 
  Each puncture carries an $SU(N)$ global symmetry, thereby leading to an $SU(N)^3$ flavor symmetry.
  It is an $\CN=2$ SCFT and it admits $U(1)_{\CN=2} \times SU(2)_{R}$ $R$-symmetry.
 When we describe it as an $\CN=1$ SCFT, we use the $\CN=2$ $R$-symmetry to write $R_0$ and $\CF$ as
    \bea
    R_{0}
     =     \frac{1}{2} R_{\CN=2} + I_{3}, ~~~
    \CF
     =   - \frac{1}{2} R_{\CN=2} + I_{3}
    \eea
  where $R_{\CN=2}$ and $I_{3}$ are generators of $U(1)_{\CN=2}$ and 
  the diagonal $U(1)$ of the $SU(2)_{R}$ respectively.
  This theory has chiral operators $\mu_{i}$ 
  ($i$ labels the three $SU(N)$ flavor symmetries) which are the moment maps of the $SU(N)$ flavor symmetries. 
  It also has operators $Q^{(k)}$ transforming in the $k$-th antisymmetric representation of all three $SU(N)$ symmetries
  \cite{Gaiotto:2008nz,Gadde:2011uv,Maruyoshi:2013hja}.
  Their $R_{0}$ and $\CF$ charges are
    \bea
    &&
    R_{0}(\mu_{i})
     =     \CF(\mu_{i})
     =     1,~~~~
    R_{0}(Q^{(k)})
     =     \CF(Q^{(k)})
     =     \frac{k(N-k)}{2}. 
    \eea

  \begin{figure}[t]
	\centering
	\includegraphics[width=3.0in]{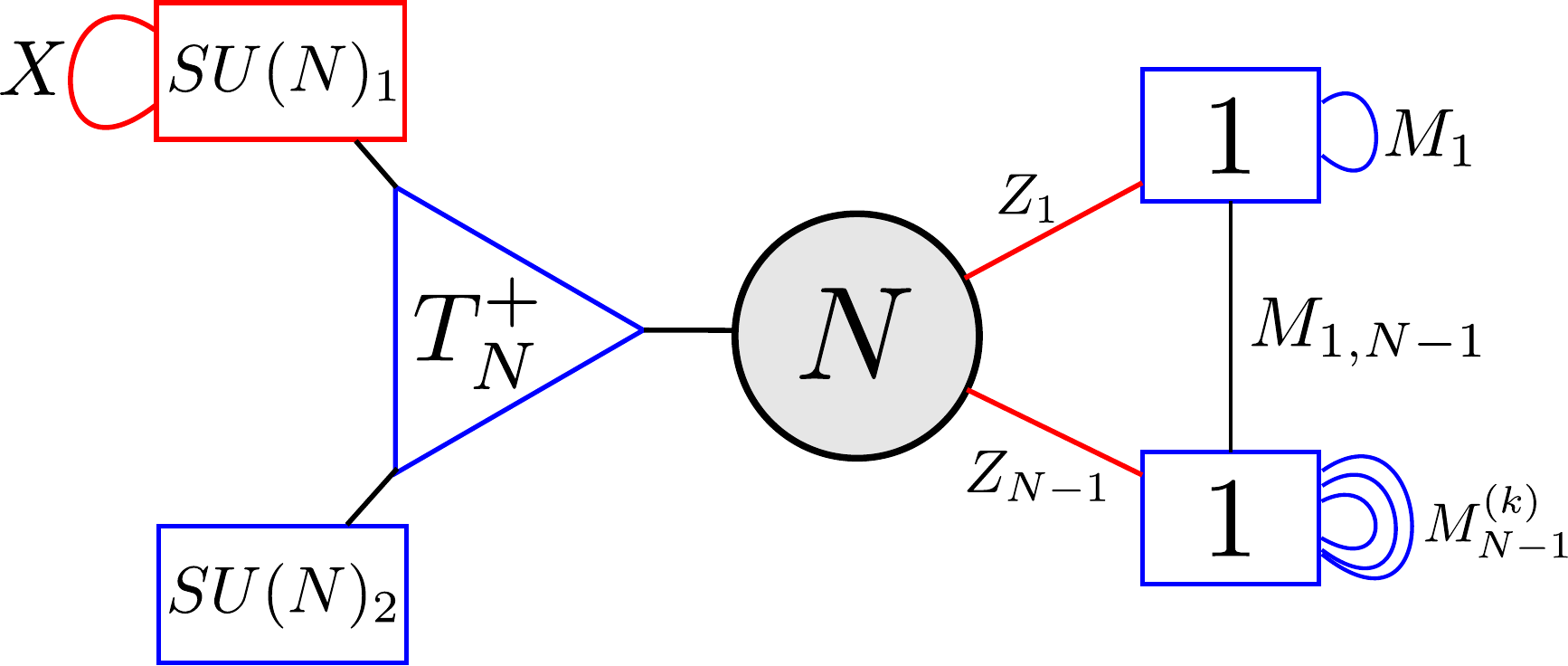}
	\caption{Analog of Argyres-Seiberg dual to the $\CN=1$ $SU(N)$ SQCD with $2N$ flavors. }
	\label{fig:SQCDAS}
\end{figure}

  The results of section \ref{sec:fan} tell us that figure \ref{fig:ASDualSQCD} represents
  an $SU(N)$ gauge theory coupled to the Fan with $\sigma=-1$ labelled by $(N, 0)$ and 
  a partition $N=1+(N-1)$, i.e., $\ell =N-1$, $n_1 = n_{N-1}=1$ and $n_i=0$ otherwise. 
  It is coupled to the $T_{N}$ theory by gauging an $SU(N)$ flavor symmetry.
  Furthermore, a chiral field $X$ transforming in the adjoint representation 
  of $SU(N)_1$ flavor symmetry of the $T_{N}$ theory is added.
 $SU(N)_{1,2}$ are the flavor symmetries of the $T_{N}$ theory which are not gauged.   

The dual theory is described by the quiver in figure \ref{fig:SQCDAS}.  The matter content is summarized in table \ref{table:mag}.  
For convenience of the discussion, we write fields from the Fan as $M_{1} :=M_{1,1}^{(0)}$, $M_{N-1}^{(k)} := M_{N-1,N-1}^{(N-1-k)}$ and $z:= Z_{N-1}$.  
  
The important data needed in including the $T_N$ in these quivers is its contribution to the anomalies.  These are described in section \ref{sec:anomaly}.  For the purpose of the quiver in \ref{fig:SQCDAS}, the contribution of $T_N$ to the chiral anomalies $(R_0 SU(N)^2,\CF SU(N)^2)$ is the same as $N$ fundamental $(J_+, J_-) = (1,0)$ hypermultiplets.   
    \begin{table}[h]
    \begin{center}
    \begin{tabular}{|c||c|c|c|c|c|c|c|}
    \hline
        & $SU(N)_{g}$ & $SU(N)_{1}$ & $SU(N)_{2}$ & $U(1)_{R_0}$ & $U(1)_{\CF}$ & $U(1)_1$ & $U(1)_{N-1}$ \\ \hline \hline
    $(Z_{1},\widetilde{Z}_1)$ & $(\Box,\bar{\Box})$ & $\cdot$ & $\cdot$ & 1/2 & $-1/2$ & $(-1,1)$ & $\cdot$ \\ \hline
    $(z,\tilde{z})$ & $(\Box,\bar{\Box})$ & $\cdot$ & $\cdot$ & $\frac{3-N}{2}$   & $\frac{1-N}{2}$ & $\cdot$ & $(-1,1)$ \\ \hline
    $M_{1}$ &      $\cdot$     & $\cdot$ & $\cdot$ & 1   & $ 1$ & $\cdot$ & $\cdot$          \\ \hline
    $(M_{1,N-1}$,$M_{N-1,1}$) &  $\cdot$  & $\cdot$ & $\cdot$ & $N/2$ & $ N/2$ & $(1,-1)$ & $(-1,1)$      \\ \hline
    $M_{N-1}^{(k=1, \cdots, N-1)}$ &  $\cdot$   & $\cdot$ & $\cdot$ & $k$  & $k$ & $\cdot$ &  $\cdot$          \\ \hline
    $X$     & $\cdot$  & adj & $\cdot$ & 1   & $-1$ & $\cdot$ & $\cdot$ \\ \hline
    \end{tabular}
    \caption{Charges of matter multiplets in the dual theory,
    where $M_{1}:=M_{1,1}^{(0)}$ and $M_{N-1}^{(k)} := M_{N-1,N-1}^{(N-1-k)}$ and $z= Z_{N-1}$. }
    \label{table:mag}
    \end{center}
    \end{table}
  
Finally, one linear combination of $M_1$ and $M_{N-1}^{(1)}$ must be projected out.  We denote the combination that survives as $\hat{M}_1$.  We can then write the superpotential as  
    \bea
    W_{m}
    &=&    \hat{M}_{1} ( \tr z \mu_{g}^{N-2} \tilde{z} + \tr Z_{1} \widetilde{Z}_{1}) + 
          \sum_\alpha \sum_{k=2}^{N-1} \mathcal{M}^{(k),\alpha}_{N-1} \tr z \mu_{g}^{N-1-k} \tilde{z}
               \nonumber \\
    & &  + M_{1,N-1} \tr Z_{1} \tilde{z} + {M}_{N-1, 1} \tr z \widetilde{Z}_{1}
          + \tr \mu_{1} X + \tr Z_{1} \mu_{g} \widetilde{Z}_{1} + \tr z \mu_{g}^{N-1} \tilde{z},
    \eea
  where $\mu_{1}$ and $\mu_{g}$ are the moment maps 
  of $SU(N)_{1}$ and $SU(N)_{g}$ symmetries respectively.  The set of operators, $\mathcal{M}^{(k),\alpha}_{N-1}$ correspond to all possible composite operators with charge $(2k,0)$.  
  
  Note that this is reminiscent of the Argyres-Seiberg duality \cite{Argyres:2007cn} 
  of $\CN=2$ $SU(3)$ SQCD with six flavors.
 Indeed, if we consider the analogous UV curve in the $\CN=2$ setting without color assignments, this dual frame is exactly that of Argyres-Seiberg when $N=3$. 
  The duality presented here is an $\CN=1$ analog of that. It will be interesting to derive this duality through the technique of inherited duality \cite{Argyres:1996eh, Argyres:1999xu}.
  
  We identify $U(1)_A$ and $U(1)_B$ of the SQCD as
  \begin{equation}
  U(1)_A = U(1)_1 + U(1)_{N-1}, \qquad U(1)_B = (N-1) U(1)_1 -  U(1)_{N-1}.  
  \end{equation} 
  It is a straightforward calculation to show that all the anomaly coefficients of the flavor symmetries 
  agree on both sides of the duality.
  We will see this in section \ref{sec:anomaly}.
  In section \ref{sec:index}, we will also see the agreement of the superconformal index of both theories. 
  This will be the strongest check of the duality.


\section{Anomalies and central charges }
\label{sec:anomaly}

  In this section we compute the 't Hooft anomaly coefficients of various objects.
  In section \ref{subsec:anomalyfan}, we start with computing those of the Fan 
  introduced in section \ref{subsec:fan}.
  We then interpret the results in terms of a sphere with punctures
  and give a concise expression for the anomaly coefficients of the class $\CS$ theories in general, 
  in section \ref{subsec:anomalyS}.

\subsection{Anomalies of the Fan}
\label{subsec:anomalyfan}
  The matter content of the Fan labelled by $(N, N'=0)$ and a partition $N = \sum_{k=1}^\ell k n_k$ with $\sigma = -1$ is given in the table \ref{table:fan}.
  One can choose $\s = +1$ by swapping $J_+$ and $J_-$ charges. 
  In evaluating the anomalies, it is useful to write them in terms of
    \begin{equation}
    N_i
     =     \sum_{k=1}^i n_k k + i \sum_{k=i+1} n_k, 
           \label{Ni}
    \end{equation} 
  and to notice the following identity
    \begin{equation}
    N^2
     =     2 \sum_{i=1}^\ell N_i \sum_{j=i}^\ell n_j - \sum_{i=1}^\ell N_i n_i.
           \label{N^2}
    \end{equation}
  We find the 't Hooft anomaly coefficients of $SU(n_{i})$ and $U(1)_{i}$ for $i\geq 2$ are
    \bea &&
    \mbox{Tr} T_i^{2} R_0
     =     \sigma \mbox{Tr} T_i^{2} \CF
     =   - \frac{1}{2} N_i,
            \nonumber \\ &&
    \mbox{Tr} U_{i}^{2} R_0 
     =   - N_i -i  (i-1)  n_i ,~~~~~
    \sigma \mbox{Tr} U_{i}^{2} \CF 
     =    - N_i +i  (i+1)   n_i,
    \eea
  where $T_{i}$ and $U_{i}$ are the generators of $SU(n_{i})$ and $U(1)_{i}$ respectively.
  The other anomaly coefficients are given by
    \bea
    \mbox{Tr}R_0 &=& - \sum_{i=1}^\ell N_i \sum_{j=i}^\ell n_j, ~~~~~
    \sigma \mbox{Tr}\CF = \sum_{i=1}^\ell N_i \sum_{j=i+1}^\ell n_j + 1,
    \label{anomalyfan1}
    \\
    \mbox{Tr}\CF^a R_0^{3-a} 
    &=&  - \frac{(-\sigma)^a}{4} \sum_{i=1}^\ell n_i \left[\sum_{j=1}^{i} n_j \left( i^3 j - f_{a} (i,j) \right)
         + \sum_{j=i+1}^{\ell} n_j \left( i^3 j - f_{a} (j,i) \right) \right],
    \label{anomalyfan2}
    \eea 
  where
    \begin{equation}
    f_{a}(i,j)
     =     \frac{1}2 \sum_{p=0}^{j-1} \left(i+j-2p -2\right)^{3-a} \left(i+j-2p \right)^a.
    \end{equation} 
  Writing explicitly,
    \begin{align}
    \mbox{Tr} R_0 \CF^2 
    &=     \frac{1}{4} \sum_{i=1}^\ell \left( N^2-N_i^2 \right)
         - \frac{1}{4} \sum_{i=1}^\ell N_i \sum_{j=i}^\ell n_j,    
           \\ 
    \sigma \mbox{Tr} R_0^2 \CF 
    &=     \frac{1}{4} \sum_{i=1}^\ell \left( N^2-N_i^2 \right)
         + \frac{1}{2} N^2 - \frac{1}{4} \sum_{i=1}^\ell N_i \sum_{j=i}^\ell n_j.  
    \end{align}  
  We found that the rest can be obtained from
    \begin{align}
    \mbox{Tr} \CF^3 = \mbox{Tr} \CF - 3 \mbox{Tr} \CF R^2_0, ~~~~
    \mbox{Tr} R_0^3 = \mbox{Tr} R_0 - 3 \mbox{Tr} \CF^2 R_0.
    \label{anomalyrelation}
    \end{align}

\paragraph{From Linear quiver}
  The above anomalies can also be obtained by a rather indirect way.
  The idea is to use the duality: 
  as we saw in section \ref{subsec:fanduality}, the Fan was obtained by taking various Seiberg dualities 
  to the linear quiver theories with $\CN=1$ $SU(N)$ gauge theory coupled to $\CN=2$ quiver tail 
  labelled by partitions of $N$: $N= \sum_{k=1}^\ell n_k k$.
  Thus, let us first focus on this original theory. 
  This theory has gauge symmetry $G = \prod_{k=1}^\ell SU(N_k)$ with \eqref{Ni}.
  Notice that $N_\ell =N$.
  All the gauge groups except for the $\ell$-th one are $\CN=2$.
  In addition to the bifundamentals, 
  there are $n_i$ fundamental hypermultiplets attached to the $SU(N_i)$ gauge group. 
  The tail has a label $\sigma=\pm 1$ depending on the $\CF$-charge of the matter fields $\CF =\sigma/2$.
  (The $\CF$-charge of the chiral adjoint multiplets of the gauge symmetry is $-\sigma$.)  
  We end the quiver by adding $N$ fundamental hypers 
  with $R_{0} = 1/2$ and $\CF = - \sigma/2$ to $SU(N_\ell)$ gauge group.  
  We further attach a chiral multiplet ($R_{0}=1$ and $\CF=\sigma$) 
  in the adjoint representation of the $SU(N)$ flavor symmetry of $N$ hypers.
  
  Then, the 't Hooft anomaly coefficients of $R_0$ and $\CF$ of this theory are given as
    \bea
    \mbox{Tr}R_0 
    &=&  - \ell -\sum_{i=1}^\ell N_i \sum_{j=i}^\ell n_j, ~~~
    \mbox{Tr}R_0^3 
     =     \frac{1}{4} \mbox{Tr}R_0 + \frac{3}{4} \sum_{i=1}^\ell \left(N_i^2 -1\right) ,
           \label{anomalyN=2quiver1} \\
    \mbox{Tr}\CF 
    &=&  - \sigma(2 + \mbox{Tr} R_0), ~~~
    \mbox{Tr}\CF^3 
     =     \frac{1}{4} \mbox{Tr}\CF - \frac{3\sigma}{4} \left[\sum_{i=1}^{\ell} \left(N_i^2-1 \right) 
         - 2  \left(N^2 -1 \right) \right],
           \label{anomalyN=2quiver2} \\
    \mbox{Tr} R_0 \CF^2 
    &=&    \frac{1}{4} \mbox{Tr}R_0 - \frac{1}{4} \sum_{i=1}^\ell (N_i^2 -1), ~~~
    \mbox{Tr} R^2_0 \CF 
     =   - \sigma \left( \mbox{Tr} R_0 F^2 + \frac{1}{2} N^2 \right). 
           \label{anomalyN=2quiver3} 
    \eea
  Again we note that they satisfy \eqref{anomalyrelation}.
 
  After repeatedly applying the Seiberg dualities, we end up with 
  an $\mathcal{N}=2$ linear quiver attached to the Fan.
  The quiver has $\ell$ gauge nodes with $SU(N)$ gauge groups linked together by bifundamentals
  with $R_{0} = 1/2$ and $\CF = \sigma/2$.
  All gauge groups are $\mathcal{N}=2$ vector multiplets except for the one at $k=1$.
  The Fan (with $- \sigma$) is attached to this $k=1$ $\CN=1$ gauge node.
  By subtracting the contribution of this quiver except for the Fan from \eqref{anomalyN=2quiver1},
  \eqref{anomalyN=2quiver2} and \eqref{anomalyN=2quiver3}, 
  we reproduce the anomaly coefficients \eqref{anomalyfan1} and \eqref{anomalyfan2}.
  Note that for $\Tr \CF$, $\Tr \CF^{3}$ and $\Tr R_{0}^{2} \CF$, there are overall sign differences from 
  \eqref{anomalyfan1} and \eqref{anomalyfan2}.
  This is because the Fan appeared here is specified by $- \sigma$. 

\subsection{Anomalies of class $\CS$ theories}
\label{subsec:anomalyS}
  So far we have computed the anomaly coefficients of the Fan.
  In the class $\CS$ point of view, the Fan with $\sigma=+1$ is associated to a sphere ($p=1, q=0$)
  with a maximal puncture with $\sigma =+1$, a minimal puncture with $\sigma=+1$
  and a puncture labeled by $Y$ with $\sigma = -1$ or the opposite choice.
  Here we will show that the anomaly coefficients can be given in terms of the data of
  the Riemann sphere and the punctures.
  By generalizing this observation, 
  we will conjecture that the anomaly coefficients of the class-$\CS$ theories can be written down 
  as a sum of contributions from the following: 
    \begin{itemize}
    \item Background contribution from the curve: $\CC_{g, n}$ with normal bundle 
          $\CL(p) \oplus \CL(q)$ specified. Here $p+q = 2g-2+n$ is imposed. 
    \item Local contributions from each puncture $(\rho, \s)_{i=1, \ldots n}$. 
    \end{itemize}
  If we write the number of punctures with color $\s$ to be $n_\s$,  $n=n_+ + n_-$ is the total number of punctures. 
  We will first summarize the case of $\CN=2$ theories, which have been worked out in full generality 
  by \cite{Chacaltana:2012zy}, and then give a generalization to the $\CN=1$ theories. 

  In the $\CN=2$ case, we always set $q = 0$ and $n_- = 0$  
  so that the total space becomes the cotangent bundle of the Riemann surface $\CC_{g, n}$. 
  All the punctures have the same color, thus they are specified entirely by the embedding of $SU(2)$ 
  into $\Gamma$ labeling the class $\CS$ theory. 
  For these $\CN=2$ theories, the number of effective vector multiplets $n_v$ and hypermultiplets $n_h$ 
  can be used to determine the anomaly coefficients of the $\CN=2$ $R$-symmetries:
    \bea
    \Tr R_{\CN=2}
     =     \tr R_{\CN=2}^{3}
     =     2(n_{v} - n_{h}), ~~~~
    \Tr R_{\CN=2} I_{3}^{2}
     =     \frac{n_{v}}{4}.
           \label{N=2anomaly}
    \eea
  The quantities $n_v, n_h$ are well-defined in the case of Lagrangian theories, 
  but it is useful book-keeping device to use for non-Lagrangian theories as well.  

  For a given punctured Riemann surface, we can separate the contribution 
  from the background Riemann surface and the punctures. 
  For $\Gamma = A_{N-1}$, the background contribution for a genus $g$ Riemann surface with $n$ punctures is given by
    \be
    n_h (\CC_{g, n}) &=& \frac{2}{3} (2g-2 +n)  N (N^2 - 1) , \\
    n_v (\CC_{g, n}) &=& \frac{1}{6} (2g-2 +n) (4N^3 - N - 3). 
    \ee
  Note that the definition of the background contribution is slightly different from the one
  in the literature by the terms including $n$.
  The factor $2g-2+n$ is the number of the pairs of pants,
  and this definition is more convenient to proceed to $\CN=1$ class $\CS$ theories.

  For a puncture labeled by a Young diagram $Y$, (called regular punctures)
    \be
    n_h (Y) &=& \half \sum_r l_r^2 + \sum_{k=2}^N (2k-1) p_k  - \frac{1}{6} (4N^{3} - N), \\
    n_v (Y) &=& \sum_{k=2}^N (2k-1) p_k - \frac{1}{6} (4N^{3} - N - 3), 
    \ee
  where $p_k$ labels the structure of the poles at the puncture 
  (which can be read off from $Y$) \cite{Gaiotto:2009we}
  and $l_r$ is the length of the $r$-th row of $Y$. 
  For example, the maximal puncture has the pole structure $p_{\textrm{max}} = (0, 1, 2, \cdots, N-1)$ 
  and the minimal puncture has $p_{\textrm{min}} = (0, 1, 1, \cdots, 1)$. 
  Note again that the last terms are absent in the definition in the literature.
  These compensate the changes in the background contributions.
  In general, one can also have irregular punctures as well, but we will not consider them here. 
  For example, the maximal puncture has
    \be
    n_h (Y_{\textrm{max}}) = 0, ~~~
    n_v (Y_{\textrm{max}}) = - \frac{1}{2} (N^{2}-1),
    \ee
  and the minimal puncture has
    \bea
    n_h (Y_{\textrm{min}})
    &=&   - \frac{1}{6}(4N^{3} - 6N^{2} - 4N), \\
    n_v (Y_{\textrm{min}})
    &=&   - \frac{1}{6} (4N^{3} - 6N^{2} - N + 3).
    \eea
  By summing altogether, $n_{h}$ and $n_{v}$ are
    \bea
    n_{h}
     =     n_{h}(\CC_{g, n}) + \sum_{i} n_{h}(Y_{i}),~~~~
    n_{v}
     =     n_{v}(\CC_{g, n}) + \sum_{i} n_{v}(Y_{i}).
           \label{nvnh}
    \eea
  
  Also, the flavor central charge of an $\CN=2$ theory is defined by
    \bea
    k \delta^{ab}
     =   - 2 \tr R_{\CN=2} T^{a} T^{b}
           \label{flavorcentralcharge}
    \eea
  where $T^{a}$ is the generator of the flavor symmetry.

  We now define the $\CN=1$ version of $n_{h}$ and $n_{v}$.
  Let $\sigma_{i}$ be the sign of the $i$-th puncture.
  They are given by
    \bea
    \hat{n}_{h}
     =     \hat{n}_{h}(\CL^{p, q}) + \sum_{i} \sigma_{i} n_{h}(Y_{i}) \ , \qquad
    \hat{n}_{v}
     =     \hat{n}_{v}(\CL^{p, q}) + \sum_{i} \sigma_{i} n_{v}(Y_{i}) \ ,
           \label{nvnhhat}
    \eea
  where 
    \bea
    \hat{n}_h (\CL^{p, q}) &=& \frac{2}{3} (p-q)  N (N^2 - 1) \ , \\
    \hat{n}_v (\CL^{p, q}) &=& \frac{1}{6} (p-q) (4N^3  - N - 3) \ .
    \eea
  Since we are considering $\CN=1$ theories, $\hat{n}_{h}$ and $\hat{n}_{v}$ 
  do not have the interpretation of the effective numbers of hyper and vector multiplets.
  However, we continue to use these letters.
  In terms of these, our proposal for the 't Hooft anomaly coefficients are as follows:
    \bea
    \Tr R_{0}
    &=&    n_{v} - n_{h}, ~~~
    \Tr R_{0}^{3}
     =     n_{v} - \frac{n_{h}}{4},
           \\
    \Tr \CF
    &=&  - (\hat{n}_{v} - \hat{n}_{h}), ~~~
    \Tr \CF^{3}
     =   - \hat{n}_{v} + \frac{\hat{n}_{h}}{4},
           \\
    \Tr R_{0} \CF^{2}
    &=&  - \frac{n_{h}}{4}, ~~~
    \Tr R_{0}^{2} \CF
     =     \frac{\hat{n}_{h}}{4},
           \label{anomalyformula}
    \eea
  where $n_{h}$ and $n_{v}$ are \eqref{nvnh} with $2g-2+n = p+q$.
  
  In an $\CN=1$ theory which can be obtained from the $\CN=2$ one, we identify the $R$-symmetries as
  \cite{Tachikawa:2009tt}
    \bea
    R_{0}
     =     \frac{1}{2} R_{\CN=2} + I_{3}, ~~~~
    \CF
     =   - \frac{1}{2} R_{\CN=2} + I_{3}.
    \eea
  With these, the above anomaly coefficients (without hats) can be obtained by using \eqref{N=2anomaly}.
  Then we changed $n_{v}$ and $n_{h}$ into $\hat{n}_{v}$ and $\hat{n}_{h}$ for the anomalies involving
  odd power of $\CF$.
  We are proposing these formulae, however, for the theories which do not necessarily have the $\CN=2$ origin,
  like the Fan.
  
  Let us check these formulae are indeed correct for a few theories.
  
\paragraph{Fan}
  The Fan with $\sigma=+1$ is associated with a sphere with $p=1$ and $q=0$
  and three punctures, maximal, minimal and the one specified by $Y$.
  Therefore, we get from \eqref{nvnh} and \eqref{nvnhhat}, 
    \bea
    n_{v}
    &=&    \sum_{k=2}^N (2k-1) p_k - \frac{1}{6} (4N^{3} - 3 N^{2} - N),~~~
    n_{v} - n_{h}
     =   - \frac{1}{2} (N^{2} + \sum_{r} l_{r}^{2}),
           \nonumber \\
    \hat{n}_{v}
    &=&  - \sum_{k=2}^N (2k-1) p_k + \frac{1}{6} (4 N^{3} + 3 N^{2} - N - 6), ~~~
    \hat{n}_{v} - \hat{n}_{h}
     =   - \frac{1}{2} (N^{2} + 2 - \sum_{r} l_{r}^{2}).
           \nonumber
    \eea
  It is straightforward to see that the anomaly coefficients obtained 
  by substituting these into \eqref{anomalyformula}
  agree with the ones from the direct computation \eqref{anomalyfan1} and \eqref{anomalyfan2},
  by using the identities $\sum_{r} l_{r}^{2} = \sum_{i=1}^{\ell} N_{i} n_{i}$,
  $\sum_{k=2}^N (2k-1) p_k = \frac{N}{6}(4N^{2} - 3N - 1) - \sum_{i}(N^{2} - N_{i}^{2}) $,
  and \eqref{N^2}.

\paragraph{$SU(N)$ SQCD with $N_{f} = 2N$}
  Now let us try to see how the formulae work in other class $\CS$ theories.
  A simple example is SQCD with $N_{f} = 2N$ considered in section \ref{sec:ASduality} 
  which is associated with a sphere 
  with two maximal punctures with $\sigma=+1$ and $\sigma = -1$
  and two minimal punctures with $\sigma=+1$ and $\sigma = -1$ and also with $p=q=1$. 
  The anomalies are given by
    \bea
    \Tr R_{0}
    &=&    -N^{2}-1, ~~~~
    \Tr R_{0}^{3}
     =     \frac{N^{2}}{2} -1, ~~~~
    \Tr R_{0} \CF^{2}
     =   - \frac{N^{2}}{2}, \\
    \Tr \CF
    &=&    \Tr \CF^{3}
     =     \Tr R_{0}^{2} \CF
     =     0.
    \eea
  These can also be computed directly from the matter content of the SQCD
  as in the table \ref{table:ele}.
  
  For completeness, let us compute the anomaly coefficients of non-Abelian symmetry.
  For the gauge symmetry, we have $\Tr R_{0} T_{g}^{2} = \Tr \CF T_{g}^{2} =0$ 
  indicating the vanishing exact beta function and anomaly-free $U(1)_{\CF}$.
  The anomalies which involves $SU(N)$ flavor symmetries are as follows:
    \bea
    \tr R_{0} T^{2}_{1}
     =     \tr R_{0} T^{2}_{2}
     =     \tr \CF T^{2}_{1}
     =   - \tr \CF T^{2}_{2}
     =   - \frac{N}{2},
           \label{flavorSU}
    \eea
  where $T_{1,2}$ are the generators of $SU(N)_{1,2}$.
  Since there is no non-baryonic $U(1)$ symmetry, the $U(1)_{R_{0}}$ is the true $R$-symmetry in the IR.

\paragraph{Linear quiver}
  We have computed in the previous section the 't Hooft anomaly coefficients of the linear quiver
  with $\CN=2$ tail \eqref{anomalyN=2quiver1}, \eqref{anomalyN=2quiver2} and \eqref{anomalyN=2quiver3}.
  Let us reproduce these results from our formulae.
  The quiver (we fix $\sigma=1$) is associated with a sphere with $p=\ell$ and $q=1$ and 
  $\ell+1$ minimal punctures with $\sigma=+1$, one maximal puncture with $\sigma=+1$
  and a puncture specified by $Y$ with $\sigma=-1$.
  It is easy to get
    \bea
    n_{v}
    &=&    \sum_{i=1}^{\ell} (N_{i}^{2} - 1), ~~~
    n_{v} - n_{h}
     =   - \ell - \sum_{i=1}^{\ell} N_{i} \sum_{j=i}^{\ell} n_{j},
           \\
    \hat{n}_{v}
    &=&    \sum_{i=1}^{\ell} (N_{i}^{2} - 1) - 2N^{2} +2, ~~~~
    \hat{n}_{h}
     =   - \ell + 2 - \sum_{i=1}^{\ell} N_{i} \sum_{j=i}^{\ell} n_{j}.
    \eea
  These reproduce \eqref{anomalyN=2quiver1}, \eqref{anomalyN=2quiver2} and \eqref{anomalyN=2quiver3}.

\paragraph{$\CN=1$ gauging}
  Let us consider a pair of class $\CS$ theories, $\CT_{1}$ and $\CT_{2}$, 
  each of which has an $SU(N)$ flavor symmetry.
  Let the colors of the maximal punctures be different
  and $\CT_{1}$ and $\CT_{2}$ be associated to a pair-of-pants decompositions
  where each color of the maximal puncture is the same as that of the pair-of-pants 
  to which the puncture attached. 
  Then let us think of gluing these punctures. 
  This corresponds to the $\CN=1$ gauging of the diagonal $SU(N)$ symmetry
  of two $SU(N)$ flavor symmetries of $\CT_{1}$ and $\CT_{2}$.
  The resulting theory is again in class $\CS$.
  
  The 't Hooft anomaly coefficients of the resulting theory are written
  as the sum of those of $\CT_{1}$ and $\CT_{2}$, and of $\CN=1$ vector multiplet.
  The anomalies of the latter can be computed as
    \bea
    \Tr R_{0}
    &=&    \Tr R_{0}^{3}
     =     N^{2} -1, ~~~
    \Tr \CF
     =     \Tr \CF^{3}
     =     \Tr R_{0}^{2} \CF
     =     \Tr R_{0} \CF^{2}
     =     0.
           \label{anomaliesN=1gauging}
    \eea
  These can be obtained from our formulae.
  Indeed from the Riemann surface point of view, the $\CN=1$ gauging corresponds to
  subtracting two maximal punctures with different signs.
  Thus, we have
  $\delta n_{v} = N^{2} -1$, $\delta \hat{n}_{v} = \delta n_{h} = \delta \hat{n}_{h} = 0$. 
  These reproduce \eqref{anomaliesN=1gauging}.
  
\paragraph{$\CN=2$ gauging}
  Instead, let us consider the gauging by an $\CN=2$ vector multiplet.
  Namely, consider $\CT_{1}$ and $\CT_{2}$ with maximal punctures whose colors are the same.
  $\CT_{i}$ is associated to pants decomposition 
  where the colors of the maximal puncture and of the pair-of-pants 
  to which the puncture is attached are the same.
  Let us suppose the color is $\sigma=+$.
  In this case the 't Hooft anomalies are the sum of those of $\CT_{1}$, $\CT_{2}$
  and of $\CN=2$ vector multiplet where the gauge adjoint chiral field has $R_{0} = - \CF = 1$.
  The latter contributes to the anomalies as
    \bea
    \Tr R_{0}
    &=&    \Tr R_{0}^{3}
     =   - \Tr \CF
     =   - \Tr \CF^{3}
     =     N^{2}-1, ~~~
    \Tr R_{0}^{2} \CF
     =     \Tr R_{0} \CF^{2}
     =     0.
    \eea 
  Again this can be obtained from the formulae with
  $\delta n_{v} = \delta \hat{n}_{v} = N^{2} - 1$ and $\delta n_{h} = \delta \hat{n}_{h} = 0$.

\paragraph{A theory coupled to an adjoint}
  Let us consider the Riemann surface with a maximal puncture such that $\sigma_{Y_{max}}$ is 
  different from the sign of the background.
  In \cite{Gadde:2013fma}, it was noticed that this represents a theory
  (associated to the same Riemann surface where the maximal puncture has the same sign as the bulk) coupled to 
  a chiral multiplet $M$ transforming in the adjoint representation of the $SU(N)$ flavor symmetry
  associated to the maximal puncture, by the superpotential
  $W = \tr \mu M$ where $\mu$ is the moment map of the $SU(N)$.
  The charges of $M$ are $R_{0} = 1$ and $\CF=\sigma_{Y_{max}}$.
  (When the Riemann surface is a sphere with two maximal and a minimal punctures,
  this boils down to the Fan with $Y$ is maximal.)
  Let us see this is consistent with our formula.
  
  Suppose that the sign of the background is $+1$ and $\sigma_{Y_{max}} = -1$.
  Compared to the case where $\sigma_{Y_{max}} = +1$, $\hat{n}_{v}$ increases by $\delta \hat{n}_{v} = N^{2} - 1$,
  while $n_{v}$, $n_{h}$ and $\hat{n}_{h}$ are kept intact.
  Therefore the contribution of changing $\sigma_{Y_{max}}$ from $+1$ to $-1$ to the anomaly coefficients is
    \bea
    \delta \Tr \CF
     =     1 - N^{2}, ~~~~
    \delta \Tr \CF^{3}
     =     1 - N^{2}.
    \eea
  while other coefficients remain to be the same. 
  These are exactly the contribution of a chiral multiplet in the adjoint representation of $SU(N)$
  with $\CF=-1$ and $R_{0}=1$.

\paragraph{$\CN=1$ Argyres-Seiberg dual theory}
  The dual theory is an $\CN=1$ $\SU(N)$ gauge theory coupled to the Fan specified by $Y_{min}$ 
  and to the $T_{N}$ theory where a adjoint chiral multiplet is attached to a maximal puncture.
  By the class $\CS$ interpretation of the anomaly coefficients,
  it is almost trivial to see that the anomalies of this dual theory agree with those of the SQCD,
  because they are represented by the same Riemann surface.
  Actually, we already show above that the anomaly coefficients of the Fan satisfies the formulae,
  and that attaching an adjoint field is interpreted as changing the sign of the puncture.
  Also, the anomalies of the $T_{N}$ theory itself are written by using \eqref{nvnh}:
  $n_{v}^{T_{N}} = \frac{2N^{3}}{3} - \frac{3N^{2}}{2} - \frac{N}{6} +1$, 
  $n_{h}^{T_{N}} = \frac{2N^{3}}{3} - \frac{2N}{3}$.
  
  For the anomaly coefficients of non-Abelian symmetries, 
  we use the result of the contribution of the Fan and the $T_{N}$ theory ($k^{T_{N}} = 2N$) 
  to the flavor central charge
  \eqref{flavorcentralcharge}.
  It is easy to show that these cancel for $\Tr R T_{g}^{2}$ and $\Tr \CF T_{g}^{2}$.
  For the anomalies involving the flavor $SU(N)$, the Fan part does not contribute, 
  thus reproducing \eqref{flavorSU} upon using \eqref{flavorcentralcharge}.

\section{Superconformal index}
\label{sec:index}
In this section, we compute the superconformal indices of the $\CN=1$ class $\CS$ theories. The two-dimensional generalized TQFT structure ensures the invariance of index under various dualities we described. The generalized TQFT structure of the $\CN=1$ class $\CS$ theories has been shown in \cite{Beem:2012yn}, generalizing the $\CN=2$ case studied in \cite{Gadde:2009kb, Gadde:2011ik, Gadde:2011uv, Gaiotto:2012xa}. In \cite{Gadde:2013fma, Agarwal:2013uga}, the prescription of adding adjoint chiral field for oppositely colored punctures has been shown to be consistent with the generalized TQFT structure of the $\CN=1$ class $\CS$ theories. In this section, we show that the matter content and charges of the Fan can be obtained by assuming the TQFT structure. 

\subsection{Review}
The superconformal index for $\CN=1$ theories is defined as 
\be
 I(\fp, \fq, \xi; \vec{z}) = \Tr (-1)^F \fp^{j_1 + j_2 + R_0/2} \fq^{j_2 - j_1 + R_0/2} \xi^{\CF} \prod_{i} z_i^{F_i} \ , 
\ee
where $j_1, j_2$ are the Cartans of the Lorentz group $SU(2)_1 \times SU(2)_2$ and 
$F_i$ are generators of flavor symmetries. 
Strictly speaking, $R_0$ in the index has to be the exact $R$-charge in the IR. However in our case, we can simply keep it as $R_0$, as long as we keep the fugacity $\xi$ turned on. After determining the amount of mixing through a-maximization, we can simply replace $\xi \to \xi (\fp \fq)^{\e/2}$ to get the correct $R$-charge $R = R_0 + \e \CF$. 

A good thing about the superconformal index is that it can be computed purely in terms of the matter content in the UV. The contribution for a chiral multiplet in a representation $\Lambda$ of certain flavor or gauge group is given by
\be \label{eq:Ichi}
 I_{\textrm{chi}} (\fp, \fq, \xi; \vec{z}) = \textrm{PE} \left[ \frac{(\fp \fq)^{R_0/2} \xi^{\CF} \chi_{\Lambda} (\vec{z}) - (\fp \fq)^{1-R_0/2} \xi^{-\CF} \chi_{\bar{\Lambda}}(\vec{z}) }{(1-\fp)(1-\fq)}\right] \ , 
\ee
where $\chi_{\Lambda} (z)$ is the character of the representation $\Lambda$. The $R_0$ is the $R_0$-charge of the scalar in the chiral multiplet.

The chiral multiplet index \eqref{eq:Ichi} can be written in terms of elliptic Gamma function as
\be
 I_{\textrm{chi}} (\fp, \fq, \xi; \vec{z}) = \prod_{v \in \Lambda} \Gamma( (\fp \fq)^{R/2} \xi^{\CF} \vec{z}^v; \fp, \fq) \ , 
\ee
where $\Lambda$ is the weight lattice of the representation. We use the notation 
$ \vec{z}^v = \prod_i z_i^{v_i} $. 
Also, 
\be
 \Gamma (z; \fp, \fq) = \prod_{m, n \ge 0} \frac{ 1 - z^{-1} \fp^{m+1} \fq^{n+1}}{1-z \fp^m \fq^n} \ , 
\ee
is the elliptic Gamma function. For a vector multiplet, it contributes
\be
 I^{\CN=1}_{\text{vec}} (\fp, \fq; \vec{z}) = \left( (\fp; \fp) (\fq; \fq) \right)^r \prod_{\a \in \Delta(G)} \Gamma (\vec{z}^\a; \fp, \fq)^{-1} \ , 
\ee
where $\Delta(G)$ is the set of all roots for the gauge group $G$, $r$ being the rank of $G$ and $(z; q) \equiv \prod_{m=0}^{\infty} (1- z q^m)$ is the $q$-Pochammer symbol.\footnote{Here we also included the Haar measure factor to the $I_{\textrm{vec}}^{\CN=1}(\fp, \fq; \vec{z})$.}  

\paragraph{Generalized TQFT structure of the index}
The superconformal index of a class $\CS$ theory given by a UV curve can be written in terms of pair-of-pants (or three-punctured sphere) and cylinders connecting them. For a pair-of-pants (or three-punctured sphere) with maximal punctures with colors $\s_i$, we can write the index as
\be
 I_{(\s, \s_i)} (\vec{a_1}, \vec{a_2}, \vec{a_3}) = \sum_{\vec \lambda} C^\s_{\vec\lambda} (\fp, \fq, \xi) \psi^{\s_1}_{\vec \lambda} (\vec{a_1}) \psi^{\s_2}_{\vec \lambda} (\vec{a_2}) \psi^{\s_3}_{\vec \lambda} (\vec{a_3}) \ , 
\ee
for the $\s$-colored pair-of-pants. Here the sum is over the representations $\vec{\lambda}$ of $\Gamma \in ADE$ labeling the class $\CS$ theory. The functions $C^\s_{\vec\lambda} (\fp, \fq, \xi)$ and $\psi_{\vec \lambda}^{\s_i} (\vec{a_i})$ are called the `structure constant' and the wave function of the TQFT respectively. We omitted its dependence on $(\fp, \fq, \xi)$. One of the key relation we use for the wave function is 
\be
 \psi^{\s}_{\vec{\lambda}} (\vec{a}) = M^\s (\vec{a}) \psi^{-\s}_{\vec{\lambda}} (\vec{a}) \ , 
\ee
where 
\be
 M^\s(\vec{a}) = \textrm{PE} \left[ \frac{(\xi^{\s} - \xi^{-\s}) \sqrt{\fp \fq}}{(1-\fp)(1-\fq)} \chi_{\textrm{adj}} (\vec{a}) \right] 
 = \Gamma(t_\s; \fp, \fq)^{r} \prod_{ i \neq j} \Gamma(t_\s a_i a_j^{-1}; \fp, \fq) \ ,  
\ee
where $t_\s = \xi^{\s} \sqrt{\fp \fq}$ and $r$ is the rank of group $\Gamma$. 
It was shown in \cite{Beem:2012yn} that this wave function is essentially determined by the $\CN=2$ counterpart, which is given by an eigenfunction of elliptic Ruijsenaars-Schneider model \cite{Gaiotto:2012xa}. More precisely, the relation between $\CN=1$ and $\CN=2$ version of the wave function is given as
\be
 \psi^{\s}_{\vec \lambda} (\vec{a}; \fp, \fq, \xi) = \psi_{\vec \lambda} (\vec{a}; \fp, \fq, t=t_\s) .
\ee
Also, the structure constant can be simply fixed from that of the $\CN=2$ couterpart as $C_{\vec \lambda}^\s (\fp, \fq, \xi) = C_{\vec \lambda} (\fp, \fq, t=\xi^{\s} \sqrt{\fp \fq}) $. 

The wave function can be written as $\psi^\s_{\vec \lambda} (\vec{a}) = K^\s (\vec{a}) \Psi_{\vec \lambda} (\vec{a})$ where $K^\s (\vec a)$ is a prefactor which does not depend on $\vec \lambda$ and $\Psi_{\vec \lambda} (\vec{a})$ is another function which depends on the representations $\vec \lambda$ of the group $\Gamma$. 
The prefactor is given by
\be
K^\s (\vec{a}) = \textrm{PE} \left[ \frac{\xi^{\s} \sqrt{\fp \fq} - \fp \fq}{(1-\fp)(1-\fq)} \chi_{\textrm{adj}} (\vec{a}) \right] \ . 
\ee
Note that the function $\Psi_{\vec\lambda}$ does not depend on color $\s$. 
In terms of these functions, we can write the index for a three-punctured sphere as
\be
 I_{(\s, \s_i)} (\vec{a_1}, \vec{a_2}, \vec{a_3}) = \frac{K^{\s_1} (\vec{a_1}) K^{\s_2} (\vec{a_2}) K^{\s_3} (\vec{a_3}) }{K^\s (t_\s^\rho)} \sum_{\vec \lambda} \frac{ \Psi_{\vec \lambda} (\vec{a_1}) \Psi_{\vec \lambda} (\vec{a_2}) \Psi_{\vec \lambda} (\vec{a_3})}{  \Psi_{\vec \lambda} (t_\s^\rho) } \ , 
\ee
where $t^{\rho}_{\s} = ((t_\s)^{\rho_1}, (t_\s)^{\rho_2}, \cdots, (t_\s)^{\rho_r} )$ with $\rho$ being the Weyl vector of the group $\Gamma$.
When we glue the pair-of-pants by a gauge group, we integrate over the gauge fugacities with vector multiplet measure. Since we have
\be
 \int [da] I^{\s \s'}_{\textrm{vec}} (p, q; \vec{z}) \psi^{\s}_{\vec{\lambda}}(\vec{z}) \psi^{\s'}_{\vec{\lambda'}} (\vec{z}) = \delta_{\vec \lambda \vec \lambda'} \ , 
\ee
where $I^{\s \s'}_{\textrm{vec}} (\fp, \fq; \vec{z})$ is $\CN=2$ vector multiplet when $\s = \s'$ and $\CN=1$ otherwise, we can write the superconformal index for any UV curve with colored full punctures as
\be
  I(\vec{a_i}, \vec{b_j}; \fp, \fq, \xi) = \sum_{\vec \lambda} \frac{\prod_{i=1}^{n_+}\psi^{+}_{\vec\lambda} (\vec{a_i}) \prod_{j=1}^{n_-} \psi^{-}_{\vec\lambda} (\vec{b_j})}{\left( \psi^{+}_{\vec\lambda} (t^{\rho}_+) \right)^p \left( \psi^{-}_{\vec\lambda} (t^{\rho}_-) \right)^{q}} \ , 
\ee
where $(n_+, n_-)$ are the number of punctures of each color, and $(p,q)$ are the degrees of the normal bundles satisfying $2g-2+ (n_+ + n_-) = p+q$ and $\vec{a_i}, \vec{b_j}$ are the flavor fugacities. As we see clearly, the index only depends on the topological data. 

Now, if we choose the punctures to be non-maximal, we replace the fugacities in the wave function appropriately. The prescription is to replace
\be
 \Psi_{\vec\lambda}^\s (\vec{a}) \to \Psi^\s_{\vec\lambda} (\vec{u} t_\s^\Lambda) \ , ~~~ 
 K^\s (\vec a) \to K^\s_{\Lambda} (\vec {u}) = \textrm{PE} \left[ \sum_j \frac{t_\s^{1+j} - \fp \fq t_\s^j}{(1-\fp)(1-\fq)} \chi_{R_j} (\vec u) \right] \ , 
\ee
for a puncture labelled by the $SU(2) \hookrightarrow \Gamma$ embedding $\Lambda$ which decomposes $\textrm{adj} \to \bigoplus_j R_j \otimes V_j$ where $R_j$ is the representation of the commutant of $\Lambda(SU(2))$ in $\Gamma$ and $V_j$ is the spin-$j$ representation of $SU(2)$. The notation $\vec{u} t^\Lambda$ means replacing the flavor fugacity appropriately in accordance with the broken flavor symmetry. See \cite{Mekareeya:2012tn} for a detailed discussion on this notation and its physical meaning. We will give an example in the section \ref{subsec:ASidx}, and then the full expression in \ref{subsec:FanIdx}.  

\subsection{$\CN=1$ Argyres-Seiberg duality} \label{subsec:ASidx}
The agreement of the index for the Argyres-Seiberg duality can be checked using the TQFT language as done in \cite{Gadde:2013fma, Agarwal:2013uga}. In the SQCD frame as in figure \ref{fig:ElectricSQCDQuiver} or figure \ref{fig:ElectricSQCD}, the index can be written as
\be \label{eq:IdxSQCD}
 I (\vec{x}_1, \vec{x}_2, a, b) = \frac{K_{\star}^{-}(a) K^{-}(\vec{x}_1) K_{\star}^{+}(b)  K^{+} (\vec{x}_2)  }{ K_\varnothing^- K_\varnothing^+}\sum_{\vec \lambda} \frac{ \Psi_{\vec\lambda}(a t^\star_-) \Psi_{\vec\lambda} (\vec{x}_1) \Psi_{\vec \lambda}(b t^\star_+) \Psi_{\vec \lambda}(\vec{x}_2)    }{ \Psi_{\vec \lambda} (t_-^\varnothing) \Psi_{\vec \lambda} (t_+^\varnothing) } \ , \quad
\ee 
where $\star$ denotes the embedding associated to the minimal puncture. 
Here, all the $+$ colored contributions are coming from the functions with $+$ labels and vice versa since the color of the pair-of-pants is the same as the punctures. Here we denote fugacities of the flavor symmetry $SU(3)_1 \times SU(3)_2 \times U(1)_A \times U(1)_B$ to be $\vec{x}_1, \vec{x}_2, a, b$ respectively. 

Now, we need to show that this index is the same in the Argyres-Seiberg frame as in the figure \ref{fig:ASDualSQCD}. There, we have punctures with different color from the pair-of-pants. On the left-side of the figure, we have maximal punctures with each color and thus an adjoint chiral field $N$.
On the right-hand side, we have two minimal punctures with each color,
corresponding to an adjoint field but with a nilpotent vev imposed, giving a number of components that survive according to the $SU(2)$ embedding labelled by the puncture. In the case of generic $\rho$ being $\textrm{adj} \to \bigoplus_j R_j \otimes V_j$, this contribution to the index is
\be \label{eq:IdxMeson}
 M_\rho^\s (\vec{u}) = \textrm{PE} \left[ \sum_j \frac{t_\s^{1+j} - \fp \fq t_\s^{-1-j} }{(1-\fp)(1-\fq)} \chi_{R_j} (\vec{u}) \right] \ . 
\ee
This is coming from the shift of $R$-charge $R_0 \to R_0 + 2 \rho(\s_3)$ under the Higgsing. 
Thus $M_{\star}^{+}$ represents the components appearing in the dual frame.

The index in the Argyres-Seiberg dual frame can be written as 
\be \label{eq:IdxAS}
 M^{-} (\vec{x}_1) M^{+}_{\star} (b) \frac{K_\star^{-}(a)  K^{-}(b t_+^\star) K^{+} (\vec{x}_1) K^{+}(\vec{x}_2)}{ K_\varnothing^+ K_\varnothing^-} 
 \sum_{\vec\lambda} \frac{ \Psi_{\vec \lambda} (a t^\star_-) \Psi_{\vec\lambda}(b t^\star_-) \Psi_{\vec\lambda} (\vec{x}_1) \Psi_{\vec\lambda}(\vec{x}_2) }{ \Psi_{\vec\lambda} (t_+^\varnothing) \Psi_{\vec \lambda} (t_-^\varnothing) }  \ . \quad
\ee
The first two terms are coming from the additional fields in the dual theory and the signs of $K^\s$s are determined by the color of the pair-of-pants. 
From the identity \cite{Gadde:2013fma}
\be
 M^\s_\Lambda (\vec{u}) K^{-\s} (\vec{u} t_\s^\Lambda) = K^\s_\Lambda (\vec{u}) \ , 
\ee
we see that the \eqref{eq:IdxSQCD} and \eqref{eq:IdxAS} are equal. This shows consistency of the TQFT description of the superconformal index of class $\CS$ theories. 

\begin{figure}[h]
	\begin{subfigure}[h]{3in}
	\centering
	\includegraphics[width=2.0in]{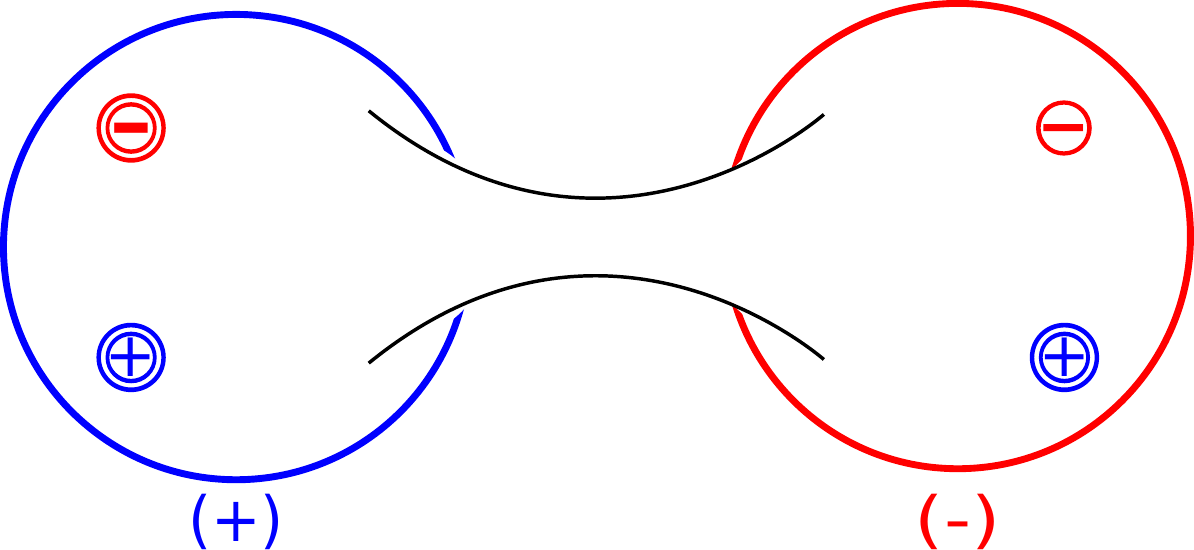}
	\end{subfigure}
	\begin{subfigure}[h]{3in}
	\centering
	\includegraphics[width=2.3in]{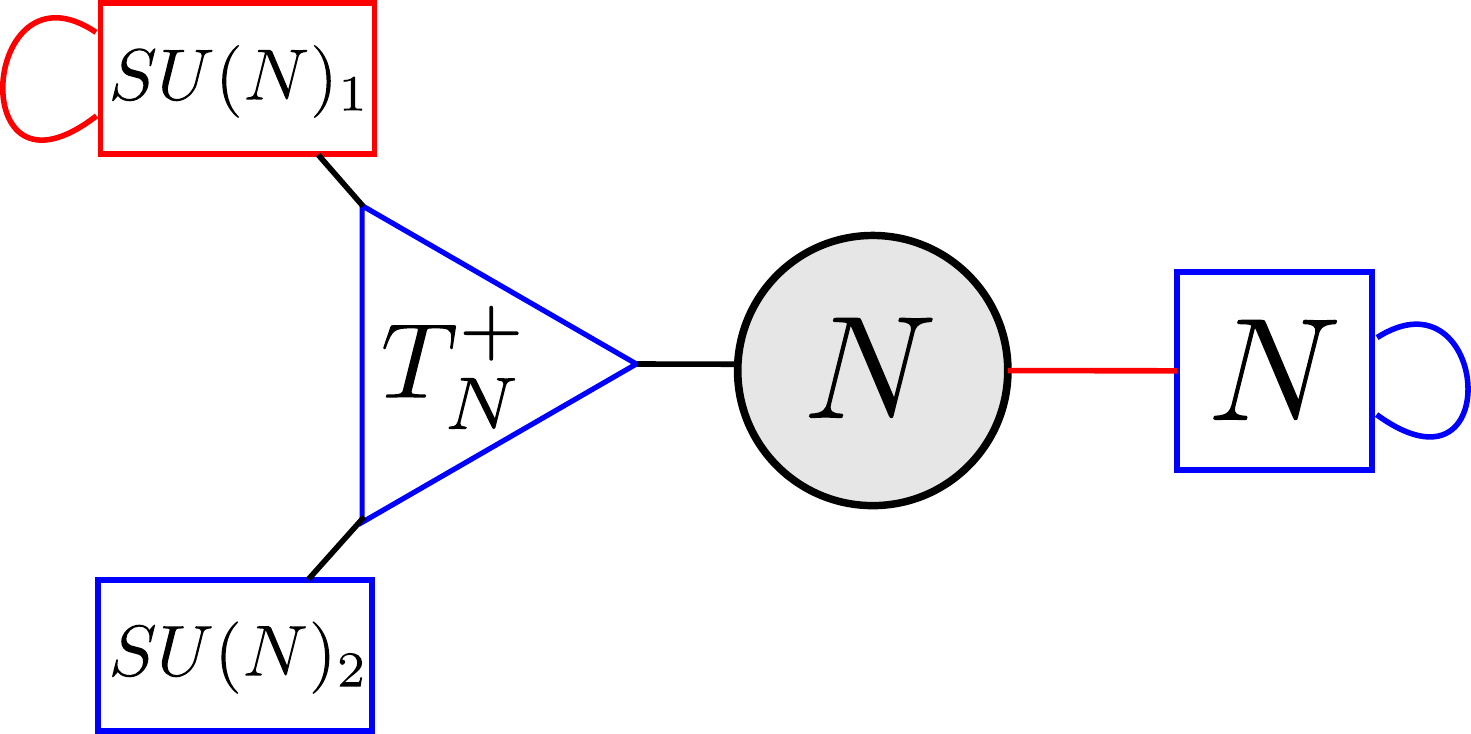}
	\end{subfigure}
	\caption{Unhiggsed SQCD in the Argyres-Seiberg frame} 
	\label{fig:SQCDASCurveUnhiggs}
\end{figure}

This agreement from the TQFT was quite formal, and works for any kind of puncture. We should be able to calculate this index in the Argyres-Seiberg dual frame using the matter content we found in the previous section. This can be done by looking at the index of the unhiggsed theory and Higgsing to get the Argyres-Seiberg frame. Let us consider the Argyres-Seiberg frame before Higgsing the dual meson $M$, which is realized by two maximal punctures on the left with each color, and one maximal puncture with $+$ color and one minimal puncture with $-$ color as in the figure \ref{fig:SQCDASCurveUnhiggs}. 
This realizes $T_N$ theory with one of $SU(N)$ gauged by $\CN=1$ vector multiplet, and $N$ fundamentals attached to it. We also have an adjoint field $N$ associated to one of $SU(N)$ flavor symmetries on the $T_N$ side, and another adjoint field $M$ attached to the fundamentals. The index of this theory can be written as
\be \label{eq:idxbeforeHiggs}
 I(\vec{x}_1, \vec{x}_2, \vec{y}, b) &=& \oint \prod_{i=1}^{N-1} \frac{dz_i}{2\pi i z_i} \Delta(\vec{z}) I^{\CN=1}_{\textrm{vec}}(\vec{z})  I^+_{T_N} (\vec{x}_1, \vec{x}_2, \vec{z}) 
 \prod_{i, j=1}^{N} \Gamma ( t_-^\half  (z_i y_j b)^\pm)  \\  
 &{ }&  \times  \left( \Gamma (  t_- )^{N-1} \prod_{i \neq j} \Gamma( t_-  x_{1,i} x_{1, j}^{-1}) \right) 
 \left( \Gamma (t_+ )^{N-1} \prod_{i \neq j} \Gamma ( t_+  y_i y_j^{-1}) \right) \ ,   \nn
\ee
where we used the short-hand notation $\Gamma(z) = \Gamma(z; \fp, \fq)$, and $\prod_{i=1}^N z_i = 1$ is assumed. The symbol $I_{T_N}$ refers to the index of the $T_N$ block and $I^{\CN=1}_{\textrm{vec}}$ is the $\CN=1$ vector multiplet contribution to the index and $\Delta(\vec{z})$ is the Haar measure of the gauge group. The last term in the first line is the contribution from the fundamental quarks with $R_0 = \half, \CF = -\half$. The second line corresponds to the contributions from the fields $N$ and $M$ respectively. 

Now, upon Higgsing, we specialize the fugacity $\vec{y}$ to the ones determined from the $SU(2)$ embedding $N \to (N-1) + 1$. In our case, we put $\vec{y} = (a t_+^{\frac{N}{2}-1}, a t_+^{\frac{N}{2}-2}, \cdots, a t_+^{1-\frac{N}{2}}, a^{-N+1})$. Then the last term in the first line of \eqref{eq:idxbeforeHiggs} becomes
\be \label{eq:ASIdxquarkHiggsed}
 \prod_{i=1}^N \left[ \left(\prod_{m=1}^{N-1} \Gamma ((\fp \fq)^{\frac{1}{4}} \xi^{-\half} ( z_i a (\xi \sqrt{\fp \fq})^{\frac{N}{2} - m} b )^{\pm}) \right) \Gamma((\fp \fq)^{\frac{1}{4}} \xi^{-\half} (z_i a^{-N+1} b)^\pm) \right] \ , 
\ee
where the terms in the parenthesis can be written as 
\be
\prod_{m=1}^{N-1} \Gamma((\fp \fq)^{\frac{1+N-2m}{4}} \xi^{\frac{-1+N-2m}{2}} z_i a b)  
 \prod_{m'=1}^{N-1} \Gamma((\fp \fq)^{\frac{1-N+2m'}{4} } \xi^{\frac{-1-N+2m'}{2} } (z_i a b)^{-1})    \ . 
\ee
Due to the identity $\Gamma(z; \fp,\fq) \Gamma(\frac{\fp \fq}{z}; \fp, \fq) = 1$, all the terms with $m=m'-1$ are cancelled. The only remaining terms are the ones with $(m, m') = (N-1, 1)$. Therefore, \eqref{eq:ASIdxquarkHiggsed} can be written as
\be
 \prod_{i=1}^N \Gamma ((\fp \fq)^{\frac{3-N}{4}} \xi^{\frac{1-N}{2}} (z_i a b)^\pm) \Gamma((\fp \fq)^{\frac{1}{4}} \xi^{-\half} (z_i a^{-N+1} b)^\pm)  \ , 
\ee
which is the contribution from the quarks of $(J_+, J_-) = (2-N, 1), (0, 1)$ or $(R_0, \CF) = (\frac{3-N}{2}, \frac{1-N}{2}), (1, -1)$. We see that the index can be used to extract the matter content and the charges of the Higgsed theory. 

Contribution from $M$ upon Higgsing is determined through the minimal $SU(2)$ embedding 
\be \label{eq:adjtoU1}
 {\bf \rm{adj}} &\to& \left( \bigoplus_{m=1}^{N-1} V_{m-1}^{0}  \right) \oplus V_{\frac{N-2}{2}}^{-N} \oplus V_{\frac{N-2}{2}}^{N} \ ,  
\ee
where the supersubscript means the charge of the commuting $U(1)$. From this, we get 
\be
 M_\star^{+} (a) &=& \textrm{PE} \left[ \sum_{m=1}^{N-1} \frac{ (\fp \fq)^{\frac{m}{2}} \xi^{m} - (\fp \fq)^{1-\frac{m}{2}} \xi^{-m}} {(1-\fp)(1-\fq)} + \frac{(\fp \fq)^{\frac{N}{4}} \xi^{\frac{N}{2}} - (\fp \fq)^{1 - \frac{N}{4}} \xi^{-\frac{N}{2}} }{(1-\fp)(1-\fq)} (a^N+a^{-N}) \right]  . ~~~~~~~~ ~~
\ee
From here, we see that we have mesons with $(J_+, J_-) = (2m, 0)$ or $(R_0, \CF) = (m, m)$ with $m=1, \cdots, N-1$ and two mesons with $(J_+, J_-) = (N, 0)$ or $(R_0, \CF) = (N/2, N/2)$ which are exactly the same as that of our result in the section \ref{sec:ASduality}.

\subsection{Index of the Fan} \label{subsec:FanIdx}
We can repeat the similar procedure for a generic Fan as in the section \ref{subsec:ASidx}. Consider a partition of $N$ given by $\sum_{k=1}^\ell k n_k$ labelled by a Young diagram $Y$. For this partition, the flavor fugacity for the puncture is given as
\be \label{eq:IdxFanFug}
 \vec{u} t^\Lambda_\s &=& (\vec{u}_1 t^{\Lambda_1}_\s , \vec{u}_2 t^{\Lambda_2}_\s, \cdots, \vec{u}_\ell t^{\Lambda_\ell}_\s  ) \ ,  \nn \\ 
\vec{u}_k t^{\Lambda_k}_\s &=& (\vec{u}_k t_\s^{\frac{k - 1}{2}} , \vec{u}_k t_\s^{\frac{k-3}{2}}, \cdots \vec{u}_k t_\s^{\frac{1 - k}{2}} ) \ , 
\ee
where $\vec{u}_k = (u_{k, 1}, u_{k, 2}, \cdots, u_{k, n_k})$ is an $n_k$-dimensional vector for the $U(n_k)$ fugacities. We also impose the condition $\prod_{k=1}^\ell \prod_{i=1}^{n_k} {u}_{k, i} = 1$ so that the flavor symmetry is $S \left[ \prod_{k=1}^\ell U(n_k) \right]$. 

Plugging \eqref{eq:IdxFanFug} into the index formula for $N$ fundamental quarks, we get
\be
 \prod_{\a, \b=1}^{N} \Gamma ( \xi^{-1/2} (\fp \fq)^{1/4}  (z_\a y_\b b)^\pm)  \to 
  \prod_{\a=1}^N \prod_{k=1}^\ell \prod_{i=1}^{n_k} \prod_{m=1}^k \Gamma(\xi^{-1/2} (\fp \fq)^{1/4} (z_\a  u_{k, i} t_+^{\frac{k-2m+1}{2}} b )^\pm )    \ , 
\ee
where we assumed that the Fan is of the type $\s = -$ as in the previous example for simplicity. As in the section \ref{subsec:ASidx}, we see cancellations among upon Higgsing. The above equation can be written as 
\be
\prod_{\a=1}^N \prod_{k=1}^\ell \prod_{i=1}^{n_k} \left[ \prod_{m=1}^k \Gamma(\xi^{\frac{k-2m}{2}} (\fp \fq)^{\frac{2+k-2m}{4}} z_\a  u_{k, i}  ) \prod_{m'=1}^k \Gamma(\xi^{\frac{-2+2m'-k}{2}} (\fp \fq)^{\frac{2m'-k}{4}} (z_\a  u_{k, i} )^{-1} ) \right ] . \quad
\ee
We can see that the terms with $m' = m+1$ are cancelled so that only terms with $m=k$, $m'=1$ contribute. Therefore, we get
\be \label{eq:idxHiggsedQuarks}
I^{\textrm{quarks}} (\vec{z}, \vec{u}) = \prod_{\a=1}^N \prod_{k=1}^\ell \prod_{i=1}^{n_k}  \Gamma (\xi^{-\frac{k}{2}} (\fp \fq)^{\frac{2-k}{4}} (z_\a u_{k, i} b)^\pm ) \ ,  
\ee
which is the contribution from the quarks of desired charges $(J_+, J_-) = (1-k, 1)$ or $(R_0, \CF) = (\frac{2-k}{2}, -\frac{k}{2})$. 

The contribution from the adjoint fields are given as \eqref{eq:IdxMeson}. In the current case, the adjoint representation will decompose in to the form written as \eqref{eq:adjDecomp}. Therefore, the index for the resulting components can be written as 
\be \label{eq:idxHiggsedMeson}
 M^\s_{Y} (\vec{u}) &=& \prod_{i < j} \prod_{k=1}^i \textrm{PE} \left[ \frac{t_\s^{\half (j-i+2k)} - (\fp \fq) t_\s^{-\half (j-i+2k)} }{(1-\fp)(1-\fq)} \left( \chi_{R_i} (\vec{u}_i) \chi_{\bar{R}_j} (\vec{u}_j) +  \chi_{\bar{R}_i} (\vec{u}_i) \chi_{R_j} (\vec{u}_j)  \right) \right] \nn \\
 &{ }& \times \left(\prod_{i=1}^\ell \prod_{k=1}^i \textrm{PE}\left[ \frac{t_\s^k - (\fp \fq) t_\s^{-k} }{(1-\fp)(1-\fq)} \chi_{\textrm{adj}}^{U(n_i)} (\vec{u}_i) \right] \right)
  \times \textrm{PE} \left[ \frac{t_\s - \fp \fq t_\s^{-1}}{(1-\fp)(1-\fq)}\right]^{-1} \ , 
\ee
where the first term is coming from the bifundamentals of $U(n_i) \times U(n_j)$ and the second term is coming from the adjoints of $U(n_i)$ and the last piece takes care of the traceless condition. One can rearrange the first term by taking $i \to k-p$ so that the $R$-charges are given by $(R_0, \CF) = \half (i+j-2p, i+j-2p)$ with $p=0, \cdots, \textrm{min}(i, j)-1$. Likewise, the second term gives the adjoint fields of charge $(R_0, \CF) = (i-p, i-p)$ with $p=0, \cdots, i-1$ which agrees with the charges of the table \ref{table:fan}. 

Therefore, we find all the matter fields and charges as given in the table \ref{table:fan} for $N'=0$ case. Now, the index can be written in a contour integral form as 
\be \label{eq:IdxIntegral}
 I(\vec{x}_1, \vec{x}_2, \vec{y}, \vec{u}) = M^- (\vec{x}_1) \int \prod_{i=1}^{N-1} \frac{dz_i}{2\pi i z_i} \Delta(\vec{z}) I_{\textrm{vec}}(\vec{z}) I_{T_N} (\vec{x}_1, \vec{x}_2, \vec{z}) I^{\textrm{quarks}} (\vec{z}, \vec{u}) M^\s_Y (\vec{u}) , \qquad
\ee 
where $I^{\textrm{quarks}}(\vec{z}, \vec{u})$ and $M^\s_{Y}(\vec{u})$ are given by \eqref{eq:idxHiggsedQuarks} and \eqref{eq:idxHiggsedMeson} respectively, representing the components of the Fan.

\paragraph{Contour of the index integral}
Let us make a comment on the integration contour of equation \eqref{eq:IdxAS} and \eqref{eq:IdxIntegral}. Normally, for the purpose of evaluating the superconformal index, it is assumed that $|\fp|, |\fq| < 1$ and all the flavor fugacities to be unimodular $|a|=1$. Typically, the poles are of the form $z = a (\fp \fq)^{r/2} \fp^m \fq^n$ with $m, n \in \IZ$ and $R$-charge of the chiral multiplet being $r >0$. Therefore we pick all the poles with $m, n \ge 0$. But if we use this prescription in the current case, we may hit a pole along the contour of integration. Therefore we need to find a good contour to get away with this situation, because the usual contour of integration is not well-defined. 

In order to understand the situation, let us go back to the procedure of evaluating the index. When we evaluate the index, we first count all the (gauge non-invariant) operators satisfying certain shortening condition formed out of elementary quarks and various matter multiplets in the theory. Then, we impose the gauge invariance condition or the Gauss law constraint by integrating over the gauge group with the Haar measure. From this perspective, we have to include contributions from every elementary field regardless of its $R$-charges. This Gauss law constraint should be imposed after rescaling $a$ such that $|a (\fp \fq)^{r/2-1}| = 1$ for any chiral multiplet of $R$-charge $r$ with global symmetry fugacity $a$. 

Higgsing procedure is consistent with this prescription. Prior to Higgsing, all the flavor fugacities were assumed to be unimodular. 
But when we Higgs, the flavor fugacities are dressed with $\fp, \fq$ and quite often it contributes negative powers in $\fp \fq$. Superficially, this makes us think that some of the poles with $m=n=0$ are outside of the unit circle. As we have seen in the previous paragraph, due to the cancellation among the integrands, some of the poles are gone and the remaining poles under the Higgsing are those coming from the quarks in the Fan. But, note that all the Higgsed flavor fugacities $\vec{u} t^\Lambda_\s$ have to be unimodular. Even though superficially the poles appear to be outside of the unit circle, it is actually $a(\fp \fq)^{\frac{r-1}{2}}$ that has modulus 1 with $a$ being the flavor fugacity. Therefore, we have to include all the poles of the form $z=a (\fp \fq)^{r/2} $ even for negative or zero $r$.

\section{Conclusion}

We studied nilpotent Higgsing in $\mathcal{N}=1$ linear quiver gauge theories of class $\CS$.  In the case of $\mathcal{N}=2$ theories such Higgsing yields regular punctures that can be associated to quiver tails labelled by partitions of $N$.  Surprisingly, in $\mathcal{N}=1$ linear quiver gauge theories, such Higgsing yields a new type of quiver dubbed as the Fan.  This object is labelled by two integers $N$ and $N'$, and a partition of $N-N'$.  We provided further evidence of the Fan by ``Higgsing" the superconformal index.  

Armed with the Fan, we constructed many new SCFTs.  These provide various field theoretic descriptions of M5-branes wrapped on punctured Riemann surfaces.  Under Seiberg duality, quivers with Fans will transform to new quivers with different Fans.  Geometrically, this corresponds to different colored pair-of-pants decomposition of Riemann surface. Using the Fan, we find a new dual frame of $\CN=1$ $SU(N)$ SQCD with $2N$ flavors which is analogous to the Argyres-Seiberg duality. This dual frame is described by a $T_N$ theory coupled to the Fan and chiral multiplets. 

In our discussion, we only considered the UV curve with locally $\CN=2$ regular punctures. In $\CN=1$ class $\CS$ theories, one could have much more general punctures. In terms of generalized Hitchin system \cite{Xie:2013gma,Bonelli:2013pva}, we only considered the case where only one of two Hitchin fields become singular at a given puncture. It should be possible to consider the case where two Hitchin fields have singularities at the same point. This will yield genuinely $\CN=1$ punctures that we expect to be associated with a variation of the Fans. This is a work in progress. 

We hope to find an intersecting brane realization of these new SCFTs in type IIA string theory, which can be uplifted to M-theory. It will also be interesting to find a gravity dual description of the Fan and its variations in M-theory by using the system of \cite{Bah:2013qya}. This is also a work in progress. 

In this paper, we have not studied the detailed phase structure of the theory. The spectral curve approach from the generalied Hitchin system as done in \cite{Bonelli:2013pva, Xie:2013rsa} should be useful. 
It would be also interesting to identify the Fan for the $D, E$-type theories of class $\CS$, also possibly with outer-automorphism twists.

\acknowledgments
We would like to thank Rodrigo Alonso, Francesco Benini, Nikolay Bobev, Abhijit Gadde, Ken Intriligator, Jack Kearney, John McGreevy, Yuji Tachikawa, Hagen Triendl, Nick Warner, Brian Wecht and Dan Xie for various helpful discussions. We especially thank Nikolay Bobev who participated in the early stage of this work.
We also thank the referee for pointing out various grammatical mistakes and typos in the previous version of the paper and making numerous suggestions to improve the presentation. 
IB is grateful for the hospitality and work space provided by the UCSD Physics Department.  
KM, JS thank the hospitality of Simons Center for Geometry and Physics and the organizers of the 2014 Summer Simons Workshop in Mathematics and Physics. 
PA, IB, JS thank the hospitality of KITP and the organizers of the program New Methods in Nonperturbative Quantum Field Theory. 
PA thanks the organizers of PiTP-2014 for their kind hospitality. 
IB and JS also thank the Perimeter Institute for Theoretical Physics for the hospitality. 
KM would like to thank the organizers of Kavli IPMU-FMSP workshop ``Supersymmetry in Physics and Mathematics" for hospitality. KM would like to thank Nagoya University, Osaka City University, Rikkyo University, University of Tokyo, Komaba and Hongo, and Yukawa Institute for Theoretical Physics, for hospitality.
JS thanks the organizers of ``Exact Results in SUSY Gauge Theories in Various Dimensions" at CERN and also CERN-Korea Theory Collaboration funded by National Research Foundation (Korea), for the hospitality and support.  
The work of PA and JS is supported by DOE grant DOE-FG03-97ER40546. 
IB is supported in part by the DOE grant DE-FG03-84ER-26-40168, ANR grant 08-JCJC-0001-0, and the ERC Starting Grants 240210-String-QCD-BH, and 259133-ObservableString.
The work of KM is supported by a JSPS fellowship for research abroad.

\appendix

\section{The superpotential for the Fan} \label{sec:appFanW}

  We now derive the superpotential that is obtained after integrating out the massive modes in section \ref{subsec:derivation}. 
  Before integrating these out, the superpotential is given by
\be
W_1 = \tr q_0 \rho^+ \tilde{q}_0 + \tr q_0 M \tilde{q}_0 + \tr \tilde{\mu}_1q_0 \tilde{q}_0 \ ,
\label{eq:uvsup}
 \ee
  where $\rho^\pm= \rho(\sigma^{\pm})$. Here we write only those terms in the superpotential that are relevant to Higgsing.  
  Recall that $\rho^+$ here is also the raising operator for the $SU(2)$ embedding specified by the partition of $N$. Also, $\tilde{\mu}_1$ is the quark bilinear given by $\tilde{\mu}_1= \tilde{q_1} q_1 - \frac{1}{N}\tr q_1 \tilde{q_1}$.
 
Let $P$ and $\widetilde{P}$ be the projection matrices that project on to the massive modes of $q_0$ and $\tilde{q}_0$ respectively i.e.
\be
\begin{split}
&\chi = q_0 P \ , \\
&\widetilde{\chi} = \widetilde{P} \tilde{q}_0 \ ,
\end{split}
\ee 
where $\chi$ and $\widetilde{\chi}$ represent the massives chiral fields. It is easy to check that 
\be
\begin{split}
&\widetilde{P} = \rho^- \rho^+ \ , \\
&P= \rho^+ \rho^- \ .
\end{split}
\ee
These projection operators satisfy $\widetilde{P} \widetilde{P}= \widetilde{P}$ and $P P= P$ as is expected.  The massless modes are given by $Z = q_0 (\mathbb{1} - P)$ and $\widetilde{Z} = q_0 (\mathbb{1} - \widetilde{P})$.

The superpotential in \eqref{eq:uvsup} can now be expanded in terms of the massive and massless modes, such that the equation of motion for $\chi$ can be written as 
\be
\rho^+ \widetilde{\chi}  +  M \widetilde{\chi} +  \widetilde{\chi} \tilde{\mu}_1 + M \widetilde{Z} +   \widetilde{Z} \tilde{\mu}_1 = 0\ . 
\label{eq:eomchi}
\ee
Note that since $\tilde{\mu}_1$ in the above equation is contracted through the color indices, therefore it can be treated as a scalar multiplier in the above equation. This equation of motion can be simplified by multiplying it on the left with $\rho^-$ reducing it to the following form
\be
 \widetilde{\chi}  + \rho^- M \widetilde{\chi} +  \rho^-\widetilde{\chi} \tilde{\mu}_1 +\rho^- M \widetilde{Z} +  \rho^- \widetilde{Z} \tilde{\mu}_1 = 0\ . 
\ee
The solution for $\widetilde{\chi}$ is 
\be
\widetilde{\chi} = (\mathbb{1} - A)^{-1} A \widetilde{Z}\ ,
\ee
where 
\be
A= -(\rho^- M + \tilde{\mu}_1 \rho^- ) \ .
\ee
Recall that here we are treating $\tilde{\mu}_1$ as a scalar multiplier and will appropriately contract it using its color indices at a later stage. Notice that $A$ is a nilpotent matrix such that $A^\ell = 0$. This follows from the fact that $A^\ell \propto (\rho^-)^\ell (M+\tilde{\mu}_1 \mathbb{1})^\ell$ and $(\rho^-)^\ell=0$ since it is the lowering operator of $SU(2) \hookrightarrow SU(N)$. Here we have also used the commutation relation $[ \rho^-, M ] = 0$ which is due to the elements of $M$ being in the lowest weight state of their respective $SU(2)$ representations. Thus 
\be
\widetilde{\chi} = \sum_{n=1}^{\ell-1}  A^{n} \widetilde{Z}\ .
\ee
Substituting this back in \eqref{eq:uvsup} we find that the low energy superpotential is 
\be
W_{\textrm{eff}} &=& \Tr Z \widetilde{Z} \tilde{\mu}_1 + \Tr Z M \widetilde{Z} 
 + \sum_{n=1}^{\ell-1} \Tr Z MA^{n} \widetilde{Z} +  \sum_{n=1}^{\ell-1}  \Tr Z A^{n} \widetilde{Z} \tilde{\mu}_1 \ . 
\ee


\paragraph{An example for the $SU(6)$ quiver}

As an example of our previous derivation, let us study the nilpotent Higgsing of the linear quiver with $SU(6)$ symmetries. Consider the partition $6 \rightarrow 3+2+1$. This implies
\begin{equation}
\langle M_0 \rangle = \rho^+ =  \left( \begin{array}{cccccc}
0 & 1 & 0  & 0 & 0 & 0\\
0 & 0 & 1 & 0 & 0 & 0\\
0 & 0 & 0 & 0 & 0 & 0\\
0 & 0 & 0 & 0 & 1 & 0\\
0 & 0 & 0 & 0 & 0 & 0\\
0 & 0 & 0 & 0 & 0 & 0 
\end{array} \right) \ .
\end{equation} 
The components of the (anti-)quark matrices can be written as  
\begin{equation*}
q_0 = \left( 
\begin{array}{cccccc} 
\chi_1 & \chi_2 & Z_3 &\chi_3 & Z_2 & Z_1  
\end{array} \right) 
\hspace{0.5cm} \text{and} \hspace{0.5cm} \tilde{q}_0 = \left( \begin{array}{c} \widetilde{Z}_3 \\  \widetilde{\chi}_1 \\ \widetilde{\chi}_2 \\ \widetilde{Z}_2 \\ \widetilde{\chi}_3 \\ \widetilde{Z}_1 \end{array} \right) \ . 
\end{equation*}
with $\widetilde{\chi}_1$,  $\widetilde{\chi}_2$, $\widetilde{\chi}_3$, $\widetilde{Z}_1$, $\widetilde{Z}_2$,  and $\widetilde{Z}_3$ being row vectors, each of which corresponds to an anti-fundamental of $SU(6)_1$;  similarly,  $\chi_1$, $\chi_2$, $\chi_3$, $Z_1$, $Z_2$, and $Z_3$ are column vectors, each of which corresponds to a fundamental of $SU(6)_1$.  The vev for $M$ gives mass to $\widetilde{\chi}_1$,  $\widetilde{\chi}_2$, $\widetilde{\chi}_3$,  $\chi_1$, $\chi_2$ and $\chi_3$.  The fluctuations $M$ (around the vev $\rho^+$) that stay coupled to the theory are found by using the argument in \cite{Gadde:2013fma}. These are
\be
M = \left( \begin{array}{cccccc}
 M_{33}^2 & 0 & 0 & 0 & 0 & 0 \\
M_{33}^1 &  M_{33}^2 & 0 & {M}_{32}^1 & 0 & 0 \\
M_{33}^0 & M_{33}^1 &  M_{33}^2  & {M}_{32}^0  &  {M}_{32}^1 &  {M}_{31}^0    \\
M_{23}^1 & 0 & 0 & M_{22}^1 & 0 & 0 \\
M_{23}^0 & M_{23}^1 & 0 & M_{22}^0 & M_{22}^1 &  {M}_{21}^0 \\
M_{13}^0 & 0 & 0 & M_{12}^0 & 0 & -(3 M_{33}^2 + 2 M_{22}^1 ) \end{array} \right) \ .
\label{eq:SU(6)mes}
\ee
Upon integrating out the massive chiral fields and including the fluctuations \eqref{eq:SU(6)mes}, the effective superpotential becomes
\be
\begin{split}
W_{\textrm{eff}} ~=~  &  \tr \tilde{\mu}_1Z_1 \widetilde{Z}_1 -3 \tr Z_1 M_{33}^2 \widetilde{Z}_1 - 2  \tr Z_1 M_{22}^1 \widetilde{Z}_1 +  \tr Z_2 M_{22}^0 \widetilde{Z}_2 +  \tr Z_3 M_{33}^0 \widetilde{Z}_3   \\
& - \tr  Z_2 \widetilde{Z}_2 (\tilde{\mu}_1)^2- 2 \tr  Z_2 M_{22}^1 \widetilde{Z}_2 \tilde{\mu}_1 - \tr Z_2 (M_{22}^1)^2 \widetilde{Z}_2 + \tr Z_3 (M_{33}^2)^3 \widetilde{Z}_3  \\
&  + 3 \tr Z_3 (M_{33}^2)^2 \widetilde{Z}_3 \tilde{\mu}_1+ 3 \tr Z_3 M_{33}^2\widetilde{Z}_3 (\tilde{\mu}_1)^2 +  \tr Z_3 \widetilde{Z}_3 (\tilde{\mu}_1)^3 -  2 \tr Z_3 M_{33}^1 \widetilde{Z}_3 \tilde{\mu}_1\\
&  -2 \tr Z_3 M_{33}^1 M_{33}^2 \widetilde{Z}_3 - \tr Z_3 M_{23}^1{M}_{32}^1 \widetilde{Z}_3 + \tr Z_1 M_{12}^0 \widetilde{Z}_2 + \tr Z_1 M_{13}^0 \widetilde{Z}_3\\
& + \tr Z_2 {M}_{21}^0 \widetilde{Z}_1 + \tr Z_2 M_{23}^0 \widetilde{Z}_3 - \tr Z_2 M_{22}^1 M_{23}^1 \widetilde{Z}_3 - \tr Z_2  M_{33}^2M_{23}^1 \widetilde{Z}_3 -2 \tr Z_2 M_{23}^1 \widetilde{Z}_3 \tilde{\mu}_1\\
& + \tr Z_3 {M}_{31}^0 \widetilde{Z}_1 + \tr Z_3 {M}_{32}^0 \widetilde{Z}_2 - \tr Z_3 M_{22}^1 {M}_{32}^1 \widetilde{Z}_2 - \tr Z_3 M_{33}^2 {M}_{32}^1 \widetilde{Z}_2 -2 \tr Z_3 {M}_{32}^1 \widetilde{Z}_2 \tilde{\mu}_1\\
& + \tr \mu_2 \phi +  \tr \tilde{\mu}_2 \phi \ .
\end{split}
\ee
This matches exactly with what one would write for the Fan corresponding to the partition $6 \rightarrow 3+2+1$.

\section{Higgsing $\CN=2$ quiver theories} \label{sec:N2higgs}

Consider the linear quiver in $\mathcal{N}=2$ class $\mathcal{S}$ theories of type $A_{N-1}$ with the gauge group  
\be
G= \prod_{i=1}^{N-1} SU(N)_i \ .
\ee
The matter content of the theory consist of hypermultiplets $H_i = (Q_i, \widetilde{Q}_i)$ of $SU(N)_i \times SU(N)_{i+1}$. In addition to this we also have $N$ hypermulitplets $H_0 = (Q_0, \widetilde{Q}_0)$ transforming in the fundamental representation of $SU(N)_1$ and $N$ hypermultiplets $H_{N-1} = (Q_{N-1}, \widetilde{Q}_{N-1})$ transforming in the fundamental representation of $SU(N)_{N-1}$. Thus at each of the quiver there is an SU(N) flavor symmetry acting on the hypermultiplets $H_0$ and $H_{N-1}$ respectively.  We denote the flavor symmetry of $H_0$ by $SU(N)_0$ and that of $H_{N-1}$ by $SU(N)_{N}$. 

In order to avoid introducing too many indices labeling the symmetries under which $Q_i$ and $\widetilde{Q}_i$ transform, we will treat them as $N \times N$ matrices such that $Q_i\widetilde{Q}_i$ will be an invariant of $SU(N)_i$ while $\widetilde{Q}_iQ_i$ will be an invariant of $SU(N)_{i+1}$. Thus the superpotential of this quiver will be given by 
\be
W= \sqrt{2}\sum_{i=1}^{N-1}  \Tr \left( \widetilde{Q}_{i-1} \Phi _i Q_{i-1} -  Q_i \Phi_i \widetilde{Q}_i \right) \ . 
\ee

We now wish to consider an $SU(N)$ linear quiver and Higgsing its leftmost full puncture down to a puncture given by the Young's tableau  corresponding to the following partition of $N$
\be
N= n_1 + 2 n_2 + \hdots + \ell n_\ell \ .
\label{eq:part}
\ee
This breaks $SU(N)_0$ down to $S[U(n_1) \times U(n_2) \times \hdots U(n_\ell)]$. The corresponding vev for $\mu_0= \widetilde{Q}_0Q_0  - \frac{1}{N}\tr{\widetilde{Q}_0Q_0}$ that does the job for us is given by 
\be
\vev{\mu_0} = J_1^{\oplus n_1} \oplus  J_2^{\oplus {n_2}} \oplus  \hdots \oplus J_\ell^{\oplus n_\ell} \ ,
\ee
 where $J_k$ is the Jordan cell of size $k$. This can then be decomposed into the following vevs for $Q_0$ and $\widetilde{Q}_0$:
\be
\vev{\widetilde{Q}_0} = J_1^{\oplus n_1} \oplus  J_2^{\oplus {n_2}} \oplus  \hdots \oplus J_\ell^{\oplus n_\ell} \ ,
\ee
and
\be
\vev{{Q}_0} = J_1^{\oplus{n_1}} \oplus (J_1 \oplus I_1)^{\oplus^{n_2}} \oplus \hdots \oplus (J_1 \oplus I_{\ell-1})^{\oplus n_\ell} \ .
\ee
Here $I_k$ is the identity matrix of size $k$. It is straight forward to see that this breaks $SU(N)_1$ down to $SU(n_1 + n_2 + \hdots + n_k)$. The D-term constraints are trivially satisfied while the F-term for $\Phi_1$ gives us
\be
Q_0 \widetilde{Q}_0 -\frac{1}{N} \tr Q_0\widetilde{Q}_0 =  \widetilde{Q}_1 Q_1 -\frac{1}{N} \tr Q_1\widetilde{Q}_1   \ .
\label{eq:f1}
\ee
This chiral ring relation then forces us to have
\be
\vev{\widetilde{Q}_1 Q_1} = J_1^{\oplus{(n_1 + 2 n_2)} } \oplus (J_1 + J_2)^{\oplus{n_3} } \oplus \hdots \oplus (J_1 \oplus J_{\ell-1})^{ \oplus{n_\ell}} \ ,
\ee
which decomposes into 
\be
\vev{\widetilde{Q}_1} = J_1^{\oplus{(n_1 + 2 n_2)} } \oplus (J_1 + J_2)^{\oplus{n_3} } \oplus \hdots \oplus (J_1 \oplus J_{\ell-1})^{ \oplus{n_\ell}} \ ,
\ee
and
\be
\vev{{Q}_1} = J_1^{\oplus{(n_1 + 2 n_2)}} \oplus (J_1 \oplus J_1 \oplus I_1)^{\oplus{n_3}} \oplus \hdots \oplus (J_1 \oplus J_1 \oplus I_{\ell-2})^{\oplus{n_\ell} } \ ,
\ee
thereby breaking $SU(N)_2$ down to $SU(n_1 + 2n_2 + 2n_3 + \hdots +2 n_k)$.  Application of chiral ring relation at each node then gives us the general pattern of the vevs, which are found to be
\be
\begin{split}
\vev{\widetilde{Q}_{i-1} Q_{i-1}} =  &  J_1^{\oplus {(n_1 + 2 n_2 + \hdots + i n_i)} } \oplus \hdots \oplus ( J_1^{\oplus{(i-1)}} \oplus J_{k-i+1})^{\oplus n_k} \\
& \oplus \hdots \oplus ( J_1^{\oplus{(i-1)}} \oplus J_{\ell-i+1})^{\oplus{n_\ell}} \ ,
\end{split}
\label{eq:i-thvev}
\ee
such that
\be
\begin{split}
\vev{\widetilde{Q}_{i-1}} = &  J_1^{\oplus {(n_1 + 2 n_2 + \hdots + i n_i)} } \oplus \hdots \oplus ( J_1^{\oplus{(i-1)}} \oplus J_{k-i+1})^{\oplus n_k} \\
& \oplus \hdots \oplus ( J_1^{\oplus{(i-1)}} \oplus J_{\ell-i+1})^{\oplus{n_\ell}} \ ,
\end{split}
\ee
and 
\be
\begin{split}
\vev{Q_{i-1}} =& J_1^{ \oplus{(n_1 + 2 n_2 + \hdots + i n_i)}} \oplus\hdots \oplus( J_1^{\oplus{i}} \oplus I_{k-i})^{\oplus{n_k} } \\
&\oplus\hdots \oplus(J_1^{\oplus{i}} \oplus I_{\ell-i})^{\oplus{n_\ell} } \ . 
\end{split}
\ee
To check that these vevs do satisfy \eqref{eq:i-thvev} we use the rules that $J_k \cdot (J_1 \oplus I_{k-1}) = J_k$ and $(J_1 \oplus I_{k-1}) \cdot J_k=J_1 \oplus J_{k-1}$. The structure of these vevs imply that $SU(N)_i$ gets broken down to $SU(n_1 + 2n_2 + 3n_3 + \hdots + i n_i + i n_{i+1} \hdots + i n_k )$. Also $SU(N)_{\ell-1}$ gets broken down to $SU(N-n_l)$ while all the gauge groups from $SU(N)_\ell$ onwards remain unbroken. Thus we see that the gauge symmetry of the low energy theory obtained after Higgsing is given by 
\be
G' = \prod_{i=1}^{\ell-1} SU(N_i) \times \prod_{j=1}^{N-\ell} SU(N)_j \ ,
\ee
where $N_i = n_1 + 2n_2 + 3n_3 + \hdots + i n_i + i n_{i+1} \hdots + i n_\ell$. Apart from hypermultiplets $H_i$ transforming as the bifundamental of $SU(N_{i-1}) \times SU(N_i)$, there will be $m_i$ fundamentals at the gauge group $SU(N_i)$. Superconformality requires that
\be
m_i + N_{i-1} + N_{i+1} = 2 N_i \ ,
\ee
which then leads to $m_i = n_i$. This is coherent with the fact that the flavor symmetry of the Higgsed puncture corresponds to the symmetry associated with the additional $n_i$ fundamentals attached to $SU(N_i)$. 

\begin{figure}[h]
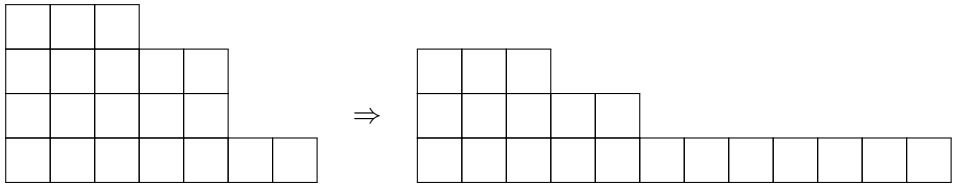

\centering
\be
\ytableausetup{aligntableaux=bottom}
\ydiagram{3,5,5,7} \quad \begin{ytableau}\none \\ \none \\  \none[\Rightarrow]   \\ \none \end{ytableau} \quad \ydiagram{3,5,12}
\nn
\ee
\caption{Collapsing of a Young tableau}
\label{fig:YoungTabCollap}
\end{figure}
Notice that the vev $\vev{\mu_i}= \vev{\widetilde{Q}_iQ_i}  - \frac{1}{N}\vev{\tr{\widetilde{Q}_iQ_i}}$ can be understood as the vev corresponding to partitioning $N$ as $N=(N_{i-1} + n_i) + 2 n_{i+1} + \hdots + (\ell-i+1)n_\ell $. The section of the quiver tail from the $i$-th node onward can then be thought of as being obtained from a linear $SU(N)$-quiver whose left puncture has been Higgsed according to this partition.  This implies that the propagation of vevs along the tail can also be neatly encoded into the process of collapsing the Young's tableau at each step. Thus if we start with the partition $N= n_1 + 2 n_2 + \hdots + \ell n_\ell$, then the Young's tableau at the next step in the quiver tail is obtained in the following manner: We remove the highest box from each column of boxes in the tableau. The boxes that were removed are stacked against  the residual tableau in a single row. For example if we consider the partition $20= 1+1+3+3+4+4+4 $, then at the next step in the quiver tail, its tableau collapses into the partition as described in figure \ref{fig:YoungTabCollap}. 

\paragraph{The massive and massless matter fields}

In order to obtain the number of fundamentals at the $i$-th node of the tail, we had invoked superconformality of the low energy theory, however, we should be able to derive this without resorting to an a priori assumption that the low energy theory is superconformal. To do this  we now focus on the various matter fields that get massive in the process of giving vevs. Once again we consider the case of partial Higgsing (given by the partition of $N$, as in \eqref{eq:part}) of a full-puncture of the $SU(N)$ linear quiver. We will make use of the following rules of decomposition:
\be
\begin{split}
SU(N)_i &\rightarrow SU(N_i) \\ 
{N} &\rightarrow {N_i} \oplus {1}^{\oplus{(N-N_i)}} \ ,\\
\textrm{adj} &\rightarrow \textrm{adj} \oplus {N_i}^{ \oplus^{(N-N_i)} }  \oplus {\bar{N}_i}^{\oplus{(N-N_i)}} \oplus {1}^{ \oplus{(N-N_i)^2}} \ .
\end{split}
\label{eq:vec_decomp_gen}
\ee
Also note that $H_{i-1}$ transforms as a bifundamental of $SU(N)_{i-1} \times SU(N)_i$ and can be decomposed into irreducible representations of $SU(N_{i-1}) \times SU(N_i)$ as
\begin{equation*}
\begin{split}
SU(N)_{i-1} \times SU(N)_i &\rightarrow SU(N_{i-1}) \times SU(N_i)\\
Q_{i-1} : ({\bar{N}}, {N}) &\rightarrow  ({\bar{N}_{i-1}}, {N_i}) \oplus ({\bar{N}_{i-1}}, {1})^{\oplus{(N-N_i)}} \oplus ({1}, {N_i})^{\oplus{(N-N_{i-1})}} \oplus ({1}, {1})^{\oplus{(N-N_i)(N-N_{i-1})}} \ , \\
\widetilde{Q}_{i-1} : ({N}, {\bar{N}}) &\rightarrow  ({N}_{i-1}, {\bar{N}_i}) \oplus ({N}_{i-1}, {1})^{\oplus {(N-N_i)}} \oplus ({1}, {\bar{N}_i})^{\oplus{(N-N_{i-1})}} \oplus ({1}, {1})^{\oplus{(N-N_i)(N-N_{i-1})} } \ .
\end{split}
\end{equation*}
From \eqref{eq:vec_decomp_gen} we see that upon Higgsing $SU(N)_i \rightarrow SU(N_i)$ via vevs for $H_{i-1}$ and $H_i$, the vector multiplets of $SU(N)_i$ that end up getting a mass will need to eat $2(N-N_i)$ chiral multiplets transforming as the ${N_i}$-dimensional representation of $SU(N_i)$. There are $(N-N_{i-1})$ such chirals in $H_{i-1}$ and $(N-N_i)$ such chirals in $H_i$. Thus we are left behind with $2(N-N_i) -(N-N_{i-1}) - (N-N_i) = n_i$ chiral super fields that transform as fundamentals of $SU(N_i)$. We will similarly be left with $n_i$ chiral multiplets transforming as the anti-fundamental of $SU(N_{i})$. These will together give us $n_i$ hypers transforming in the fundamental of $SU(N_i)$.  We also end up eating some of the singlets. The number of singlet hypers that are left behind (these are the hypers that decouple from the rest of the quiver) is then given by
\be
\sum_{i=1}^{k} (N-N_i)(N_i - N_{i-1}) \hspace{1 cm} \textrm {where $N_0 = 0$} \ .
\ee
These decoupled hypers are the Goldstone multiplets that we expect upon spontaneously breaking the global symmetry. It can be easily checked that the number of the Goldstone chiral superfields in these hypers is same as the number of generators of the complexified $SU(N)$ that are broken by $\vev{\mu}$ i.e. the Goldstone chiral superfields  are in one-to-one correspondence with the generators $X$ of $SL(N,\mathbb{C})$ which obey
\be
[X,\vev{\mu_0}] \neq 0 \ .
\ee
Apart from these there will of course be massless hypers that transform as bifundamentals of $SU(N_{i-1}) \times SU(N_i)$. We thus obtained the desired low energy quiver.

As an explicit example of the above pattern of massive and massless matter fields, we consider an $SU(4)$ linear quiver and Higgs its left full-puncture down to a simple puncture. We give appropriate vevs to $H_0$ and $H_1$, Higgsing $SU(4)_1 \times SU(4)_2$ down to $SU(2) \times SU(3)$. The decomposition of vector multiplets into irreps. of the low energy gauge symmetry is given by 
\be
\begin{split}
 SU(4)_1 \times SU(4)_2 & \rightarrow SU(2) \times SU(3) \\
 V_1 : (\textrm{adj},{1}) &\rightarrow  (\textrm{adj},{1}) \oplus  ({2},{1}) \oplus ({2},{1}) \oplus ({\bar{2}},{1})  \oplus ({\bar{2}},{1})   \oplus ({1},{1})^{ \oplus 4} \ , \\
 V_2: ({1}, \textrm{adj}) &\rightarrow ({1}, \textrm{adj}) \oplus ({1},{3}) \oplus ({1},{\bar{3}}) \oplus ({1},{1}) \ ,
 \end{split} 
\ee
while the hypers $H_0$ and $H_1$ decompose as
\be
\begin{split}
 SU(4)_1 \times SU(4)_2 & \rightarrow SU(2) \times SU(3) \\
 (Q_0)_i : ({4}, {1}) &\rightarrow ({2},{1}) \oplus ({1},{1})^{\oplus 2} \ ,\\
  (\widetilde{Q}_0)_i : ({\bar{4}}, {1}) &\rightarrow ({\bar{2}},{1}) \oplus ({1},{1})^{\oplus 2}\ , \\
  Q_1: ({\bar{4}},{4}) &\rightarrow  ({\bar{2}},{3}) \oplus ({1},{3})^{\oplus 2} \oplus  ({\bar{2}},{1}) \oplus ({1},{1})^{\oplus 2} \ ,\\
  \widetilde{Q}_1: ({4},{{4}}) &\rightarrow  ({2},{\bar{3}}) \oplus ({1},{\bar{3}})^{\oplus 2} \oplus  ({2},{1}) \oplus ({1},{1})^{\oplus 2} \ ,
\end{split} 
\ee
The various chiral multiplets that get eaten via Higgsing are: 4 copies transforming as $({2},{1})$, 4 copies of $({\bar{2}},{1})$, 2 copies each of $({1},{3})$ and $({1},{\bar{3}})$ and 10 copies of $({1},{1})$. We are thus left behind with a chiral multiplet for each of $({2},{1})$, $({\bar2},{1})$, $({2},{\bar{3}})$ and $({\bar{2}},{3})$ along with 10 chirals which are singlets and hence decouple from the rest of the theory. These can then be organized as a hyper transforming in the fundamental of $SU(2)$, another hyper transforming as the bifundamental of $SU(2) \times SU(3)$ and 5 decoupled hypers.

\bibliographystyle{ytphys}
\bibliography{refs}

\end{document}